\documentclass[aps,pra,onecolumn,superscriptaddress,showpacs,preprintnumbers,amsmath,amssymb]{revtex4-2}
    \usepackage[utf8]{inputenc}
    \usepackage[english]{babel}
    \newtheorem{theorem}{Theorem}

    \usepackage{float}

    \usepackage{setspace}
    \usepackage{booktabs}
    \usepackage{multirow}
    \usepackage{amsmath,amssymb}
    \usepackage{qcircuit}
    \usepackage{lipsum}
    \usepackage{tabularx}
    \usepackage{natbib}
    \usepackage{stfloats}
    \usepackage{graphicx}
    \usepackage{rotating}
    \usepackage{ulem}

    \usepackage{makecell}
    \setcellgapes{2.3pt}

    \usepackage{physics}

    \usepackage{xcolor}
    \usepackage{subcaption}
    \usepackage{hyperref}

    \usepackage{adjustbox,expl3,etoolbox}
    \letcs\replicate{prg_replicate:nn}
    \newcommand{\tabgap}{\makebox[0.55cm]{}}
    
    \makeatletter
\newcommand{\sectionnotoc}[1]{%
  \begingroup
  \def\addcontentsline##1##2##3{}%
  \section{#1}%
  \endgroup
}

\newcommand{\starsectionnotoc}[1]{%
  \begingroup
  \def\addcontentsline##1##2##3{}%
  \section*{#1}%
  \endgroup
}

\newcommand{\subsectionnotoc}[1]{%
  \begingroup
  \def\addcontentsline##1##2##3{}%
  \subsection{#1}%
  \endgroup
}

\makeatother
    
\begin{document} 
    
    \title{Analog photonic simulator for large-scale transport}


\author{Mengyu Zhao}
\altaffiliation{These authors contributed equally to this work.}
\affiliation{State Key Laboratory of Quantum Optics Technologies and Devices, Institute of Opto-Electronics, Shanxi University, Taiyuan 030000, China}
\author{Xuezhi Zhu}
\altaffiliation{These authors contributed equally to this work.}
\affiliation{State Key Laboratory of Quantum Optics Technologies and Devices, Institute of Opto-Electronics, Shanxi University, Taiyuan 030000, China}
\author{Nikita Guseynov}
\altaffiliation{These authors contributed equally to this work.}
\affiliation{Global College, Shanghai Jiao Tong University, Shanghai 200240, China}
\author{Yewei Yuan}
\altaffiliation{These authors contributed equally to this work.}
\affiliation{Global College, Shanghai Jiao Tong University, Shanghai 200240, China}
\author{Na Wang}
\affiliation{State Key Laboratory of Quantum Optics Technologies and Devices, Institute of Opto-Electronics, Shanxi University, Taiyuan 030000, China}
\author{Meihong Wang}
\affiliation{State Key Laboratory of Quantum Optics Technologies and Devices, Institute of Opto-Electronics, Shanxi University, Taiyuan 030000, China}
\affiliation{Collaborative Innovation Center of Extreme Optics, Shanxi University, Taiyuan 030000, China}
\author{Yunyun Cao}
\affiliation{State Key Laboratory of Quantum Optics Technologies and Devices, Institute of Opto-Electronics, Shanxi University, Taiyuan 030000, China}
\author{Shi Jin}
\affiliation{Institute of Natural Sciences, Shanghai Jiao Tong University, Shanghai 200240, China}
\affiliation{School of Mathematical Sciences, Shanghai Jiao Tong University, Shanghai 200240, China}
\author{Nana Liu}
\email{nana.liu@quantumlah.org}
\affiliation{Global College, Shanghai Jiao Tong University, Shanghai 200240, China}
\affiliation{Institute of Natural Sciences, Shanghai Jiao Tong University, Shanghai 200240, China}
\affiliation{School of Mathematical Sciences, Shanghai Jiao Tong University, Shanghai 200240, China}
\author{Changde Xie}
\affiliation{State Key Laboratory of Quantum Optics Technologies and Devices, Institute of Opto-Electronics, Shanxi University, Taiyuan 030000, China}
\author{Kunchi Peng}
\affiliation{State Key Laboratory of Quantum Optics Technologies and Devices, Institute of Opto-Electronics, Shanxi University, Taiyuan 030000, China}
\author{Xiaolong Su}
\email{suxl@sxu.edu.cn}
\affiliation{State Key Laboratory of Quantum Optics Technologies and Devices, Institute of Opto-Electronics, Shanxi University, Taiyuan 030000, China}
\affiliation{Collaborative Innovation Center of Extreme Optics, Shanxi University, Taiyuan 030000, China}

\begin{abstract}
Transport equations describe how physical quantities -- such as mass, energy, momentum, concentration, probability, or fields -- are carried, propagated, or redistributed through space and time, forming a foundational class of partial differential equations across science and engineering. However, high-dimensional partial differential equations are difficult to represent on digital grids because the number of degrees of freedom grows exponentially with dimension. Continuous-variable quantum photonics on the other hand can represent and evolve these large-scale fields without first discretizing space into a discrete grid. We demonstrate a large-scale analog photonic simulator for the constant-coefficient advection equation, a transport equation that is a fundamental benchmark for scientific computing. The solution of a $d$-variable advection equation is encoded into $d$ optical modes, so that the partial differential equation evolution maps directly to programmable phase-space displacements generated by optical quadrature momenta. Using a time-domain continuous-variable quantum photonic platform, we validate programmable control with $20,000$ single-mode squeezed states and $20,000$ two-mode squeezed states, and implement transport dynamics on a $20,000$-mode cluster-state resource. Homodyne measurements then verifies mode-resolved displacement control,  which can provide first and second-order moment information of the solution to the advection equation, with final achievable relative error as low as $0.8\%$ and $0.92\%$ for first and second-order moment observables respectively. Our results establish continuous-variable photonics as a suitable programmable
analog platform for large-scale advection equations. 
    \end{abstract}

     \maketitle
    
    \sectionnotoc{Introduction}
    Partial differential equations (PDEs) serve as a fundamental language underpinning the laws of nature. They govern the spatiotemporal evolution of diverse physical systems, with applicability spanning from the quantum realm to vast cosmological expanses, as well as complex human-engineered systems, and they remain the cornerstone of modern scientific computing. They are also substantially more difficult to simulate than ordinary differential equations, where a single scalar PDE can be considered as an infinite number of scalar ordinary differential equations. PDEs are notoriously difficult to solve when the number of variables is large \cite{han2018solving,bungartz2004sparse,sirignano2018dgm}. A digital computer must usually represent the solution on a grid, and the number of grid points grows exponentially with dimension. This “curse of dimensionality” limits simulations in wide-ranging applications from finance, climate modelling to machine learning.
    
   Quantum computers, on the other hand, have been proposed as an alternative computational paradigm to help address this curse of dimensionality. While in theory, this does have the potential to alleviate the classical curse of dimensionality problem \cite{childs2021high, jin2024quantum, an2023linear, lubasch2025quantum,hu2024quantum,sato2024hamiltonian}, it is still difficult to demonstrate this in practice. One reason is that most quantum algorithms for PDEs rely on first discretising space and time, and use digital algorithms based on qubit-based devices. In these cases, each simulation of a PDE require the application of a large number of error-corrected gates, which currently forbids successful direct simulation even for the simplest equation like the advection equation. Future fault-tolerant quantum computers can obviate the issue, but despite tremendous progress in recent years, the current status of digital quantum computers is still not capable of large-scale error-correction or fault tolerance \cite{google2025quantum, sales2025experimental,bluvstein2026fault,butt2026demonstration}.
    
    A different route is to represent the function physically using continuous-variable (CV) quantum optical systems, where the information can instead be encoded in an infinite-dimensional Hilbert space with a continuum of eigenvalues \cite{vanloockQIwithCV,GaussianQI,QStheoryCV,Andersen2015hybrid}. The quadratures of light are continuous degrees of freedom, and many optical modes can be generated, controlled, and measured \cite{su2013gate,2014_cluster_PfisterO,2017_cluster_TrepsN,2019_2D_cluster_Furusawa,2019_2D_cluster_Andersen,2025_cppd_cluster,2025_3D_cluster,2026_path_CVcluster,2025_QML_AndersenUL}. A photonic device can therefore store an analog state directly in its optical field and evolve that state through physical transformations, rather than first discretizing the state into qubits or classical grid points. The CV quantum optical device can also leverage deterministic generation of entangled quantum states and encode the solution of the partial differential equations in the quadratures of the bosonic modes. 
    
We demonstrate this approach by using the advection equation as a starting point. It is one of the most basic components in applied mathematics and appears as a building block in fluid dynamics, atmospheric and ocean transport, plasma physics, traffic models, conservation laws, and wave propagation. Although the constant-coefficient advection has a simple analytical solution and it is only effective by itself in ballistic, drift-dominated, or short-time regimes, it provides a clean and important benchmark. The task for the CV quantum platform is then the preparation of a CV quantum state whose amplitudes correspond to the solution of a large-scale advection equation, where the first and second-moments of the solution can be extracted from the resulting quantum state. The evolution itself, describing the transport, can be implemented directly by optical displacements. 

To establish the system's capability for this large-scale quantum simulation, we first validate time-bin-resolved Gaussian-state preparation and synchronized manipulation in the time domain by using 20,000 single-mode squeezed states and 20,000 two-mode squeezed states, and apply variable displacement on each mode. These correspond to simulations of ensembles of 20,000 independent one-dimensional advection equations and ensembles of 20,000 independent two-dimensional advection equations, respectively. Next, we take a time-domain multiplexed CV cluster state with 20,000 entangled modes and precisely control the displacement on each quantum mode (qumode) and perform homodyne detection. We note that although the entanglement structure is not fully connected, it forms a continuous and irreducible chain and the covariance matrix confirms a banded structure, where we show that the measured quantum correlations can extend across four modes. Unlike the one and two-mode squeezed initial states, the 20,000-mode CV cluster state is a multimode connected resource, thus providing a direct physical implementation of a large-scale advection equation with sparsely correlated initial conditions. We then extract the first and second-order moments which can be mapped to important observables of the PDE solution. We then show that the resulting moments can be extracted from the quantum state to high precision, with relative errors as low as $0.8 \%$ and $0.92\%$ for the first and second-order moments respectively. On the other hand, we find that the cost of a comparable simulation task using qubit-based quantum implementations would require long sequences, $O(10^6)$ or larger, of error-corrected one- and two-qubit gates, which are beyond the capability of current digital quantum devices. 

The present experiment is a proof of principle demonstration, to establish that current non-fault-tolerant and non-error-corrected quantum devices are already capable of simulating certain large-scale PDEs like advection equations with Gaussian state initial conditions. Although it currently treats the constant-coefficient advection equation, for which the required evolution is Gaussian, more general PDEs will require additional ingredients, including mode coupling, nonunitary embeddings, and eventually non-Gaussian operations. Nevertheless, this work provides an important benchmark and starting point towards a scalable analog quantum photonic route for representing and evolving large-scale transport-based PDEs. 

\begin{figure}[H]
    \centering
    \includegraphics[width=\linewidth]{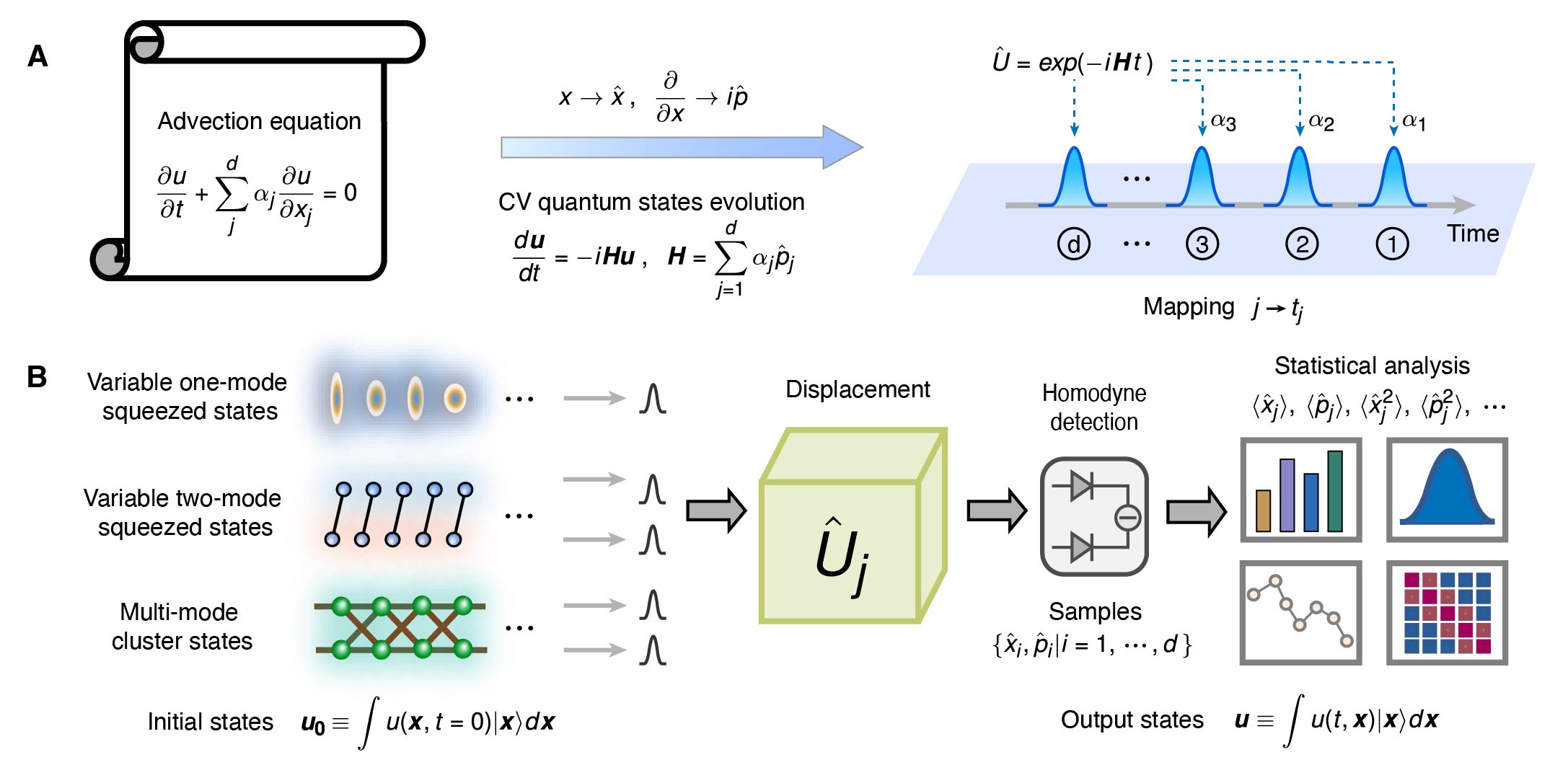}
    \caption{\textbf{Conceptual diagram and experimental procedure for simulating the advection equation on a continuous-variable quantum platform.} (A) The classical advection equation is mapped to an initial quantum state under displacement operations that act on the amplitude quadrature of the quantum states. Each dimension corresponds to an independent quantum mode (qumode). (B) Experimental workflow: the experiment begins with the preparation of the initial Gaussian quantum resources. Then controlled, variable displacement operations are applied to every quantum mode. The evolved quantum states are subsequently measured by homodyne detection. Finally, data acquisition and statistical analysis are employed to extract the first and second-order moments for each mode, enabling the reconstruction of important observables of solutions of the advection equation.}  
    \label{Fig1}
\end{figure}

\medskip\bigskip
\sectionnotoc{Analog photonic simulation of advection equation}
We consider the $d$-dimensional advection equation that describes the transport of the scalar quantity $u(t, x)$, with time $t \in [0, T]$ and spatial coordinates $x=(x_1, \cdots, x_d) \in \mathbb{R}^d$, written in the form
\begin{align} \label{eq:advection}
    \frac{\partial u(t,x)}{\partial t}+\sum_{j=1}^d \alpha_j \frac{\partial u(t,x)}{\partial x_j}=0, \qquad u_0(x)=u(0,x),
\end{align}
where $\mathbf{\alpha}=(\alpha_1, \cdots, \alpha_d) \in \mathbb{R}^d$ represents the velocity with which $u(t,x)$ is transported and $u_0(x)$ is the initial condition. Without losing generality, we can choose the $l_2$ norm $\|u_0\|=1$ since transport itself does not change this norm. We encode the field as a $d$-mode CV quantum state $|u(t)\rangle=\int u(t,x) |x\rangle dx$ also called a qumode, where $\{|x\rangle \}_{x \in \mathbb{R}^d}$ are position eigenstates. In the formalism, we can rewrite Eq.~\eqref{eq:advection} in the form $d |u(t)\rangle/dt+i\sum_{j=1}^{d} \alpha_j \hat{p}_j|u(t)\rangle=0$, where we can make the replacement $\partial/\partial x_j \rightarrow i \hat{p}_j$ and $\hat{p}_j$ is the momentum operator acting on the $j^{\text{th}}$ qumode \cite{jin2024analog}. This means that when given the $d$-qumode state $|u_0\rangle$ representing the initial condition of the advection equation, we can directly simulate $|u(t)\rangle$ by the operation $|u(t)\rangle=U(t)|u_0\rangle$, $ U(t)=\bigotimes_{j=1}^d U_j(\alpha_jt)$, $U_j(\alpha_jt)=e^{-i \hat{p}_j \alpha_jt}$, where the unitary operation $U_j(\alpha_jt)=e^{-i \hat{p}_j \alpha_jt}$ is known as the displacement operation, widely used in CV quantum optics. Since large-scale qumodes and their displacements are feasible to high accuracy, this suggests that high-$d$ advection equations can be naturally suited to CV quantum optics (Fig. ~\ref{Fig1}A).\\

In this demonstration, we consider three classes of initial states $|u_0\rangle$ (Fig. ~\ref{Fig1}B): (a) 20,000 qumode system with variable one-mode squeezing; (b) 40,000 qumode system with variable two-mode squeezing; (c) 20,000 qumode CV cluster state. By applying variable displacements to each qumode, we implement the corresponding $d$-variable dynamical evolution and obtain the output states $|u(t)\rangle$. Once the state $|u(t)\rangle$ is simulated, we can estimate observables associated with the solution of the advection equation $u(t,x)$ by measuring first-order moments like $\langle \hat{x}_j \rangle=\langle u(t)|\hat{x}_j |u(t)\rangle$ and second-order moments like $\langle \hat{x}_i \hat{x}_j \rangle=\langle u(t)| \hat{x}_i \hat{x}_j |u(t)\rangle$. For example, the  physical transported quantity is usually a concentration $q(t,x)$, which satisfies the same advection equation as $u(t,x)$  where we can use the encoding $u(t,x)=\sqrt{q(t,x)}$. Here  $\langle \hat{x}_j \rangle=\int x_j q(t,x) dx$ can correspond to mean collective coordinate or carrier density and $\langle \hat{x}_i \hat{x}_j \rangle=\int x_i x_j q(t,x) dx$ can correspond to mean carrier spatial spread or correlated conformal motion in particle transport models. A different encoding $u(t,x)=q(t,x)$ is natural for example in scenarios where $|q(t,x)|^2$ corresponds to a probability density, a signal intensity or energy concentration, so $\langle \hat{x}_j \rangle=\int x_j |q(t,x)|^2 dx$ and second-order moments like $\langle \hat{x}_i \hat{x}_j \rangle=\int x_i x_j |q(t,x)|^2 dx$ can correspond to the energy-centroid or energy-spread covariance. In our experiment with the $20,000$ multimode cluster state and displacement $|\alpha_j| \in [2.5, 20]$ (setting $t=1$ for simplicity), the relative error for $\langle \hat{x}_j \rangle$ ranges from $0.8\%$ to $7\%$, and the relative error for $\langle \hat{x}^2_j\rangle \in [52.57, 400]$ ranges from $0.92\%$ to $7\%$. We also observe non-negligible correlations $\langle \hat{x}_i \hat{x}_j\rangle$ across four modes or less. See Section II.B for more details. 


\medskip\bigskip
\subsectionnotoc{Scalable photonic platform}

\begin{figure}[t]
    \centering    
    \includegraphics[width=\linewidth]{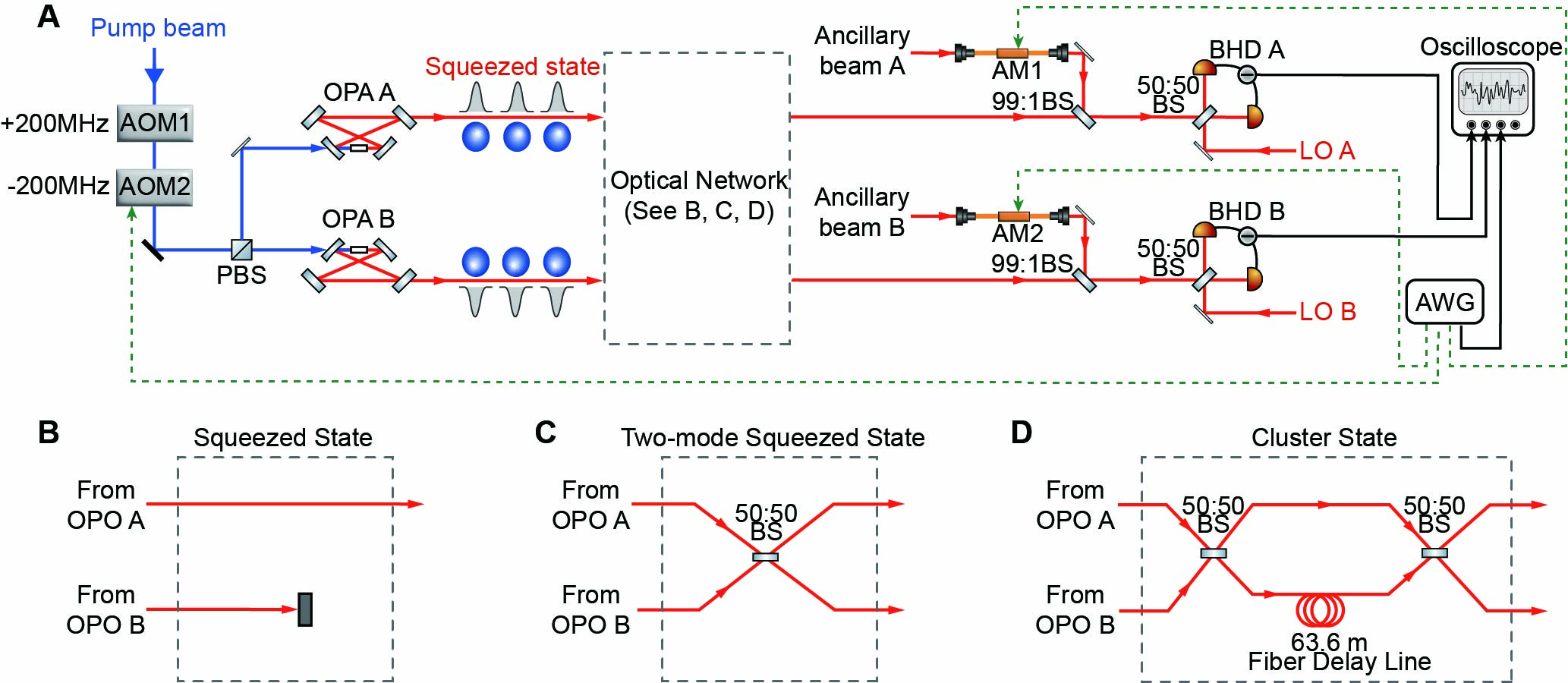}
    \caption{\textbf{Experimental setup of the continuous variable photonic platform.} (A) Main experimental setup. Two optical parametric oscillators (OPA A and OPA B) generate squeezed state sequences. The pump beam is modulated by cascaded acousto-optic modulators (AOMs), enabling dynamic control of the squeezing parameter. The generated squeezed states are injected into a reconfigurable optical network to generate the required quantum resources. Next, two amplitude-modulated (AM) ancillary beams are coupled with the signal beams to implement displacement operations. Finally, the output states are measured by two balanced homodyne detectors (BHDs). All modulators are driven by an arbitrary waveform generator (AWG). (B) Configuration for single-mode squeezed states. (C) Configuration for two-mode squeezed states. The outputs from OPA A and OPA B interfere on a beam-splitter to generate entangled modes. (D) Configuration for the cluster state. The squeezed states from OPA A and OPA B first interfere on a beam-splitter to generate the two-mode squeezed state. One output arm propagates through a fiber delay line to introduce a fixed temporal delay, and subsequently interferes again with the other output on a second 50:50 beam-splitter to generate a CV cluster state.}
    \label{Fig2}
\end{figure}

In our experiment, the initial states $|u_0\rangle$ of the advection equation are prepared with optical parametric amplifiers (OPAs), as illustrated in Fig. ~\ref{Fig2}A. Programmable control of the squeezing parameters is achieved by modulating the pump amplitudes of the OPAs with a cascaded acousto-optic modulation scheme (see Appendix~\ref{app:single_mode_squeezed} for details). Driven by arbitrary waveform generators (AWGs), the squeezing parameter can be dynamically programmed in each temporal mode. These squeezed states are subsequently routed through a reconfigurable optical network to generate different quantum resources. We prepare single-mode and two-mode squeezed states in the time domain with a wave packet width of 400 $\mu s$ (Fig. ~\ref{Fig2}B and C). These two configurations are used to verify the successful preparation of large-scale programmable temporal-mode sequences and the synchronization of the system. Furthermore, we generate a time-domain CV cluster state by implementing a time delay on one mode of the two-mode squeezed state and re-coupling the optical modes on another 50:50 beam-splitter (Fig. ~\ref{Fig2}D). The resulting cluster state contains 10,000 temporal modes with a 318 ns wave packet width, serving as the core resource for our simulation. 

The dynamical evolution of the advection equation is simulated by applying programmable displacement operations to the amplitude quadrature of individual qumodes. Experimentally, these operations are implemented by encoding uniformly distributed random amplitudes onto the ancillary beams by the amplitude modulators (AMs) and coupling the modulated ancillary beams with the time-domain qumodes on a 99:1 beam-splitter (99:1BS). By programming the amplitude of the ancillary beam, we can dynamically achieve arbitrary displacement values. The output states are measured by balanced homodyne detectors (BHDs) to extract expectation values that directly correspond to the solutions of the equation.  

In the simulation of the advection equation with squeezed states with variable squeezing as the initial condition, the greatest challenge comes from the synchronization of varying the squeezing, time bin, displacements and measurement systems, since the experiment requires up to 20,000 time bins. For the programmable single-mode and two-mode squeezed states, where each time bin has a relatively long duration (400 µs), synchronization is achieved by distributing a synchronized trigger signal from the AWGs to the acousto-optic modulators, amplitude modulators and the data acquisition system. This ensures that the displacement sequence is strictly synchronized  with the application of variable squeezing. Under this condition, timing mismatches at the nanosecond level have a negligible impact on the experiment. The requirement of synchronization is particularly stringent for the multi-mode entangled state since the width of wave packet is 318 ns. We perform a nanosecond-level timing calibration between the data acquisition signal and the displacement control signal by strictly referring to a 10 MHz clock signal. By carefully tuning the electronic delay between the displacement signal and oscilloscope, each displacement pulse is precisely aligned with its corresponding time bin, as shown in Appendix~\ref{app:temporal_synchronization}.

\medskip\bigskip
\subsectionnotoc{State preparation and displacement} \label{sec:stateprep}
Here we prepare three classes of $|u_0\rangle$: (a) 20,000 qumode system with variable one-mode squeezing. The single-mode configuration establishes time-bin resolved programmability of the initial Gaussian resource; (b) 40,000 qumode system with variable two-mode squeezing. The two-mode configuration shows that the pump modulation, spatial mode coupling, displacement, and readout remain synchronized for entangled resources; (c) 20,000 qumode multimode cluster state. 

\begin{figure}[htbp]
    \centering
    \includegraphics[width=18.4 cm]{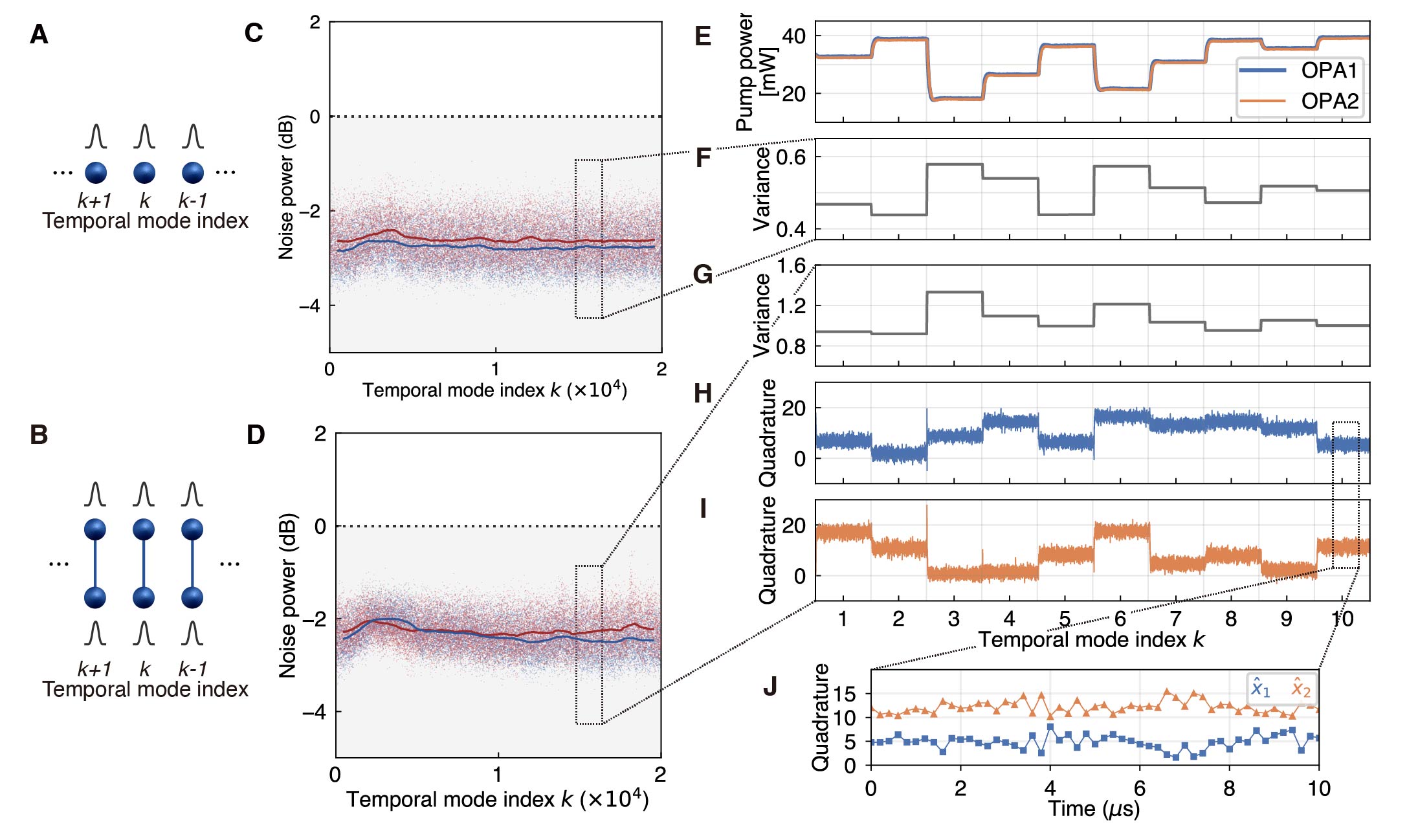}
    \caption{\textbf{Measurements on time-bin-resolved single-mode and two-mode squeezed states.} (A,B) Graphical representation of the single-mode (A) and two-mode (B) squeezed state sequences. (C,D) Noise power of the amplitude quadrature $\hat{x}$ for the single-mode squeezed states (C) and $\hat{x}_1+\hat{x}_2$ for the two-mode squeezed states (D) before and after displacement. All modes remain below the shot noise level. The red and blue scatter points represent the experimental data before and after displacement, respectively. Solid lines represent the corresponding moving averages. (E-I) Time-bin resolved sequences of modulated pump power, variance reflecting the single-mode squeezing parameter, variance reflecting the two-mode entanglement strength, amplitude quadrature measured by BHD A, and amplitude quadrature measured by BHD B. By modulating the pump power of the two OPAs, the squeezing parameter and the entanglement strength can be independently controlle across all temporal modes. Notably, all signals are precisely time-aligned. (J) Amplitude quadrature evolution within one time bin of the two-mode squeezed state. The blue and yellow curves correspond to the amplitude quadratures of two entangled modes after displacement, where clear anti-corrections are observed.}
    \label{EPR}
\end{figure}

 In the single-mode configuration (Fig.~\ref{EPR}A), the modulated pump sequence remains aligned with the measured squeezing parameter (Fig.~\ref{EPR}E,F). We dynamically control the squeezing parameter by manipulating the pump power, demonstrating the control of the initial time-bin resource. The programmed squeezing level ranges from 2.1 to 3.3 dB under random uniformly distributed pump amplitude, (Fig.~\ref{EPR}C). We then extend this control to two-mode entangled resources (Fig.~\ref{EPR}B). The pump sequence, spatial mode coupling, displacement and the readout from BHD A and B remain synchronized for the two-mode squeezed state (Fig.~\ref{EPR}E,G,H,I). This synchronization is also evidenced by the evolution of the quadrature within a representative time bin after displacement (Fig. \ref{EPR}J). To quantify the resulting entanglement, we calculate the noise power of the quadrature combinations $\hat{x}_1 + \hat{x}_2$. The entanglement strength is programmed between 1.8 and 2.7 dB throughout the sequence (Fig.~\ref{EPR}D). These measurements demonstrate synchronized control across all operations, which can be directly extended to large-scale entangled resources.

\begin{figure}[htbp]
    \centering
    \includegraphics[width=18.4 cm]{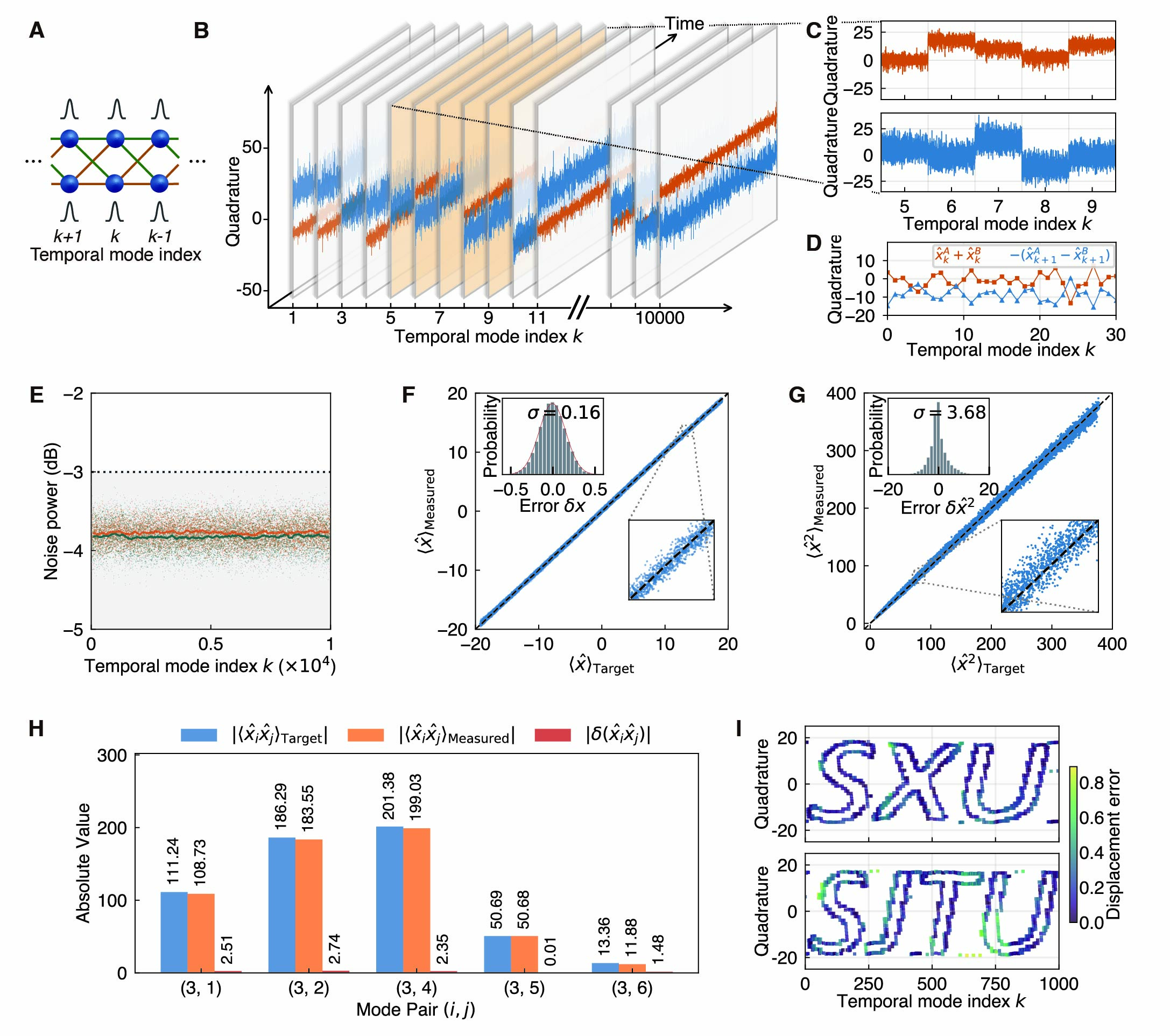}
    \caption{\textbf{Results of the analog photonic simulation with CV cluster state.} (A) Graphical representation of the CV cluster state. (B) The quadrature output after displacement crosses 10,000 temporal modes. The green and orange lines correspond to the amplitude quadrature of the two rails. (C) Amplitude quadrature values measured by BHD A and B over five consecutive temporal modes. (D) Amplitude quadrature combination evolution within one time bin. (E) $\langle  (\hat{x}_k^{A} + \hat{x}_k^{B} + \hat{x}_{k+1}^{A} - \hat{x}_{k+1}^{B})^2 \rangle$ before and after displacement. The green and orange scatter points represent the experimental data before and after displacement, respectively. Solid lines represent the corresponding moving averages. (F)  Displacement error of first-moment $\langle \hat{x}_j \rangle$ across 20,000 uniformly random displacements. The experimental data (blue scatter points) agree well with the target (black dashed line). The lower right inset provides a magnified view. The upper left insert displays the probability distribution of the displacement error $\delta x$—defined as the difference between the target displacement and the measured displacement, which presents the standard deviation error of $\sigma \approx 0.16$. (G) Displacement error of second-moment $\langle \hat{x}_j^2 \rangle$ across 20,000. (H) Correlations $\langle \hat{x}_i \hat{x}_j \rangle$ and absolute errors $|\delta(\hat{x}_i \hat{x}_j)|$ for representative mode pairs involving $i=3$. (I) Amplitude quadrature values measured by BHD A and B after patterned displacement on a 2000-mode cluster-state resource, where the colour indicates the corresponding displacement error.}
    \label{Cluster}
\end{figure}

Next, we apply this control scheme to the time-domain CV cluster state. Its resource structure is shown schematically in Fig.~\ref{Cluster}A, features a dual-rail entangled structure. The correlations can be verified by calculating the noise power of the nullifiers $\hat{X}_k = \hat{x}_k^A + \hat{x}_k^B + \hat{x}_{k+1}^A - \hat{x}_{k+1}^B$
and $\hat{P}_k = \hat{p}_k^A + \hat{p}_k^B - \hat{p}_{k+1}^A + \hat{p}_{k+1}^B$, 
where superscripts A and B denote the two spatial modes measured by two homodyne detectors, and the subscript $k$ denotes the index of the temporal mode \cite{2011_nullifier_vLP,2021_cluster_nullifier_prx}. Experimentally, the successful generation of these specific structures is confirmed by demonstrating that the measured nullifier variances are well below -3 dB noise power \cite{2003_genuine_cluster}. The key measured output of the cluster state are mode-resolved amplitude-quadrature values obtained from a 20,000-mode entangled resource (Fig.~\ref{Cluster}B). The 20,000-modes of the cluster state consists of 10,000 time bins, where there are two spatial modes per time bin. Once the output state $|u(t)\rangle$ is prepared, its corresponding first and second-order moments provide observables associated with the solution of the advection equation. We apply uniformly distributed random displacements to the amplitude quadrature of each qumode, demonstrating a programmable and mode-resolved simulation of the system. The readout from BHD A and B over five consecutive time bins shows how the output state is reconstructed from the measurements (Fig.~\ref{Cluster}C).

The success of the simulation also relies on whether the quantum state retains its non-classical properties after the displacement operation. The measured quantum correlations of the first thirty wave packets (Fig.~\ref{Cluster}D) and nullifiers across all temporal modes (Fig.~\ref{Cluster}E) demonstrate that the entanglement remains observable after displacement, as expected. This confirms that the simulated output in Fig.~\ref{Cluster}B is derived from the intended cluster-state resource.

Next, we quantify the precision of the displacement operation used in the cluster-state experiment. We used $|\alpha_j | \in [0, 20]$ (setting $t=1$ for simplicity) and find that the standard deviation error for 1 is $\sigma \approx 0.16$ for all displacement values (Fig. ~\ref{Cluster}F). The relative error in estimating $\langle \hat{x}_j\rangle$ is thus roughly $0.16/|\alpha_j|$, which increases from  $\sim 0.8\%$ to $7\%$ as $|\alpha_j|$ decreases from $20$ to $2.5$. In estimating $\langle \hat{x}^2_j\rangle$, Fig. ~\ref{Cluster}G shows that the standard deviation error is $\sigma \approx 3.68$ for all values of $\langle \hat{x}^2_j\rangle$, thus the relative error increases from $0.92 \%$ to $7\%$ as the target $\langle \hat{x}^2_j\rangle$ decreases from $400$ to $52.57$. For the correlations between modes $i$ and $j$, we find that there is non-negligible correlations $\langle \hat{x}_i \hat{x}_j\rangle$ across four modes and less. For the examples in Fig. ~\ref{Cluster}H, the relative error in estimating the covariance matrix elements across four modes range from $0.03\%$ to $11.11\%$. For the relative errors in estimating the first- and second-order moments using the single-mode and two-mode squeezed initial states, see Appendix~\ref{app:displacement_accuracy}. 
Finally, we apply especially programmed displacements to a 2,000-mode cluster state, forming patterns that spell out "SXU" and "SJTU". The measured outputs of BHD A and B reproduce this programmed displacement pattern, demonstrating that our platform supports programmable encoding beyond random target sequences (Fig.~\ref{Cluster}I).

\medskip\bigskip
\sectionnotoc{Comparison to corresponding quantum digital simulation} \label{sec:comparison}
This experiment implements the advection flow directly in continuous-variable optical modes. It is therefore useful to compare costs to two natural digital alternatives: (i) a spatial-grid qubit encoding of the PDE; (ii) a Fock-space qubit encoding of the optical modes. These methods (i), (ii) provide a direct digital simulation of the advection equation by preparing the state $|u(t)\rangle$, which is the comparable task performed by the analog quantum simulator. Then we compute the resources required to prepare $|u(t)\rangle$ with enough precision to obtain the value $\langle \hat{x}_j\rangle$ to within $7\%$ error (which is valid for any $|\alpha_j| \in [2.5, 20]$ used in our analog photonic simulation).  

The most direct digital approach (i) is to discretize the $d$-dimensional coordinate space (with each coordinate represented by $N_x$ grid points) and encode the resulting field amplitudes $u(t,x)$ into an $n$-qubit register using amplitude encoding over $N_x^d=2^n$ discretized points. The resulting state can be written $|u(t)\rangle=\hat{U}(t)|u_0\rangle$, where $\hat{U}(t)=\bigotimes_{j=1}^d \hat{U}_j(\alpha_jt)$, and $\hat{U}_j(\alpha_jt)$ is a finite-matrix approximation of $U_j(\alpha_jt)=\exp(-i \alpha_j \hat{p}_j t)$. On a digital platform $\hat{U}_j(t)$ can be approximated by a concatenation of large number $m$ of digital quantum gates \cite{hu2024quantum,sato2024hamiltonian}. In Table~\ref{tab:digital_resource_estimates}~(i), we provide conservative estimates of the total number of single and two-qubit gates to simulate $\hat{U}(t)$ (here we use periodic boundary conditions with central finite-difference scheme), where we see that even for $1000$ modes, it is already already prohibitively large (see Appendix~\ref{sec: direct} for more details). Thus, without even accounting for the extra resources required for the preparation of a qubit-based $|u_0\rangle$ and using conservative estimates, this quantum simulation is still not within the capability of current digital quantum devices. 

\begin{table}[H]
    \centering
    \caption{Digital quantum resource estimates for qubit-based direct quantum simulation strategies to prepare $|u(t)\rangle$ that achieves relative error $7\%$ for the target $\langle \hat{x_j}\rangle$: (i) Resource estimation (aggregated over $1000$ modes) for the simulating $\hat{U}(t)$  
(ii) Resource estimation for the $N$-mode cluster state (estimated with two-mode Fock space expansion), with errorless qubit evolution and maximum absolute value of displacement $|\alpha|=20$, with $t=1$ for simplicity (Cutoff $N_p$ strategy: mode one cut-off $n_1=5$, mode two cut-off: $n_2=3$). The single-qubit gate count is estimated as roughly twice the CNOT count, accounting for basis changes and local operations required in standard gate decompositions.}
    \label{tab:digital_resource_estimates}

    \footnotesize
    \setlength{\tabcolsep}{3.2pt}
    \renewcommand{\arraystretch}{1.12}

    \footnotesize
\setlength{\tabcolsep}{4.0pt}
\renewcommand{\arraystretch}{1.15}

\begin{tabular}{@{}|cccc|@{}c@{}|ccc|@{}}
    \multicolumn{4}{c}{\textbf{(i) Direct Hamiltonian simulation of displacement}}
    & \multicolumn{1}{@{}c@{}}{\tabgap} &
    \multicolumn{3}{c}{\textbf{(ii) Fock-space simulation with $\alpha=20$}} \\
    \cline{1-4}\cline{6-8}

    \makecell{Qubits\\per qumode}
    & \makecell{1-qubit\\rotations}
    & CNOTs
    & \makecell{Single-gate fidelity\\required}
    & \tabgap &
    \makecell{Qumodes\\($N$)}
    & \makecell{1-qubit\\rotations}
    & CNOTs \\
    \cline{1-4}\cline{6-8}

    5  & $6.08\times10^{6}$ & $4.80\times10^{6}$ & $1-9.68\times10^{-9}$
    & \tabgap & 128   & $2.51\times10^{8}$  & $1.26\times10^{8}$  \\

    6  & $1.77\times10^{7}$ & $1.42\times10^{7}$ & $1-3.31\times10^{-9}$
    & \tabgap & 512   & $1.01\times10^{9}$  & $5.05\times10^{8}$  \\

    7  & $4.84\times10^{7}$ & $3.94\times10^{7}$ & $1-1.20\times10^{-9}$
    & \tabgap & 2048  & $4.05\times10^{9}$  & $2.02\times10^{9}$  \\

    8  & $1.27\times10^{8}$ & $1.04\times10^{8}$ & $1-4.55\times10^{-10}$
    & \tabgap & 8192  & $1.62\times10^{10}$ & $8.09\times10^{9}$  \\

    9  & $3.23\times10^{8}$ & $2.67\times10^{8}$ & $1-1.79\times10^{-10}$
    & \tabgap & \textbf{20000} & $\boldsymbol{3.95\times10^{10}}$ & $\boldsymbol{1.98\times10^{10}}$ \\

    10 & $7.99\times10^{8}$ & $6.66\times10^{8}$ & $1-7.20\times10^{-11}$
    & \tabgap & 32768 & $6.48\times10^{10}$ & $3.24\times10^{10}$ \\
    \cline{1-4}\cline{6-8}
\end{tabular}
\end{table}

The discretisation in terms of the spatial coordinates above is the most natural form of discretisation for PDE. However, it is also possible to obtain a discretisation due to the physical constraints of the system and this leads to method (ii) mentioned above. Even though a qumode is infinite-dimensional, in physically-relevant scenarios the energy is not infinite and it is possible to obtain a discretisation of photonic systems by upper-bounding the photon number by a finite cutoff $N_p$. This means we have an approximate state $|\tilde{u}(t)\rangle=\sum_{n=0}^{N_p} \tilde{u}_n|n\rangle$, where $\{|n\rangle\}_{n=1}^{\infty}$ are photon number or Fock states. From Table~\ref{tab:digital_resource_estimates}~(ii) we see that in this case also, even without the consideration of initial state preparation, the number of one and two-qubit logical gates required  is too large to be obtainable on current digital quantum devices. See Appendix~\ref{sec:Fock} for more details. 

Since the initial states we consider are Gaussian states and displacement is also a Gaussian operation, it might be expected that an indirect simulation method might be possible to retrieve first and second-order moments of the solution by embedding the corresponding covariance matrix elements into a density matrix, with significantly less qubit resources. However, using this method, which can be done via Gaussian bosonic circuits \cite{PhysRevLett.134.070604} for example, the circuit cannot simulate the displacement operation even when $|u_0\rangle$ is already provided, which is essential for simulating the advection equation. First-order moments like $\langle \hat{x}_j \rangle$ or covariance matrix elements like $\langle \hat{x}_i \hat{x}_j \rangle$ also cannot be retrieved unless other normalisation factors dependent on the covariance matrix elements themselves (generally not available a priori) are required to be known in advance. See Appendix~\ref{sec:covariance} for more details. Finally, we note that variational quantum algorithms have also been proposed for solving partial differential equations  on digital quantum processors \cite{pool2024nonlinear,sarma2024quantum,over2025boundary}. These algorithms represent the solution using a chosen ansatz and optimized circuit parameters. While they can be potentially powerful when a compact ansatz is known, their performance depends on trainability, expressivity, and the quality of the chosen variational form, so we do not consider them here. 

\medskip\bigskip
\sectionnotoc{Summary}
These experiments establish a large-scale CV quantum photonic platform for analog simulation of transport dynamics. We implemented the constant-coefficient advection equation by mapping its evolution to programmable optical displacements, validating mode-resolved control on initial states corresponding to sequences of 20,000 single-mode squeezed states and 20,000 two-mode squeezed states, and extending the protocol to a time-domain continuous-variable cluster-state resource containing 20,000 entangled qumodes. In the cluster-state experiment with random displacements, the final state $|u(t)\rangle$ is prepared with amplitude corresponding the solution for the advection equation, whose first-and second-order moments can be extracted using homodyne measurements on $|u(t)\rangle$ with relative errors as low as $0.8\%$ and $0.92\%$ respectively. 

These results show that a physical photonic device can store and evolve a large-scale analog state directly. The present demonstration is not yet a general PDE solver: it treats constant-coefficient advection with Gaussian state initial conditions, and estimates selected quadrature observables rather than reconstructing arbitrary solutions. It does not yet implement coupled, dissipative, or nonlinear dynamics. The next step is to extend to PDEs that require genuinely richer dynamics. Variable-coefficient advection and coupled wave equations will require mode-mixing operations; diffusion, Fokker–Planck, and dissipative equations will require application of nonunitary operations. Establishing these extensions, while quantifying finite-squeezing errors, sampling costs, and comparisons with optimized digital quantum methods, will determine whether continuous-variable photonics can become a practical platform for high-dimensional scientific computing.

\bigskip
\starsectionnotoc{Acknowledgements}
The authors thank Jorg Schmiedmyer and Bill Munro for interesting and insightful discussions. This work is supported by the Quantum Science and Technology-National Science and Technology Major Project (2024ZD0302403), the National Natural Science Foundation of China (12434015, 12471411, 12341104 and U25A20525), Fund for Shanxi “1331 Project” Key Subjects Construction, the Science and Technology Commission of Shanghai Municipality (STCSM) grant no. 24LZ1401200 (21JC1402900) and the Shanghai Pilot Program for Basic Research,  Shanghai Jiao Tong University 2030 Initiative, the Shanghai Science and Technology Innovation Action Plan (24LZ1401200) and the Fundamental Research Funds for the Central Universities. 

\clearpage

\let\oldaddcontentsline\addcontentsline
\renewcommand{\addcontentsline}[3]{}
\bibliography{ref}
\let\addcontentsline\oldaddcontentsline

\clearpage
\appendix
\makeatletter
\renewcommand{\p@subsection}{\thesection.}
\renewcommand{\p@subsubsection}{\thesection.\arabic{subsection}.}
\makeatother
\onecolumngrid

\begin{center}
    \LARGE\textbf{Appendix}
\end{center}

\tableofcontents

    \section{Experiment}\label{app:experiment}
\begin{figure}[H]
    \centering
    \includegraphics[width=\linewidth]{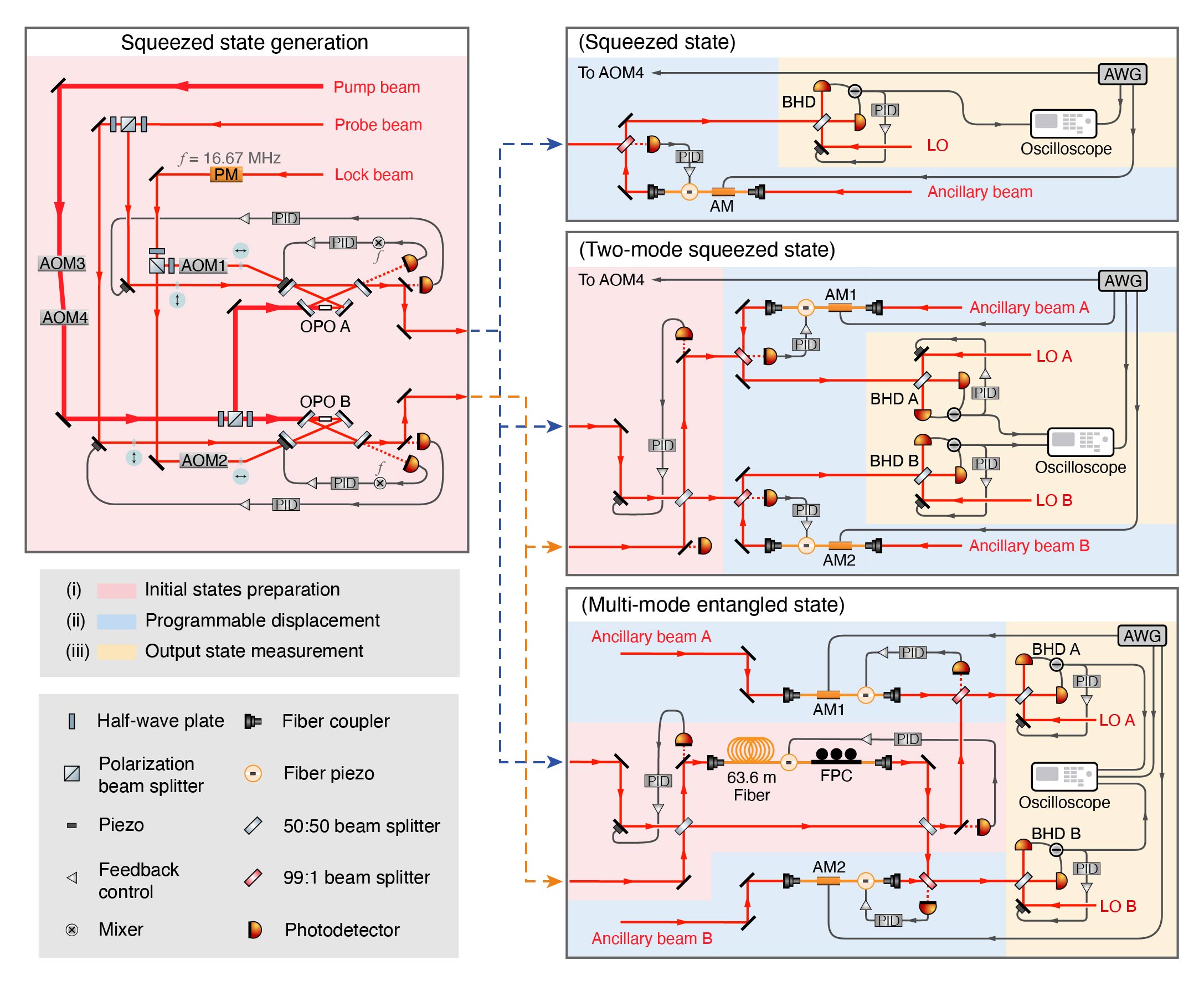}
    \caption{Experimental setup. The setup comprises three steps. First, OPA A and OPA B generate squeezed states, which are configured as single-mode squeezed state sequences, two-mode squeezed state sequences, or a time-domain continuous variable (CV) cluster state. Next, an arbitrary waveform generator (AWG) drives amplitude modulation of the ancillary beams, which is then coupled with the signal beam on a 99:1 beam-splitter. Finally, the output state is measured using balanced homodyne detection (BHD) with a local oscillator and recorded by an oscilloscope. PM, phase modulator; AOM, acousto-optic modulator; AM, amplitude modulator; FPC, fiber polarization controller.}
    \label{setup}
\end{figure}

The experimental implementation consists of three steps (Fig.~\ref{setup}), which includes initial state preparation, programmable displacement, and output state measurement. The first step generates the required quantum resources, including single-mode squeezed states, two-mode squeezed states, and a time-domain continuous-variable (CV) cluster state. The second step applies programmable displacements to the amplitude quadrature $\hat{x}$ of each qumode to simulate advection velocity transport. Finally, the output states are measured by balanced homodyne detection (BHD), and the recorded traces are processed to extract the corresponding quadrature moments.

    \subsection{Generation of initial quantum states}\label{app:initial_state_generation}
    \subsubsection{Programmable single-mode squeezed states}\label{app:single_mode_squeezed}
The experimental light source is a continuous-wave laser at 1550 nm. Quantum states are generated by two degenerate optical parametric amplifiers (OPAs) operated below threshold. Each OPA is a bow-tie traveling-wave cavity composed of two plane mirrors, two concave mirrors, and a 10 mm type-0 periodically poled KTiOPO4 (PPKTP) crystal. The bandwidth of the OPA cavity is 12.3 MHz. 
Generation of squeezed states requires active control of the relative phase between the pump field and the seed beam. A relative phase of $0$ produces an $\hat{x}$-squeezed state, whereas a relative phase of $\pi$ produces an $\hat{p}$-squeezed state. The cavity length is stabilized with a coherent locking scheme~\cite{germany}. To avoid contaminating the quantum state, the locking beam is spatially separated from the seed beam by its incident angle and is orthogonally polarized. The locking beam is frequency-shifted by acousto-optic modulators (AOMs; AOM1 and AOM2 in Fig.~\ref{setup})to maintain resonance with the seed beam. A phase modulator (PM) adds 16.67 MHz sidebands to generate the Pound-Drever-Hall (PDH) error signal.

\begin{figure}[H]
    \centering
    \includegraphics[width=8.13 cm]{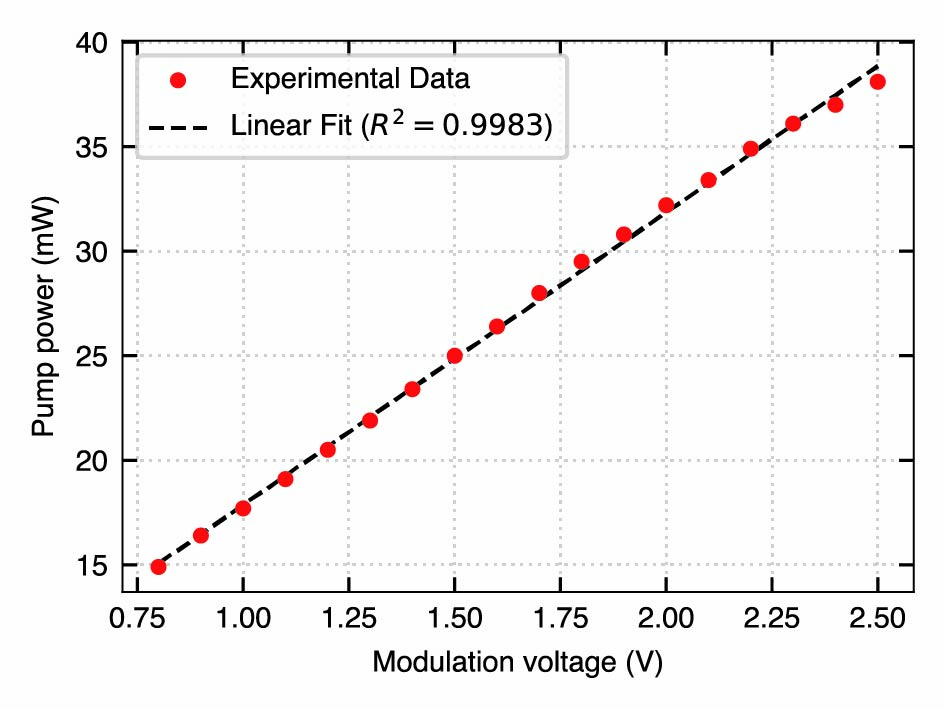}
    \caption{Calibration of pump power as a function of modulation voltage. The pump intensity is modulated by adjusting the RF drive amplitude of the AOM with an AWG. The red points represent experimental data, and the black dashed line shows a linear fit with $R^2 = 0.9983$. }
    \label{modu_pump}
\end{figure}

To program the squeezing parameter, we use cascaded AOMs on the pump path. AOM3 shifts the pump frequency by +200 MHz, and AOM4 applies a compensating -200 MHz shift. The two shifts keep the pump resonant with the OPA cavity mode. The pump power is programmed by driving the RF amplitude of AOM4 with an arbitrary waveform generator (AWG). The calibration curve is shown in Fig.~\ref{modu_pump}, and the timing sequence for single-mode squeezed states is shown in Fig.~\ref{sequence}(a). 

    \subsubsection{Programmable two-mode squeezed states}
\begin{figure}[H]
    \centering
    \includegraphics[width=16 cm]{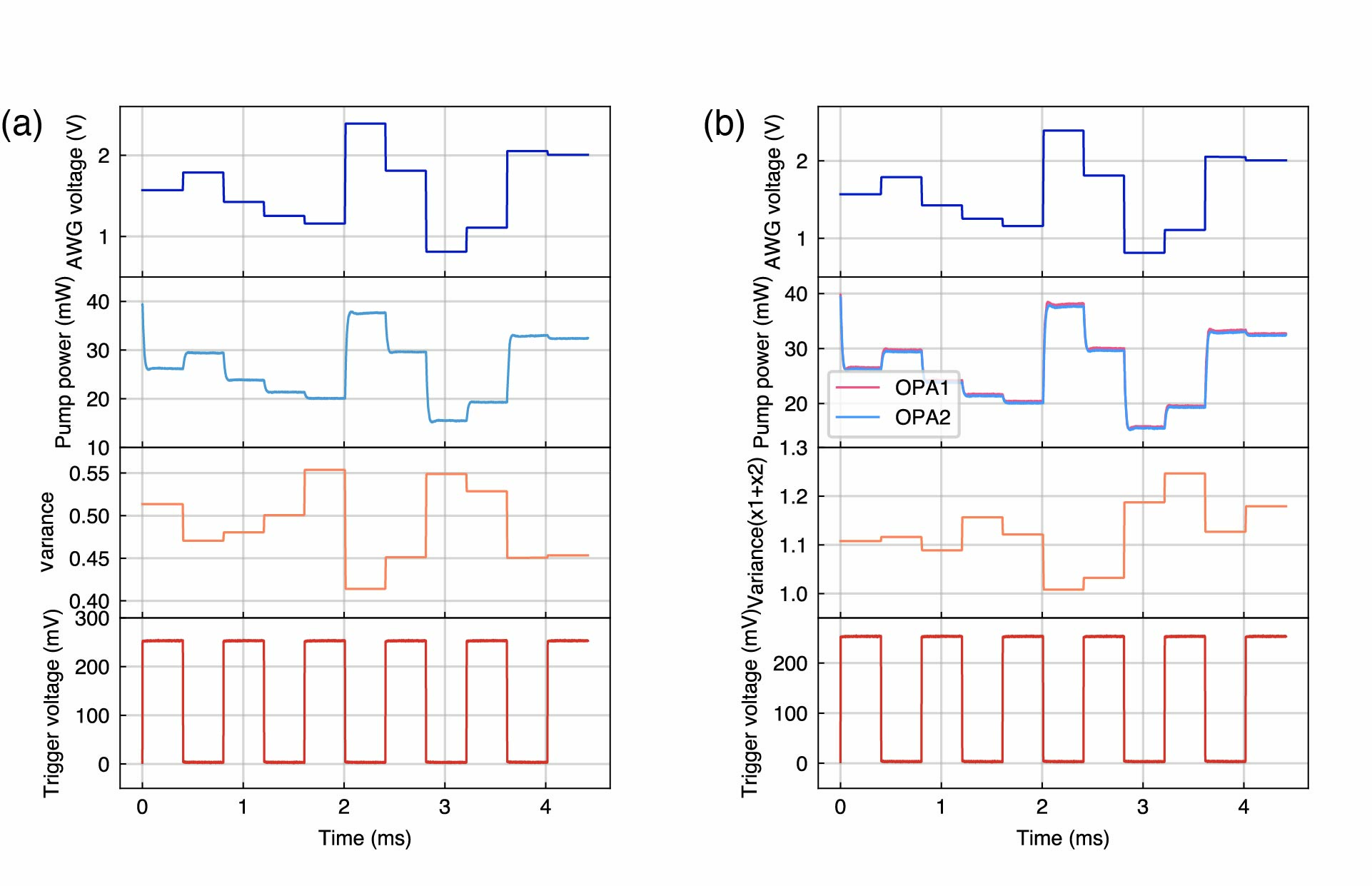}
    \caption{Timing sequences for dynamically modulated single-mode and two-mode squeezed states. (A) Timing sequence for the single-mode squeezed state. From top to bottom, the panels show the 4 MHz AWG modulation, the corresponding pump-power variation, the measured quadrature variance, and the synchronization trigger signals. (B) Timing sequence for the two-mode squeezed state. From top to bottom, the panels show the 4 MHz AWG modulation, the pump-power variations of two OPAs, the variance ${Var}(x_1+x_2)$, and the synchronization trigger pulses. These traces illustrate the temporal relationship between pump modulation, squeezing or entanglement strength, and system synchronization.}
    \label{sequence}
\end{figure}

We prepare two-mode squeezed states by interfering two independent $\hat{x}$-squeezed states on a 50:50 beam-splitter. A piezoelectric transducer (PZT) feedback loop fixes the relative phase between the two beams at $\pi/2$. In this configuration, the output spatial modes 1 and 2 satisfy $\langle \Delta(\hat{x}_1 + \hat{x}_2)^2 \rangle < 2$ and $\langle \Delta(\hat{p}_1 - \hat{p}_2)^2 \rangle < 2$, with $[\hat{x},\hat{p}]=2iI$. The entanglement can be verified using the inseparability criterion~\cite{Duan}.
To achieve programmable control of two-mode entanglement, the modulated pump beam is split and injected into both OPA cavities. Since the two OPAs shared the common modulated pump source, their squeezing parameters are modulated synchronously in each time bin, leading to a sequence of two-mode squeezed states with programmable entanglement. The corresponding timing sequence is shown in Fig.~\ref{sequence}(b). 

   \subsubsection{Time-domain CV cluster state}
We prepare a CV cluster state using a time-domain multiplexing architecture~\cite{Japan2013,Japan2016}, as shown in Fig~\ref{setup}. Two independent squeezed states with orthogonal quadratures ($\hat{x}$- and $\hat{p}$-squeezed) first interfere on a 50:50 beam-splitter with the relative phase locked to $0$. One output state passes through an optical fiber delay, while the other propagates through free space. The two states are then recombined on the second 50:50 beam-splitter with the relative phase locked to $\pi$ to generate the cluster state. The length of the delay fiber is $L = 63.6 $ m, corresponding to a temporal mode width of $T = 318$ ns. The resulting state consists of a continuous chain of temporal modes. Its entanglement structure can be described by the nullifier operators~\cite{nullifier}. For the present dual-rail time-domain structure, the nullifiers are
\begin{subequations}
\label{eq:XP_k}
\begin{align}
    \hat{X}_k = \hat{x}_k^A + \hat{x}_k^B + \hat{x}_{k+1}^A - \hat{x}_{k+1}^B, \label{eq:X_k_sub} \\
    \hat{P}_k = \hat{p}_k^A + \hat{p}_k^B - \hat{p}_{k+1}^A + \hat{p}_{k+1}^B. \label{eq:P_k_sub}
\end{align}
\end{subequations}
where $\hat{x}$ and $\hat{p}$ denote the amplitude and phase quadratures of the optical modes, and the superscripts A and B denote the two spatial modes measured by two separate balanced homodyne detectors. The subscript $k$ denotes the index of the temporal mode. 

\begin{figure}[H]
    \centering
    \includegraphics[width=8.13 cm]{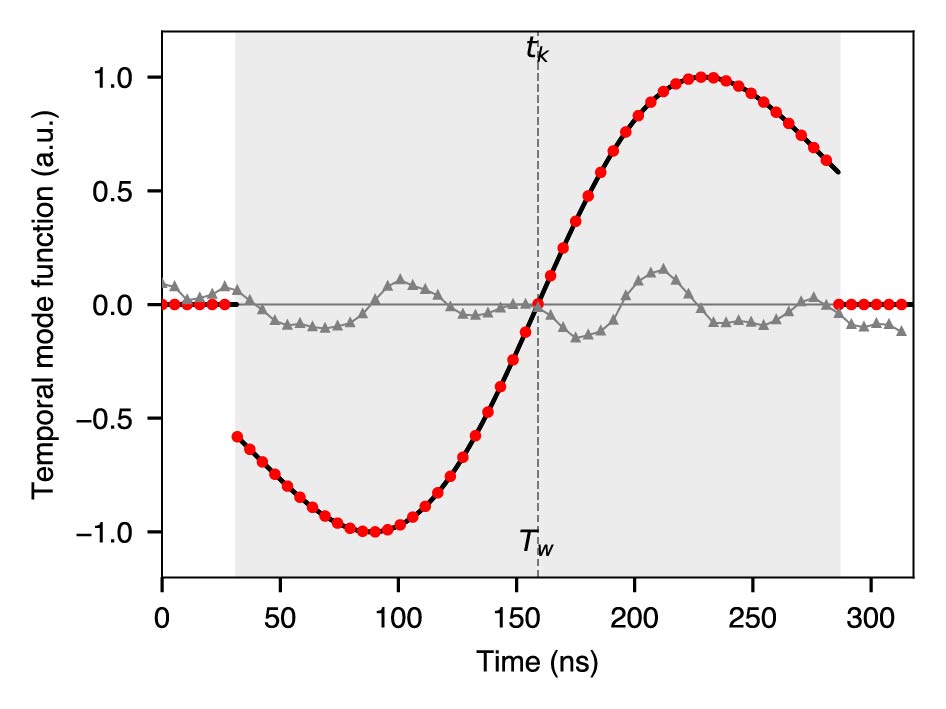}
    \caption{Temporal-mode function $g_k(t)$ and quadrature trace. The red circles connected by a black line show the mode function $g_k(t)$. The gray shaded region marks the window $T_w$. The gray line with triangles shows the quadrature trace.}
    \label{timemode}
\end{figure}

To extract the quadrature operators $\hat{x}_k$ and $\hat{p}_k$ from the continuous voltage traces recorded by the BHDs, we use temporal-mode functions $g_k(t)$. The quadrature value $\hat{q}_k$ of the $k$-th temporal mode is obtained by integrating the voltage trace $\hat{i}(t)$ with the corresponding mode function $g_k(t)$
\begin{equation}
    \hat{q}_k = \int g_k(t) \hat{i}(t) dt
\end{equation}
The mode function can suppress excess low-frequency noise while maximizing the measured squeezing. The mode function is illustrated in Fig.~\ref{timemode} and is defined as
\begin{equation}
    g_k(t) = \begin{cases} e^{-\gamma^2 (t-t_k)^2} \cdot (t - t_k), & |t-t_k| \le \frac{T_w}{2} \\ 0, & \text{otherwise} \end{cases}
\end{equation}
where $t_k = (k-1)T$ denotes the center of the $k$th temporal mode. The function consists of a Gaussian envelope with a bandwidth parameter $\gamma = 2\pi \times 1.6$ MHz and a linear factor $(t - t_k)$. The mode-function window is restricted to 80\% of the mode interval, $T_w = 0.8T$. 

\begin{figure}[H]
    \centering
    \includegraphics[width=18.4 cm]{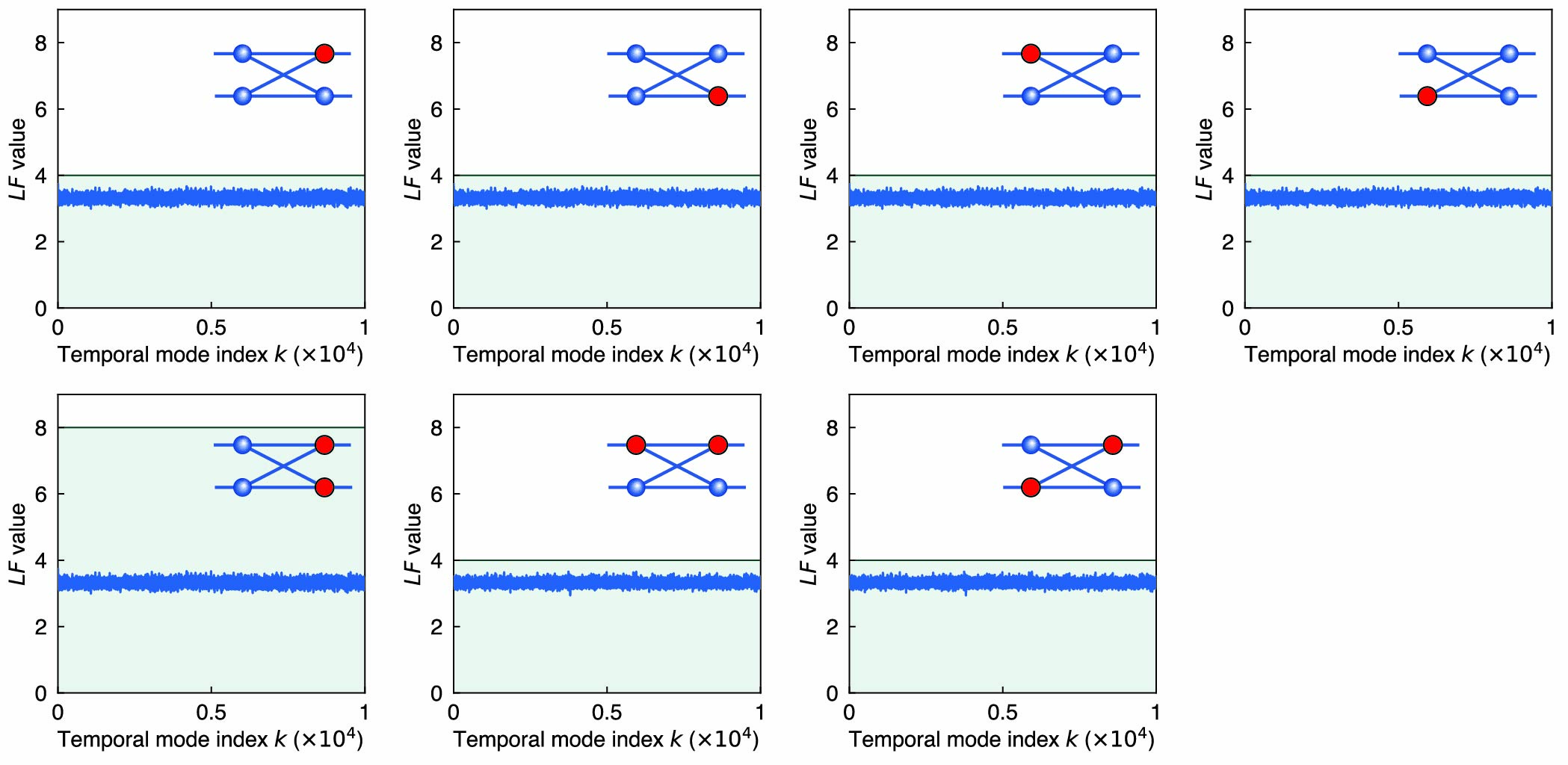}
    \caption{Van Loock--Furusawa inseparability measurement for the time-domain CV cluster state. The seven panels display the measured values $\mathrm{LF}$ for all seven possible bipartitions. Solid green lines are the corresponding separability bounds. All measured LF values are strictly below their respective bounds, indicating that the cluster state is inseparable.}
    \label{LF}
\end{figure}

Ideally, the variances of the nullifiers would be zero. However, due to finite squeeze and optical losses, the variances cannot be zero. Experimentally, we obtain $\langle \hat{X}_k^2 \rangle$ and $\langle \hat{P}_k^2 \rangle$ by measuring the corresponding quadrature combinations. 
To verify the entanglement, we use the van Loock–Furusawa inseparability criterion~\cite{van_and_furusawa}. For a given set of modes $S$, consider an arbitrary bipartition $\{S_\alpha, S_\beta\}$. We define linear combinations of quadrature operators 
\begin{equation}
    \hat{u} = \sum_{k \in S} h_k \hat{x}_k, \hspace{0.5cm}\hat{v} = \sum_{k \in S} g_k \hat{p}_k
\end{equation}
where $h_k$ and $g_k$ are real coefficients. If the quantum state is separable with respect to bipartition $\{S_\alpha, S_\beta\}$, the variances of these operators must satisfy the inequality
\begin{equation}
    \langle \Delta\hat{u}^2 \rangle + \langle \Delta\hat{v}^2 \rangle \ge 2 \left( \left| \sum_{k \in S_\alpha} h_k g_k \right| + \left| \sum_{k \in S_\beta} h_k g_k \right| \right)
\end{equation}
In our experiment, the mode set is $S_k = \{(A,k), (B,k), (A,k+1), (B,k+1)\}$. It has seven bipartitions, as shown in Fig.~\ref{LF}. We represent the van Loock–Furusawa value as $\mathrm{LF} \equiv \langle \Delta\hat{u}^2 \rangle + \langle \Delta\hat{v}^2 \rangle$. The experimentally measured LF values for all seven bipartitions are strictly below their corresponding separability bounds. The results demonstrate the full inseparability and multipartite entanglement of the generated time-domain CV cluster state.  

\begin{figure}[H]
    \centering
    \includegraphics[width=\linewidth]{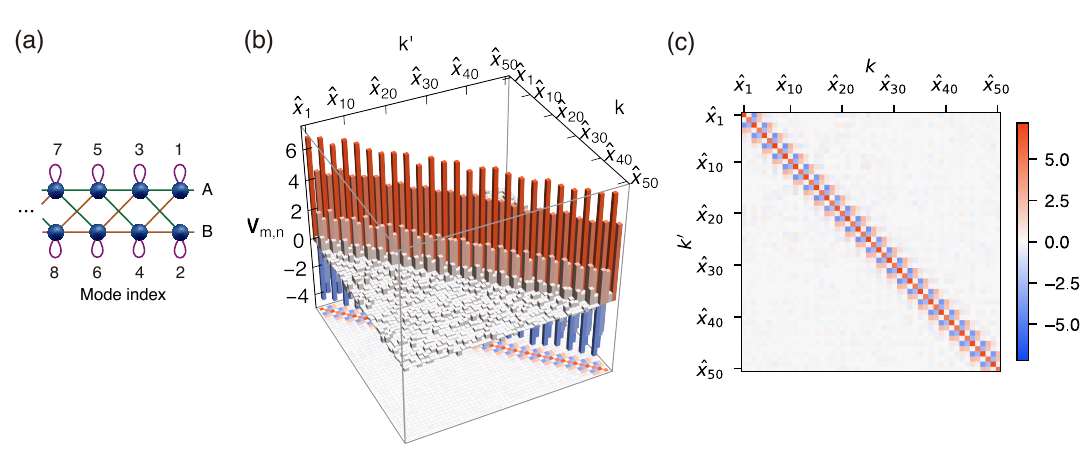}
    \caption{Covariance matrix of the time-domain CV cluster state. (a) Graphical representations of the dual-rail time-domain CV cluster state. (b) 3D representation of the amplitude-quadrature ($\hat{x}$) covariance matrix for the first 50 temporal modes, normalized to the shot-noise level. (c) 2D representation of the covariance matrix. The banded structure clearly shows the alternating positive and negative correlations between adjacent modes. }
    \label{covariance}
\end{figure}

To further show the entanglement structure of the cluster state, we calculated the full covariance matrix. Fig.~\ref{covariance} shows the amplitude quadratures ($\hat{x}$) correlations of the first 50 modes, normalized to the shot-noise level. The covariance matrix exhibits a banded structure. Specifically, the two spatial modes within the same time bin ($A_k$ and $B_k$) show very weak correlation ($\mathrm{Cov}(A_k, B_k) = +0.32$). Instead, significant correlations exist mainly between adjacent time bins ($k$ and $k+1$). These covariances exhibit an alternating positive and negative correlation structure, with an average correlation of 
\begin{gather*}
\mathrm{Cov}(A_k, A_{k+1}) = -3.84,\\
\mathrm{Cov}(A_k, B_{k+1}) = +2.42,\\
\mathrm{Cov}(B_k, A_{k+1}) = -2.57,\\
\mathrm{Cov}(B_k, B_{k+1}) = +2.04.
\end{gather*}
As the temporal separation increases to two or more time bins (e.g., between bin $k$ and bin $k+2$), the correlations decay rapidly:
\begin{gather*}
\mathrm{Cov}(A_k, A_{k+2}) = +0.87,\\
\mathrm{Cov}(A_k, B_{k+2}) = -0.38,\\
\mathrm{Cov}(B_k, A_{k+2}) = +0.42,\\
\mathrm{Cov}(B_k, B_{k+2}) = -0.01.
\end{gather*}
This rapid decay occurs because the experimental delay-loop design primarily couples temporally adjacent wave packets. Consequently, correlations for widely separated modes become negligible.
This specific correlation structure demonstrates that the cluster state is neither fully connected nor a collection of independent four-mode clusters. Instead, it is a continuous chain along time. This structure determines the dimensionality of the equation and constrains the class of PDEs that can be simulated. In our experiment, displacement operations applied across 10,000 time bins correspond to a simulation of a specific 20,000-dimensional advection equation.

    \subsection{Implementation of displacement operations}\label{app:displacement_operations}
    \subsubsection{Optical setup for displacement}
The phase-space displacement operator $\hat{D}(\alpha)$ is implemented by interfering the quantum signal field with a modulated ancillary beam. To minimize optical loss of the quantum state, we use a beam-splitter with 99\% reflectivity and 1\% transmissivity. The quantum signal propagates through the reflected port, while the ancillary beam carrying the displacement information is injected from the other port. The ancillary beam is a coherent beam modulated by a fiber-integrated electro-optic amplitude modulator (AM) with a bandwidth of 10 GHz. Adjusting the modulation voltage applied to the AM sets the displacement value of the $\hat{x}$ quadrature for each time bin.

\begin{figure}[h]
    \centering
    \includegraphics[width=17 cm]{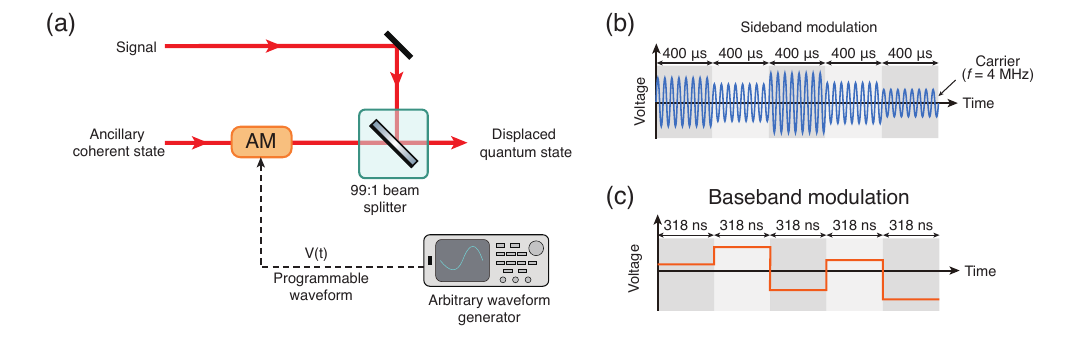}
    \caption{Experimental implementation and modulation schemes for displacement. (a) Optical implementation. The phase-space displacement operator $\hat{D}(\alpha)$ is performed by interfering the signal field with a modulated ancillary beam on a 99:1 beam-splitter. The ancillary-beam amplitude is controlled by an electro-optic amplitude modulator driven by an arbitrary waveform generator. (b) Sideband modulation for single-mode and two-mode squeezed states. The target displacement sequence $\alpha$ is encoded on a sinusoidal carrier at $f = 4~\mathrm{MHz}$, with a time-bin duration of $400~\mu\mathrm{s}$. (c) Baseband modulation for the time-domain CV cluster state. The displacement $\alpha$ is mapped directly onto the DC voltage of the modulation waveform for temporal modes with $318~\mathrm{ns}$ time-bin resolution.}
    \label{disp}
\end{figure}

We use two modulation schemes for the programmable displacement operation $\hat{D}(\alpha)$ (Fig.~\ref{disp}). For single-mode squeezed states and two-mode entangled states, we use sideband modulation to bypass low-frequency technical noise. The target displacement sequence is encoded on a 4 MHz sinusoidal carrier, and the modulation amplitude defines $\alpha$. Each time bin is 400~$\mu$s. For the time-domain CV cluster state, we use baseband modulation. In this scheme, the displacement $\alpha$ corresponds directly to the voltage level of the modulation waveform. This direct mapping is used for the 318 ns temporal modes of the cluster state.

    \subsubsection{Temporal synchronization}\label{app:temporal_synchronization}
\begin{figure}[h]
    \centering
    \includegraphics[width=8 cm]{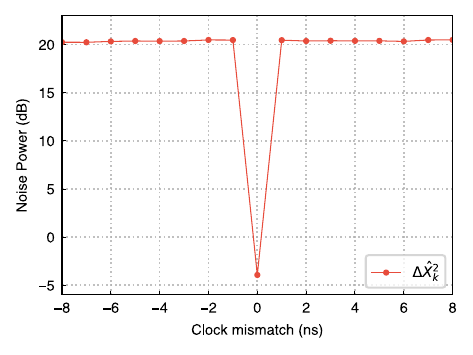}
    \caption{Impact of clock mismatch on the measured entanglement. Average nullifier noise $\langle\hat{X}_k^2\rangle$ versus clock deviation from the nominal 318 ns wave-packet width, showing the extreme sensitivity of the cluster state to timing misalignments.}
    \label{nullifierTime}
\end{figure}

\begin{figure}[h]
    \centering
    \includegraphics[width=12 cm]{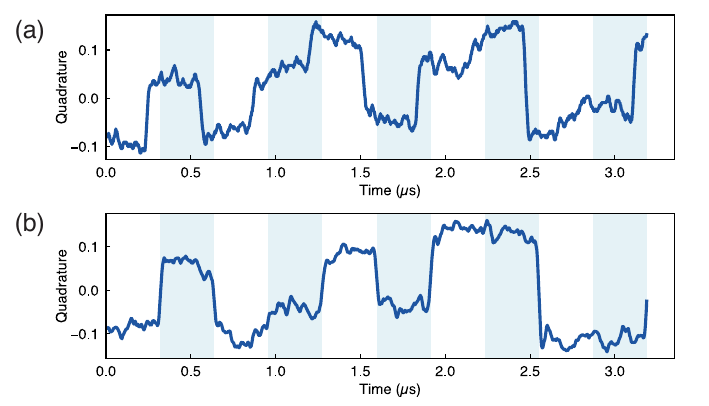}
    \caption{Synchronization between displacement control and data acquisition. (a) Measured quadrature evolution without common-clock synchronization. Timing drift accumulates and produces a mismatch between the displacement pulses and the corresponding temporal modes. (b) Measurement with common-clock synchronization. The arbitrary waveform generator and oscilloscope are locked to a common reference clock, and the electronic delay is adjusted so that each displacement pulse is aligned with its corresponding time bin.}
    \label{align}
\end{figure}

As described in the main text, the cluster state is particularly sensitive to timing mismatches due
to its short wave-packet width of 318 ns. Consequently, the AWGs driving the displacement operations, the trigger signals used for time-bin segmentation, and the oscilloscope used for data acquisition must be synchronized. Even a slight clock mismatch can affect the measured quantum state.
To illustrate this impact, Fig.~\ref{nullifierTime} presents the average noise variance of the nullifiers under clock mismatches. The result demonstrates that the measured entanglement degrades with increasing timing offset. 
To achieve precise synchronization, we perform a nanosecond-level timing calibration. Both displacement control signals and data acquisitions are referenced to a common 10 MHz clock. By adjusting the electronic delay, each displacement pulse is accurately aligned with its corresponding time bin, as illustrated in Fig. \ref{align}.

    \subsubsection{Displacement calibration}
In the experiment, due to the finite hardware responses of the AMs and the BHDs, the relationship between the digital control voltage and the actual displacement is not perfectly linear. If a simple linear mapping is used directly, it will introduce errors into the dynamical simulation. To avoid this, we calibrate the mapping between the digital control voltage $V$ and the resulting displacement amplitude $\alpha$ before data acquisition. The calibration is performed directly on the generated quantum resource state so that the fitted parameters reflect the current experimental conditions. 

First, we apply a test voltage, $V_{{test}}$, to the AM, and the resulting mean quadrature displacements, $\langle\hat{x}\rangle$, are measured using BHDs. Next, we perform a non-linear fit on the data to capture the system's actual non-linear transfer function. To ensure a linear relationship between the target and measured displacements, the inverse of the fitted non-linear function is used to generate a new voltage table, which maps the desired displacement amplitudes to the corresponding digital voltages. When the new displacement table is applied to the AMs, the displacement of each time bin is linearly related to the target displacement values, with a narrow and stable error distribution. This calibration procedure is critical for achieving high-fidelity and programmable displacement operations in dynamical simulation.
    
    \subsubsection{Displacement accuracy and linearity}\label{app:displacement_accuracy}
\begin{figure}[h]
    \centering
    \includegraphics[width=\linewidth]{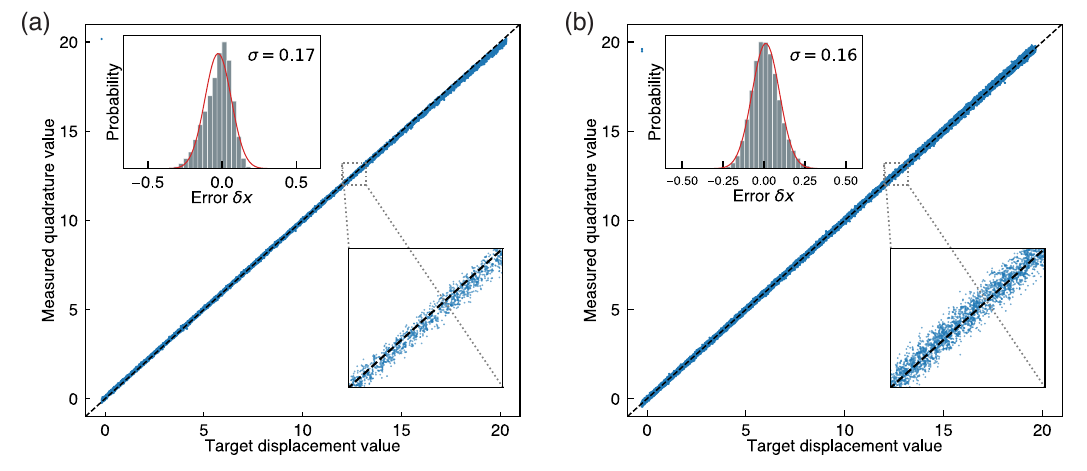}
    \caption{Displacement performance for single-mode and two-mode squeezed states. In each panel, the lower right inset provides a magnified view of the results, while the upper left inset displays the probability distribution of the displacement error $\delta x$. The solid red curves represent Gaussian fits to the error distributions, with $\sigma$ denoting the standard deviation. (a) Results for the single-mode squeezed state. The mean relative error is 0.0596. (b) Results for the two-mode squeezed state. The mean relative error is 0.0491.}
    \label{disp1and2}
\end{figure}

For single-mode and two-mode squeezed states, we applied random displacements in the range $|\alpha| \in [0, 20]$ across 20,000 time bins. As shown in Fig.~\ref{disp1and2}, we checked the linearity of the displacement by comparing the measured quadratures with the target values. For single-mode squeezed states, the mean displacement relative error $\epsilon = |\alpha_{meas} - \alpha_{set}| / |\alpha_{set}|$ is approximately 0.06. For two-mode squeezed states, the mean displacement relative error is $\epsilon \approx 0.05$. The corresponding displacement error standard deviations are $\sigma \approx 0.17$ and $\sigma \approx 0.16$, respectively. 

\begin{figure}[H]
    \centering
    \includegraphics[width=\linewidth]{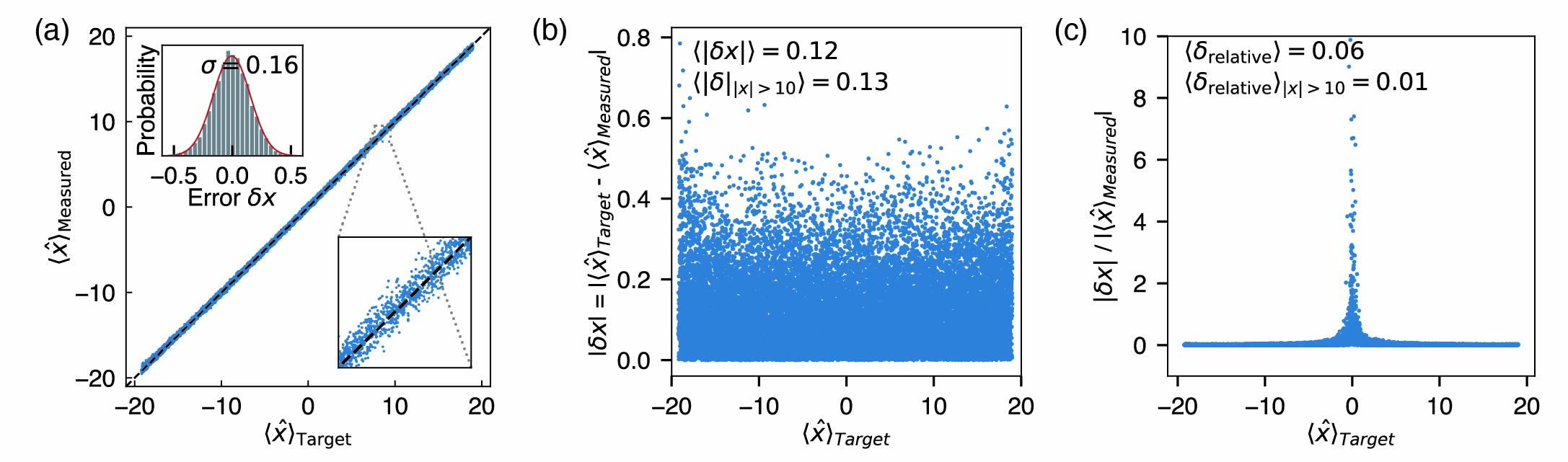}
    \caption{Displacement accuracy of first-order moments for the cluster state. (a) Correlation between the target $\langle \hat{x} \rangle$ and the experimentally measured $\langle \hat{x} \rangle$. The lower right inset provides a magnified view of the results, while the upper left inset displays the probability distribution of the displacement error $\delta x$. The solid red curves represent Gaussian fits to the error distributions, with $\sigma$ denoting the standard deviation of the error. (b) Statistical distribution of the absolute displacement error $|\delta x|$ for all 20,000 modes. (c) Statistical distribution of the relative displacement errors for all 20,000 modes. The error distribution exhibits a mean relative error of 0.06 and a standard deviation of $\sigma = 0.16$}
    \label{cluster_error_x}
\end{figure}

\begin{figure}[H]
    \centering
    \includegraphics[width=\linewidth]{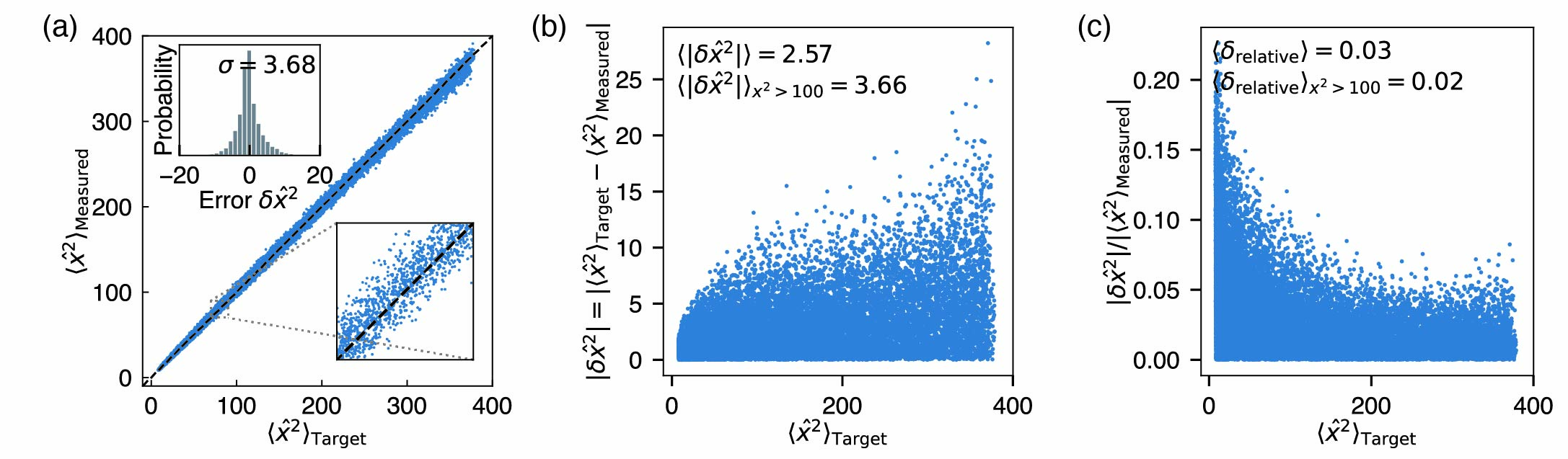}
    \caption{Displacement accuracy of second-order moments for the cluster state. (a) Correlation between the target $\langle \hat{x}^2 \rangle$ and the experimentally measured $\langle \hat{x}^2 \rangle$. The lower right inset provides a magnified view of the results, while the upper left inset displays the probability distribution of the displacement error $\delta x^2$, with $\sigma$ denoting the standard deviation of the error. (b) Statistical distribution of the absolute displacement error $|\delta x^2|$ for all 20,000 modes. (c) Statistical distribution of the relative displacement errors for all 20,000 modes. The error distribution exhibits a mean relative error of 0.03 and a standard deviation of $\sigma \approx 3.68$}
    \label{cluster_error_x_2}
\end{figure}

\begin{figure}[H]
    \centering
    \includegraphics[width=\linewidth]{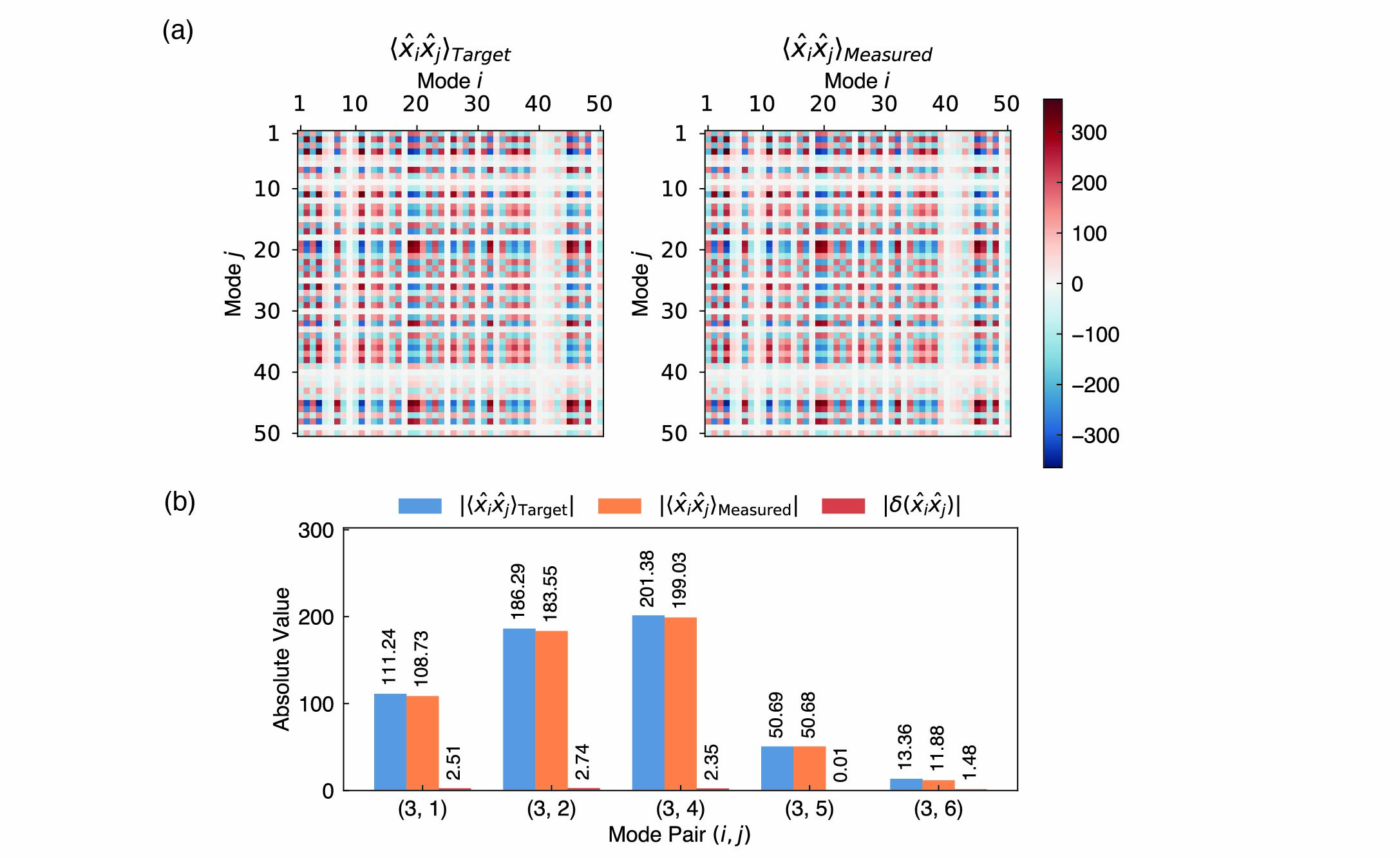}
    \caption{Displacement accuracy of cross-moments of the cluster state. (a) 2D heatmaps showing the cross-moments $\langle \hat{x}_i \hat{x}_j \rangle$ of the target (left) and experimentally measured (right) between different spatial modes. The measured correlation matrix is quite similar to the target. (c) Cross-moments and absolute errors $|\delta(\hat{x}_i \hat{x}_j)|$ for representative mode pairs involving $i=3$.}
    \label{cluster_error_xixj}
\end{figure}

For the cluster state, we applied random displacements in the range $\alpha \in [-20, 20]$  across 10,000 time bins. Fig.\ref{cluster_error_x} illustrates the accuracy of the first-order moments $\langle \hat{x} \rangle$. The measured displacements exhibit a highly linear correspondence with the target displacements, yielding a narrow Gaussian error distribution with a standard deviation of $\sigma \approx 0.16$ and a mean relative error of only 0.06. In addition, we also calculated the second-order moments $\langle \hat{x}^2 \rangle$ (Fig.\ref{cluster_error_x_2}). The experimentally measured $\langle \hat{x}^2 \rangle$ values exhibit a linear correspondence with the target distribution, maintaining a low mean relative error of 0.03 and an error standard deviation of $\sigma \approx 3.68$. Finally, we calculate the cross-moments $\langle \hat{x}_i \hat{x}_j \rangle$ between different spatial modes. As shown in the 2D heatmaps (Fig.\ref{cluster_error_xixj}a), the experimental correlation matrix of the displaced state is quite similar to the target matrix. Some specific mode pairs (Fig.\ref{cluster_error_xixj}b) show that the absolute errors $|\delta(\hat{x}_i \hat{x}_j)|$ are quite small compared to the target values. Together, these results demonstrate precise, linear, and independent control of the cluster state, which is a key requirement for simulating complex PDEs.

    \subsection{Measurement and data acquisition}
We use two data processing approaches for different quantum states. For the single-mode and two-mode squeezed states, the quantum information is encoded at a sideband frequency to avoid low-frequency technical and laser noise. The AC output of the BHDs was electronically demodulated at $\Omega = 4$ MHz using RF mixers and then low-pass filtered with a bandwidth of 1.9 MHz. The signals were sampled using a high-resolution 12-bit oscilloscope (Rigol HDO8108A). The recorded data were segmented into individual wave packets with a width of 400~$\mu$s. The displacement operation produces oscillatory distortion near the beginning and end of each wave packet. We use the central 75\% of each wave packet for further analysis. 

In contrast, the cluster state requires broadband detection to preserve its quantum information. The AC output of the BHDs was therefore collected directly and digitized by the oscilloscope at a sampling rate of 250 MSa/s. Each frame has a temporal width of 3.18 ms. A total of 1,800 frames were recorded for each quadrature to provide sufficient statistics.
After acquisition, the raw data was segmented into individual wave packets with a width of 318 ns. We use the central 90\% of each wave packet for further analysis. Under identical experimental conditions, the optical shot noise level was recorded as a normalization reference. To verify the entanglement structure of the CV cluster state, we applied digital spectral filtering to the raw data to remove technical noise introduced by the active stabilization loops. Specifically, a digital high-pass filter with a cut-off frequency of 1.2 MHz was applied to suppress phase-locking noise from FPGA-based feedback controllers. A 15 MHz low-pass filter was also used to remove modulation frequencies introduced by OPA cavity-length locking. When extracting displacement information for the simulation dynamics, no such filtering was applied and the mean quadrature values were calculated directly from the raw data in the time domain.

\section{Quantum analog versus quantum digital simulation}
In the following we present the resource costs and other requirements of three different qubit-based methods discussed in the main text: (A) a spatial-grid qubit encoding of the PDE (Appendix~\ref{sec: direct}); (B) a Fock-space qubit encoding of the optical modes (Appendix~\ref{sec:Fock}) and (C) an indirect simulation method based on Gaussian bosonic circuits (Appendix~\ref{sec:covariance}). 

\subsection{Direct qubit-based simulation of the advection equation}\label{sec: direct}
This section presents a most direct qubit-based simulation framework for
the advection equation
\begin{align}\label{eq:adv eq}
     \frac{\partial u(t,x)}{\partial t}+\sum_{j=1}^d \alpha_j \frac{\partial u(t,x)}{\partial x_j}=0, \qquad u_0(x)=u(0,x),
\end{align}
which we encode into an $n$-qubit register using amplitude encoding over $N_x$ discretized points per spatial dimension. Thus, for one dimension $d=1$, the normalized $n$-qubit state representing the field reads
\begin{equation}
\ket{u(t)}^n=\frac{1}{\sqrt{\mathcal{N}(t)}}\sum_{x=0}^{N_x-1} u(x,t)\ket{x}^n,
\label{eq:amplitude encoding direct}
\end{equation}
with $\mathcal{N}(t)=\sum_{x=0}^{N_x-1}|u(x,t)|^2$ ensuring proper normalization of
the encoded amplitude distribution.
Here, $\ket{x}^n$ labels the computational-basis state encoding spatial grid points where $x=0, 1, \cdots N_x-1$. For a general $d-$dimensional advection equation we have $N_x^d=2^n$ and $|u(t)\rangle=\hat{U}(t) |u_0\rangle$, 
and we construct the evolution operator
\begin{equation}\label{eq:circuit}
    \hat{U}(t) = \exp\!\left(-it \hat{H}\right)
    = \bigotimes_{j=1}^d \hat{U}_j(\alpha_j t), \quad \hat{H}=\sum_{j=1}^d \alpha_j \hat{p}_j, \quad   \hat{U}_j(\alpha_jt)= \exp\!\bigl(-it \alpha_j \hat{p}_j\bigr),
\end{equation}
where $\bigotimes_{j=1}^d$ denotes the tensor product. Thus we obtain $\ket{u(T)}^n=\hat{U}(T)\ket{u(0)}^n$ after applying the unitary operation to the initial state $|u(0)\rangle^n$. Here we restrict our evaluation to the cost of the evolution $\hat{U}(t)$; in particular, we count the number of fundamental computational units required to build this operator without the additional resources for state preparation or final state readout.

The operator $\hat{U}(t)$ corresponds to an independent series of applications of $\hat{U}_j(\alpha_jt)=\exp(-it \alpha_j \hat{p}_j)$ on each qumode. Thus, estimating the resource cost of the evolution 
operator $\hat{U}(t)$ scales linearly with the cost of individual terms
$\exp(-it \alpha_j \hat{p}_j)$. For the coordinate and momentum operators we introduce matrices $\hat{x}$ and $\hat{p}$ as finite-difference matrices. 
For example, for the central symmetrical finite-difference scheme for the derivative $\frac{\partial U}{\partial x}\rightarrow \frac{U_{h+1}-U_{h-1}}{2\Delta x}$; and we impose periodic spatial boundary conditions over the interval $(a,b)$:
\begin{eqnarray}
    \hat{x}=\left(\begin{array}{ccccc}
    a & 0  & \dotsm & 0 & 0 \\
    0 & a+\Delta x  &\dotsm& 0 & 0\\
    \rotatebox[origin=c]{270}{\dots}&&\rotatebox[origin=c]{-45}{\dots}&&\rotatebox[origin=c]{270}{\dots}\\ 
    0 & 0 &\dotsm & b-\Delta x &0\\
    0 & 0 &\dotsm & 0 &b\\
    \end{array}
    \right), \qquad \hat{p}=-\frac{i}{2\Delta x}\left(\begin{array}{ccccccc}
    0 & 1 & 0& \dotsm &0& 0 & -1 \\
    -1 & 0 & 1&\dotsm& 0&0 & 0\\
    \rotatebox[origin=c]{270}{\dots}&&&\rotatebox[origin=c]{-45}{\dots}&&&\rotatebox[origin=c]{270}{\dots}\\ 
    0 & 0&0 &\dotsm &-1& 0 &1\\
    1 & 0&0 &\dotsm &0& -1 &0\\
    \end{array}
    \right).
    \label{eq:x_p_cyclic_matrices}
    \end{eqnarray}

The periodic boundaries lead to a simple Hermitian form of the momentum operator and, in the present setting, also make $\hat p$ a banded sparse matrix. For the periodic second-order central-difference discretization \cite{LeVeque2007FDM} in Eq.~\eqref{eq:x_p_cyclic_matrices}, each row of $\hat p$ contains only two nonzero entries, so the sparsity is
\begin{equation}\label{eq:p_sparsity}
\mathfrak{s}=2,
\qquad
\ell_p:=\left\lceil \log_2 \mathfrak{s}\right\rceil=1.
\end{equation}
Using the explicit block-encoding framework from \cite{xlpd-fb1g,guseynov2025quantum} we construct a sparse-amplitude oracle for the two nonzero values in the first row of $\hat p$ and a banded-sparse-access unitary that converts the sparse index into the actual column index.

The sparse-amplitude oracle for $\hat p$ is
\begin{equation}\label{eq:Op_sparse}
\hat O_p^{S}\ket{0}_{1}\ket{s}_{\ell_p}
=
\frac{\hat p^{(s)}}{\sqrt{N_{\hat p}}}\ket{0}_{1}\ket{s}_{\ell_p}
+
\sqrt{1-\frac{|\hat p^{(s)}|^2}{N_{\hat p}}}\ket{1}_{1}\ket{s}_{\ell_p},
\qquad s\in\{0,1\},
\end{equation}
where $\hat p^{(s)}$ denotes the $s$-th nonzero element in the first row of $\hat p$, and $N_{\hat p}\ge \|\hat p\|_{\max}^{2}$ is a normalization constant. Since $\ell_p=1$, Appendix~B of \cite{xlpd-fb1g} yields the exact circuit cost: $2$ CNOTs and $2$ one-qubit rotations.

The second primitive is the banded-sparse-access oracle
\begin{equation}\label{eq:Op_BS}
\hat O_p^{\mathrm{BS}}
\ket{s}_{\ell_p}\ket{0}_{n-\ell_p}\ket{i}_{n}
=
\ket{r_s+i \!\!\!\!\pmod{N_x}}_{n}\ket{i}_{n},
\end{equation}
where $r_s$ is the column position of the $s$-th nonzero element in the first row of $\hat p$. For Eq.~\eqref{eq:x_p_cyclic_matrices} one has
\begin{equation}\label{eq:p_first_row_shifts}
r_0=1,
\qquad
r_1=N_x-1.
\end{equation}
The oracle factorizes as
\begin{equation}\label{eq:Op_BS_factorization}
\hat O_p^{\mathrm{BS}}
=
U_{\mathrm{SUM}}
\left(
U_p^{(\ell_p)}\otimes I^{\otimes n}
\right),
\end{equation}
where the two sub-oracles have simple meanings. First,
\begin{equation}\label{eq:Up_first_row_action}
U_p^{(\ell_p)}\ket{s}_{\ell_p}\ket{0}_{n-\ell_p}
=
\ket{r_s}_{n},
\qquad s\in\{0,1\},
\end{equation}
prepares the first-row band offset corresponding to the sparse label $s$. In the present periodic central-difference case, these offsets are precisely the two values in Eq.~\eqref{eq:p_first_row_shifts}. Since $\ell_p=1$, the unitary $U_p^{(\ell_p)}$ can be implemented with $50n-72$ CNOTs and $64n-96$ one-qubit rotations.

Next,
\begin{equation}\label{eq:Usum_action}
U_{\mathrm{SUM}}\ket{j}_{n}\ket{i}_{n}
=
\ket{j+i \!\!\!\!\pmod{N_x}}_{n}\ket{i}_{n},
\end{equation}
performs modular addition on the column register. Therefore, after $U_p^{(\ell_p)}$ prepares the first-row offset $r_s$, the action of $U_{\mathrm{SUM}}$ shifts this offset by the row index $i$ and produces the actual column position of the corresponding nonzero element. The modular adder $U_{\mathrm{SUM}}$ can be implemented with $26n-37$ CNOTs and $32n-48$ one-qubit rotations.

Combining Eqs.~\eqref{eq:Up_first_row_action} and \eqref{eq:Usum_action}, the banded-sparse-access oracle $\hat O_p^{\mathrm{BS}}$ maps the pair $(s,i)$ to the true column index $r_s+i \pmod{N_x}$ and can therefore be implemented with
\begin{equation}\label{eq:OpBS_cost}
76n-109 \ \text{CNOTs}
\qquad\text{and}\qquad
96n-144 \ \text{one-qubit rotations}.
\end{equation}

For the periodic central-difference momentum operator in Eq.~\eqref{eq:x_p_cyclic_matrices}, the two nonzero entries in the first row are
\begin{equation}\label{eq:p_nonzero_values}
\hat p^{(0)}=-\frac{i}{2\Delta x},
\qquad
\hat p^{(1)}=\frac{i}{2\Delta x}.
\end{equation}
Hence one may choose
\begin{equation}\label{eq:Np_choice}
N_{\hat p}=\frac{1}{4\Delta x^2},
\qquad
\frac{1}{\sqrt{2}\,\Delta x}=\sqrt{\mathfrak{s}N_{\hat p}}.
\end{equation}

Using the two primitives above, together with Hadamard gates on the sparse-index qubit, we define
\begin{equation}\label{eq:Up_construction}
U_p
=
\left(I^{\otimes 1}\otimes H_W\otimes I^{\otimes (2n-1)}\right)
\left(I^{\otimes 1}\otimes \left(\hat O_p^{\mathrm{BS}}\right)^\dagger\right)
\left(\hat O_p^{S}\otimes I^{\otimes n}\right)
\left(I^{\otimes 1}\otimes \hat O_p^{\mathrm{BS}}\right)
\left(I^{\otimes 1}\otimes H_W\otimes I^{\otimes (2n-1)}\right).
\end{equation}
Its action on a computational-basis state is
\begin{equation}\label{eq:Up_action}
U_p\ket{0}_{2}\ket{0}_{n-1}\ket{j}_{n}
=
\sqrt{2}\,\Delta x
\sum_{i=0}^{N_x-1}\left(\hat p_{ij}\ket{0}_{2}+J_{ij}\ket{\perp}_2\right)\ket{0}_{n-1}\ket{i}_{n},
\end{equation}
where $\bra{0}_{2}\ket{\perp}_2=0$. Since the $(n-1)$-qubit register is in the state $\ket{0}_{n-1}$ both before and after the action of $U_p$, it can be omitted from the signal subspace. Therefore,
\begin{equation}\label{eq:Up_block}
\left(\bra{0}_{2}\otimes I^{\otimes n}\right)
U_p
\left(\ket{0}_{2}\otimes I^{\otimes n}\right)
=
\sqrt{2}\,\Delta x\,\hat p.
\end{equation}
Hence $U_p$ is a $\left(\frac{1}{\sqrt{2}\,\Delta x},2,0\right)$-block-encoding of $\hat p$.

\begin{figure}[h!]
    \centering
    \includegraphics[width=0.6\linewidth]{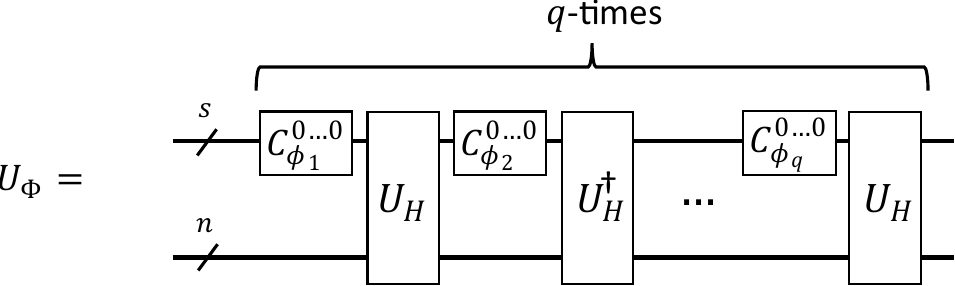}
    \caption{QSVT phase-modulation sequence built from the block-encoding $U_H$ where $H$ is a Hermitian matrix. A degree-$q$ transformation uses $q$ alternating applications of $U_H$ and $U_H^\dagger$ together with phase gadgets acting on the $s$ signal ancillas ($s=2$ for $U_p$).}
    \label{fig:qsvt_phase_modulation}
\end{figure}

A single application of $U_p$ contains one use of $\hat O_p^{S}$, one use of $\hat O_p^{\mathrm{BS}}$, one use of $(\hat O_p^{\mathrm{BS}})^\dagger$, and two Hadamard gates on the sparse-index qubit. Therefore, $U_p$ can be implemented with 
\begin{equation}\label{eq: gate count block encoding momentum}
152n-216 \ \text{CNOTs}
\qquad\text{and}\qquad
192n-284 \ \text{one-qubit rotations}.
\end{equation}
Since $\Delta x=(b-a)/(N_x-1)$ and $N_x=2^n$, Eq.~\eqref{eq:Np_choice} implies
\begin{equation}\label{eq:alpha_p_scaling}
\frac{1}{\sqrt{2}\,\Delta x}=\mathcal{O}(2^n).
\end{equation}

\begin{theorem}\label{thm:qsvt_hamiltonian}
Let $U_H$ be an $(\alpha_H,a_H,0)$-block-encoding of a Hermitian matrix $\hat H$. Then, by Corollary~18 of \cite{gilyen2019quantum}, for any $T>0$ and $\epsilon_{\mathrm{QSVT}}\in(0,1)$ there exist two phase sequences
$\Phi^{(c)}=(\phi^{(c)}_1,\ldots,\phi^{(c)}_q)$ and
$\Phi^{(s)}=(\phi^{(s)}_1,\ldots,\phi^{(s)}_q)$
such that the corresponding alternating phase-modulation sequences, as in Fig.~\ref{fig:qsvt_phase_modulation}, implement $\epsilon_{\mathrm{QSVT}}$-accurate block-encodings of $\cos(T\hat H)$ and $\sin(T\hat H)$, respectively. Consequently, by combining these two block-encodings through a linear combination of unitaries (LCU), one obtains an $\epsilon_{\mathrm{QSVT}}$-accurate block-encoding of
\begin{equation}
e^{-iT\hat H}=\cos(T\hat H)-i\sin(T\hat H).
\end{equation}
The required number $q$ of queries to $U_H$ satisfies
\begin{equation}\label{eq:qsvt_degree}
q=
\mathcal{O}\!\left(
\alpha_H T+
\frac{\log(1/\epsilon_{\mathrm{QSVT}})}
{\log\!\left(
e+\frac{\log(1/\epsilon_{\mathrm{QSVT}})}{\alpha_H T}
\right)}
\right).
\end{equation}
\end{theorem}
Consequently, for the discretized advection Hamiltonian in Eq.~\eqref{eq:circuit},
\begin{equation}\label{eq:qsvt_degree_adv}
q=
\mathcal{O}\!\left(
T\,2^n+\log(1/\epsilon_{\mathrm{QSVT}})
\right),
\end{equation}
up to problem-dependent constants independent of $n$.

Thus, the construction of $\exp(-it\alpha_j\hat p_j)$ is reduced to $q$ alternating applications of the momentum block-encoding $U_p$ and its inverse, with the dominant dependence on the spatial resolution entering through $\alpha_p=\mathcal{O}(2^n)$.
We now estimate the required gate fidelity $F_{\text{gate}}$ for different qubit counts in the Hamiltonian simulation task defined by Eq.~\eqref{eq:circuit}. We use the simple multiplicative error model \cite{Arute2019}
\begin{equation}
F = (F_{\text{gate}})^G,
\end{equation}
where $G$ denotes the total gate count and $F$ is the final-state fidelity. Within this model, the cumulative circuit fidelity decays exponentially with the number of applied gates.

Rather than prescribing a target fidelity for the final quantum state directly, we impose a target accuracy on the observable $\langle \hat{x} \rangle$, for example the mean position $\langle \hat{x} \rangle$. We define the relative error in the observable by
\begin{equation}
\varepsilon_X
:=
\frac{\bigl|\langle \hat{x} \rangle_{\text{true}}-\langle \hat{x} \rangle_{\text{approx}}\bigr|}
{|\langle \hat{x} \rangle_{\text{true}}|}.
\label{eq:relative_error_X}
\end{equation}
In the setting considered here, the wave-packet is displaced away from the origin at the terminal time, so that $|\langle \hat{x} \rangle_{\text{true}}|$ is nonzero for the instances used in the experiment, as shown in Fig.~\ref{Cluster}F of the main text.

For a general observable $X$, the expectation-value error is bounded by
\begin{equation}\label{eq:upperbound observable fidelity}
\bigl|\langle X \rangle_{\text{true}}-\langle X \rangle_{\text{approx}}\bigr|
\le
2\lVert X\rVert_{\max}\sqrt{1-F},
\end{equation}
where $\lVert X\rVert_{\max}$ denotes the operator norm of $X$. Therefore, imposing the requirement
\begin{equation}
\varepsilon_X \le r
\end{equation}
is guaranteed whenever
\begin{equation}
2\lVert X\rVert_{\max}\sqrt{1-F}
\le
r\,|\langle \hat{x} \rangle_{\text{true}}|,
\end{equation}
or equivalently,
\begin{equation}
1-F
\le
\left(
\frac{r\,|\langle \hat{x} \rangle_{\text{true}}|}{2\lVert X\rVert_{\max}}
\right)^2.
\label{eq:fidelity_bound_from_relative_error}
\end{equation}

For the advection dynamics considered in the experiment, as illustrated in Fig.~\ref{Fig1}A of the main text, the wave-packet is transported in position space and remains localized around its instantaneous mean. Let $\mathcal{D}_{\max}$ denote the maximal displacement attained during the evolution, and let $\sigma_X$ denote the characteristic spatial standard deviation of the initial condition. In order to capture the full dynamics on the chosen grid, it is sufficient to choose the position window such that
\begin{equation}
\lVert X\rVert_{\max}=b \gtrsim \mathcal{D}_{\max}+\sigma_X,
\end{equation}
consistent with the construction of the position operator in Eq.~\eqref{eq:x_p_cyclic_matrices}. At the terminal time of the evolution, the wave-packet is close to its maximally displaced location, so that its expectation value is of order
\begin{equation}
|\langle \hat{x} \rangle_{\text{true}}| \sim \mathcal{D}_{\max}.
\end{equation}
Hence,
\begin{equation}
\frac{|\langle \hat{x} \rangle_{\text{true}}|}{\lVert X\rVert_{\max}}
\sim
\frac{\mathcal{D}_{\max}}{\mathcal{D}_{\max}+\sigma_X}.
\end{equation}
For the displacement-controlled experiment, the relevant numerical scales are $\Pr(| \sigma_X|<0.5)=0.99842$, as reported in Sec.~II.B of the main text, and $\mathcal{D}_{\max}=20$, i.e.,
\begin{equation}
\sigma_X \ll \mathcal{D}_{\max},
\end{equation}
and therefore
\begin{equation}
\frac{|\langle \hat{x} \rangle_{\text{true}}|}{\lVert X\rVert_{\max}} \approx 1.
\end{equation}
Accordingly, for the purpose of resource estimation, we approximate this ratio by unity. Under this approximation, Eq.~\eqref{eq:fidelity_bound_from_relative_error} reduces to
\begin{equation}
1-F \le \left(\frac{r}{2}\right)^2.
\end{equation}

For the target relative error $r=0.07$, this yields
\begin{equation}
F_{\text{target}}
=
1-\left(\frac{0.07}{2}\right)^2
>
0.998.
\end{equation}
The corresponding required single-gate fidelity is then
\begin{equation}
F_{\text{gate}}
=
\exp\!\left(\frac{\ln F_{\text{target}}}{G}\right).
\label{eq:fidelity_required_formula}
\end{equation}
The resulting single-gate fidelity requirements are summarized in Table~\ref{table:qubit_requirements}. The estimate in Table~\ref{table:qubit_requirements} should be interpreted as a sufficient, rather than necessary, condition. The reason is that Eq.~\eqref{eq:upperbound observable fidelity} is a worst-case bound for a generic observable and does not use the detailed structure of the state, the particular form of the observable $\langle \hat{x} \rangle$, or the actual error model. Therefore, translating the $7\%$ observable-error requirement into the target fidelity $F_{\text{target}}\approx 0.998$ gives an overconservative estimate of the required single-gate fidelity. To make this dependence explicit, we also report a reference value obtained from the same multiplicative model with $F_{\text{target}}=0.5$. In other words, the column $F_{\text{single}}^{(0.5)}$ is computed as $F_{\text{single}}^{(0.5)}=\exp(\ln 0.5/G)$ and shows how close the single-gate fidelity must remain to unity even if the global target fidelity is relaxed to $50\%$.

\begin{table}[h]
    \centering
    \caption{Resource estimates for the one-dimensional Hamiltonian simulation of the advection equation for different qubit numbers using the block-encoding-based QSVT approach presented in this section. The gate counts are estimated by taking the cost of a single momentum block-encoding from Eq.~\eqref{eq: gate count block encoding momentum} and multiplying it by $2^{n+1/2}$, following Theorem~\ref{thm:qsvt_hamiltonian}. The required single-gate fidelity corresponding to a relative error of $7\%$ in $\langle \hat{x}\rangle$ is obtained from the upper bound in Eq.~\eqref{eq:upperbound observable fidelity}, which yields the target fidelity $F_{\text{target}}=0.998$. Based on this target fidelity, we compute the required single-gate fidelity $F_{\text{gate}}$. In the table, we report costs to obtain $F_{\text{gate}}^{(0.998)}$, corresponding to $F_{\text{target}}=0.998$, and $F_{\text{gate}}^{(0.5)}$, corresponding to $F_{\text{target}}=0.5$, as a reference.}
    \label{table:qubit_requirements}
    \begin{tabular}{|c|c|c|c|c|}
        \toprule
        Qubit Number & 1-Qubit Rotations & CNOTs & $F_{\text{single}}^{(0.998)}$ & $F_{\text{single}}^{(0.5)}$ \\
        \midrule
        5  & $3.06\times 10^{4}$ & $2.46\times 10^{4}$ & $1 - 2.22\times 10^{-8}$  & $1 - 1.26\times 10^{-5}$ \\
        6  & $7.86\times 10^{4}$ & $6.30\times 10^{4}$ & $1 - 8.66\times 10^{-9}$  & $1 - 4.90\times 10^{-6}$ \\
        7  & $1.92\times 10^{5}$ & $1.54\times 10^{5}$ & $1 - 3.55\times 10^{-9}$  & $1 - 2.01\times 10^{-6}$ \\
        8  & $4.53\times 10^{5}$ & $3.62\times 10^{5}$ & $1 - 1.50\times 10^{-9}$  & $1 - 8.50\times 10^{-7}$ \\
        9  & $1.05\times 10^{6}$ & $8.34\times 10^{5}$ & $1 - 6.52\times 10^{-10}$ & $1 - 3.69\times 10^{-7}$ \\
        10 & $2.37\times 10^{6}$ & $1.89\times 10^{6}$ & $1 - 2.88\times 10^{-10}$ & $1 - 1.63\times 10^{-7}$ \\
        \bottomrule
    \end{tabular}
\end{table}

\begin{table}[h!]
    \centering
    \caption{Resource estimates obtained by scaling the one-dimensional block-encoding-based QSVT estimates in Table~\ref{table:qubit_requirements} by a factor of $20000$, corresponding to $20000$ bosonic modes. The required single-gate fidelities are computed using $F_{\text{single}}=\exp(\ln F_{\text{target}}/G)$ with the scaled total gate count $G$. In the table, $F_{\text{single}}^{(0.998)}$ corresponds to $F_{\text{target}}=0.998$, and $F_{\text{single}}^{(0.5)}$ corresponds to $F_{\text{target}}=0.5$.}
    \label{tab:quantum_gate_requirements_1000modes}
    \begin{tabular}{|c|c|c|c|c|}
        \toprule
        Qubit Number & 1-Qubit Rotations & CNOTs & $F_{\text{single}}^{(0.998)}$ & $F_{\text{single}}^{(0.5)}$ \\
        \midrule
        5  & $6.12\times 10^8$   & $4.92\times 10^8$   & $1 - 1.11\times 10^{-12}$ & $1 - 6.28\times 10^{-10}$ \\
        6  & $1.572\times 10^9$  & $1.26\times 10^9$   & $1 - 4.33\times 10^{-13}$ & $1 - 2.45\times 10^{-10}$ \\
        7  & $3.84\times 10^9$   & $3.08\times 10^9$   & $1 - 1.77\times 10^{-13}$ & $1 - 1.00\times 10^{-10}$ \\
        8  & $9.06\times 10^9$   & $7.24\times 10^9$   & $1 - 7.52\times 10^{-14}$ & $1 - 4.25\times 10^{-11}$ \\
        9  & $2.10\times 10^{10}$ & $1.668\times 10^{10}$ & $1 - 3.25\times 10^{-14}$ & $1 - 1.84\times 10^{-11}$ \\
        10 & $4.74\times 10^{10}$ & $3.78\times 10^{10}$  & $1 - 1.44\times 10^{-14}$ & $1 - 8.14\times 10^{-12}$ \\
        \bottomrule
    \end{tabular}
\end{table}

The estimates in Table~\ref{tab:quantum_gate_requirements_1000modes} are obtained by applying the same fidelity model after multiplying the one-dimensional gate count by $20000$, matching the $20{,}000$-qumode cluster-state experiment described in Sec.~II of the main text. This direct upscaling is loose and should be viewed as an overconservative global-fidelity bound, rather than tight requirements for estimating a local observable. It assumes that every hypothetical gate error in every qumode contributes to the same final fidelity budget. However, the observable used in the comparison is local: an error occurring in a different qumode does not necessarily affect the measured value of $\langle \hat{x} \rangle$ on the selected qumode.

Even in the one-dimensional case, the gate overhead becomes prohibitive for near-term devices. With complexity scaling as $\mathcal{O}(2^n n)$, the resource requirements in Table~\ref{table:qubit_requirements} already exceed current hardware limits for small $n$. Maintaining the target fidelity also requires gate error rates far below state-of-the-art values: Table~\ref{table:qubit_requirements} already demands single-gate infidelities at the level of $10^{-7}$--$10^{-10}$, while the scaled estimates in Table~\ref{tab:quantum_gate_requirements_1000modes} push this requirement to $10^{-12}$--$10^{-14}$. This is clearly beyond the NISQ regime~\cite{Preskill2018quantumcomputingin}. Therefore, even without including costs in preparing $|u_0\rangle$, accurate simulation of the advection equation is beyond current devices (see Table~\ref{table:state of the art fidelities}) and would require fault-tolerant quantum simulation~\cite{PhysRevLett.77.793} to sustain the necessary coherent depths.

To make the runtime estimate, we use Google Willow~\cite{Acharya2025}, since the superconducting platform offers the fastest gate times among the devices, see Table~\ref{table:state of the art fidelities}. We also neglect topology restrictions, i.e., we assume arbitrary two-qubit gates without routing overhead, and we ignore the extra cost of error correction, so the quoted times are optimistic lower bounds. Under these assumptions, one shot of the one-dimensional circuit takes about $0.31~\mathrm{ms}$ for $n=5$ and $41.3~\mathrm{ms}$ for $n=10$. After scaling to $20000$ bosonic modes, the runtime increases to $6.1~\mathrm{s}$ per shot for $5$ qubits per qumode and to $825.7~\mathrm{s}$ per shot for $10$ qubits per qumode. Hence, even for an unrealistically perfect superconducting device, collecting useful measurement statistics would already be very costly; for example, $10^5$ shots would require about $7$ days in the first case and about $2.6$ years in the second. Consequently, the multidimensional advection equation remains prohibitive for the foreseeable future even under highly optimistic assumptions.

\begin{table}[ht!]
\centering
\caption{State-of-the-art qubit-based quantum computers’ performance. For the lifetime we use $\min(T_1,T_2)$. The first device (Google Willow) is a superconducting processor, while the second (Harvard, MIT, QuEra) is a neutral-atom processor. For neutral atoms, there are papers reporting larger systems—up to $6100$ qubits~\cite{manetsch2025tweezer}—but these do not report two-qubit gate fidelities.}
\begin{tabular}{|l|c|c|c|c|c|c|c|}
\hline
\textbf{Device} & \textbf{Qubits} & \textbf{Lifetime} & \textbf{1Q Fid.} & \textbf{2Q Fid.} & \textbf{Meas. Fid.} & \textbf{Gate time} & \textbf{Circuit depth} \\
\hline
Google Willow \cite{Acharya2025} & 105 & 68 $\mu$s & 99.965\% & 99.67\% & 99.23\% & 30 ns & 2{,}267 \\
\hline
Harvard, MIT, QuEra \cite{bluvstein2026fault} & 448 & 1--2 s & $\sim$99.9\% & 99.6\% & $\sim$99.5\% & 270 ns & $(3.7-7.4)\times 10^6$ \\
\hline
\end{tabular}
\label{table:state of the art fidelities}
\end{table}

\subsubsection{Related works}

There are several other direct, problem-specific quantum algorithms for simulating the advection equation, similar in spirit to the construction analyzed in Appendix~\ref{sec: direct}. Their technical realizations differ---QSVT, explicit time marching, Lie--Trotter product formulas, block encodings, and sparse-Hamiltonian query models---but none of them exhibit substantially different scaling in cost. In the regime relevant to our discretization, all of these approaches still contain a dominant factor proportional to $2^n$ (or worse), and therefore lead to the same basic conclusion as our current gate-count analysis: they remain beyond NISQ capabilities for the problem sizes of interest. Several of the works below state their results in terms of oracle queries rather than elementary gates; for those results, we assume that the required oracles for the discretized advection Hamiltonian have been constructed using the finite-difference representation introduced in Appendix~\ref{sec: direct}.

A problem-specific approach for constructing the time–evolution operator
$e^{-iHT}$ for the advection equation is presented in \cite{lubasch2025quantum}.
Their method uses the quantum singular value transformation (QSVT) together
with the Jacobi–Anger expansion to approximate the operator
$e^{-iT N r \sin(2 \pi \hat{k}/N_x)}$ in Fourier space. For general initial
conditions the circuit depth scales as
\[
  \mathcal{O}\!\bigl(nN( T 2^n r + \log(1/\epsilon_\text{QSVT}))\bigr),
\]
where $r$ denotes the
maximum advection velocity over all dimensions. The cost grows almost linearly with the grid size $N_x = 2^n$ per spatial dimension
and becomes exponential in $n$, similar to the gate counts in
Table~\ref{tab:quantum_gate_requirements_1000modes}. This scaling is optimal in
$\epsilon_\text{QSVT}$ but remains beyond NISQ capabilities once $n$ becomes moderate. The work also considers the case of smooth initial conditions, where the Fourier harmonics satisfy
$|k|/N_x \ll 1$, the approximation $\sin(2 \pi k / N_x) \approx 2 \pi k / N_x$
reduces the circuit depth to $\mathcal{O}(1)$ (without QFT), but this reduction applies only to strongly low-frequency initial data; in that regime, such a fine discretization is arguably unnecessary in the first place, so the practical applicability of the result is limited.

A Hamiltonian-embedded time-marching scheme for the advection equation \cite{PhysRevA.110.012430,Over_2025} discretizes the dynamics on $2^n$ spatial points per qumode, where $n$ is the number of qubits per qumode. Let $N$ denote the total number of qumodes and $T$ the total simulation time. The circuit depth derived from the iterative scheme scales as
\begin{equation}
\label{eq:related_time_marching_depth}
G=\mathcal{O}\!\left(\kappa T\,2^n\,N\,\operatorname{poly}(n)\right),
\end{equation}
where we omit $\operatorname{polylog}(1/\epsilon)$ factors for simplicity; the parameter $\kappa$ is the sparsity of the iteration matrix, which is defined from the finite-difference scheme used. The exponential factor $2^n$ immediately reproduces the prohibitive scaling trends shown in Table~\ref{tab:quantum_gate_requirements_1000modes}, so even modest values of $n$ already lead to circuit sizes and fidelity requirements far beyond present hardware. The method is nevertheless interesting conceptually because the success probability is associated with a single time step: if a step fails, the state is not destroyed, but simply remains unevolved. This improves the failure handling of the algorithm, but it does not change the asymptotic cost. Even if every time step succeeds, the scheme still requires $\sim 2^n$ time steps, so its overall scaling remains comparable to the direct construction of Appendix~\ref{sec: direct}.

Related constructions are presented in \cite{gomes2025hamiltoniansimulationadvectiondiffusionequation,hu2024quantum,sato2024hamiltonian}, where the evolution operator is implemented using Lie--Trotter product formulas. In addition to the discretization error inherent to the finite-difference approximation, these methods introduce a further product-formula error. Suppressing this additional error to $\mathcal{O}(1)$ requires
\begin{equation}
\label{eq:related_gomes_steps}
S\sim 2^{\,n(1+1/\mathfrak{q})}
\end{equation}
Trotter steps, where $\mathfrak{q}$ denotes the order of the product formula. Since each step has a gate structure comparable to the direct construction discussed above, these approaches do not remove the fundamental exponential bottleneck either.

Another recent hardware-related study is presented in \cite{li2026resource}, where the authors combine Hamiltonian embedding with Schr\"odingerization and report a trapped-ion demonstration of a two-dimensional advection equation. For the finite-difference advection setting relevant here, they provide an algorithm to retrieve an observable $\bra{\psi}\hat O\ket{\psi}$ up to error $\epsilon_{\rm obs}$ with a gate count scales as
\begin{equation}
\mathcal{O}\!\left(dn 5^p(2^n T)^{1+1/p}\log(1/\epsilon_{\rm obs})\right),
\end{equation}
where $p$ is the order of the chosen product formula. 
The Trotter number still scales at least linearly in the number $2^n$ of grid points per spatial direction. 
For their experiment with 10 grid points, they omits Richardson extrapolation and limits the Trotter--Suzuki evolution to at most two steps, corresponding to only 212 single-qubit gates and 115 two-qubit gates. 
For comparison, their Table~II shows that the case of 8 grid points already requires $767{,}438$ two-qubit gates to meet the accuracy target. Therefore, the hardware demonstration does not enforce the practical error constraints; instead, it accepts a larger Trotter error as a proof of concept.

A broader comparison is given in \cite{1fw9-h14w}, where the authors study four classes of classical methods and their quantum counterparts: linear-systems methods, time-evolution methods, quantum random walks, and QFT-based diagonalization. The key point for our purposes is that all four approaches contain an inverse-precision induced by the discretization; its contribution proportional to $1/\epsilon_\text{disc}$. In the precision regime relevant to resolving an $n$-qubit spatial register, this factor behaves as
\begin{equation}
\label{eq:related_precision_factor}
\frac{1}{\epsilon_\text{disc}}\sim 2^{2n},
\end{equation}
so the precision requirement yields an exponential runtime cost making these methods similar to one presented in this section.

An alternative line of work formulates Hamiltonian simulation in the sparse-oracle query model, where one assumes black-box access to a sparse Hamiltonian through the standard oracles
\begin{equation}
\label{eq:related_sparse_oracles}
O_H\ket{j,k,z}=\ket{j,k,z\oplus H_{jk}},
\qquad
O_F\ket{j,\ell}=\ket{j,f(j,\ell)}.
\end{equation}
Here $f(j,\ell)$ returns the column index of the $\ell$-th nonzero entry in row $j$, and $d$ is an upper bound on the row sparsity. Query complexity counts the total number of calls to $O_H$ and $O_F$. For the nearly optimal walk-plus-LCU simulator of \cite{7354428}, the number of oracle queries can be written in simplified form as
\begin{equation}
\label{eq:related_sparse_query_complexity}
Q\sim t\,d\,\|\hat H\|_{\max}.
\end{equation}
For the discretized advection Hamiltonian of Eq.~\eqref{eq:circuit} with central differences on $2^n$ grid points, we have
\begin{equation}
\label{eq:related_Hmax_advection}
\|\hat H\|_{\max}
\sim
\max_j |\alpha_j|\,\|\hat p_j\|
\sim
\frac{1}{\Delta x}
\sim
2^n,
\qquad
d\sim 3.
\end{equation}
Therefore the query count scales as
\begin{equation}
\label{eq:related_sparse_query_advection}
Q\sim t\,2^n.
\end{equation}
The same conclusion applies to the QSVT-based sparse-Hamiltonian simulator of \cite{PhysRevLett.118.010501}: although the construction is different, the dominant parameter in our setting is still $\|\hat H\|_{\max}\sim 2^n$. Hence both oracle-based approaches inherit the same exponential dependence on the spatial resolution.


Taken together, these results show that the methods may differ in the constants or dependence on $\epsilon$, but they do not remove the central bottleneck of the discretized advection problem. For the Hamiltonian in Eq.~\eqref{eq:circuit}, the dominant cost still scales exponentially with the spatial resolution through a factor of $2^n$, so the end-to-end resource requirements remain qualitatively the same as in Appendix~\ref{sec: direct}.

\subsection{Gate-based quantum simulation for multi-mode bosonic unitaries in Fock Space} \label{sec:Fock}

There is also a general qubit-based approach for simulating multi-mode bosonic unitaries. Theoretically, a single-mode Gaussian state in the infinite-dimensional Fock basis admits the expansion
\begin{equation}\label{fock_infinite}|\psi\rangle = \sum_{n=0}^{\infty} c_n|n\rangle,
\end{equation}
subject to the normalization condition $\sum_{n=0}^{\infty} |c_n|^2 = 1$. 

The overlap coefficients $c_n$ correspond to the projection of the Gaussian state onto the Fock states and are computed via the integral
\begin{equation}
\label{overlap_coeff}
c_n = \langle n | \psi \rangle = \int_{-\infty}^{\infty} \langle n | x \rangle \psi(x) \, dx,
\end{equation}
where $\psi(x) = \langle x | \psi \rangle$ is the Gaussian wavefunction in the position representation. The position-space Fock states $\langle n | x \rangle$ are given by
\begin{equation}
\langle n | x \rangle = \frac{1}{\sqrt{2^n n!}} \left( \frac{1}{\pi} \right)^{1/4} H_n\left(\frac{x}{\sqrt{2}}\right) e^{-x^2/2},
\end{equation}
with $H_n(x)$ being the Hermite polynomial of degree $n$, defined as
\begin{equation}H_n(x) = (-1)^n e^{x^2} \frac{d^n}{dx^n} e^{-x^2}.
\end{equation}
For practical qubit-based simulations, the Hilbert space must be truncated at a finite cutoff level $N_p$ (maximum photon number). This results in an approximate state
\begin{equation}
\label{fock_truncated}|\tilde{\psi}\rangle \propto \sum_{n=0}^{N_p} c_n|n\rangle,
\end{equation}
which must be renormalized such that the truncated sum of probabilities equals unity. The choice of $N_p$ determines the simulation precision by bounding the truncation error.
For a system of $N$ modes with a cutoff $N_p$, binary encoding requires a total of $n \times N$ qubits, where $n = \lceil \log_2 (N_p+1) \rceil$ is the number of qubits per mode required to represent the $N_p+1$ ($0$ to $N_p$) possible Fock states.

\begin{figure}[h]
    \centering       \includegraphics[width=0.7\linewidth]{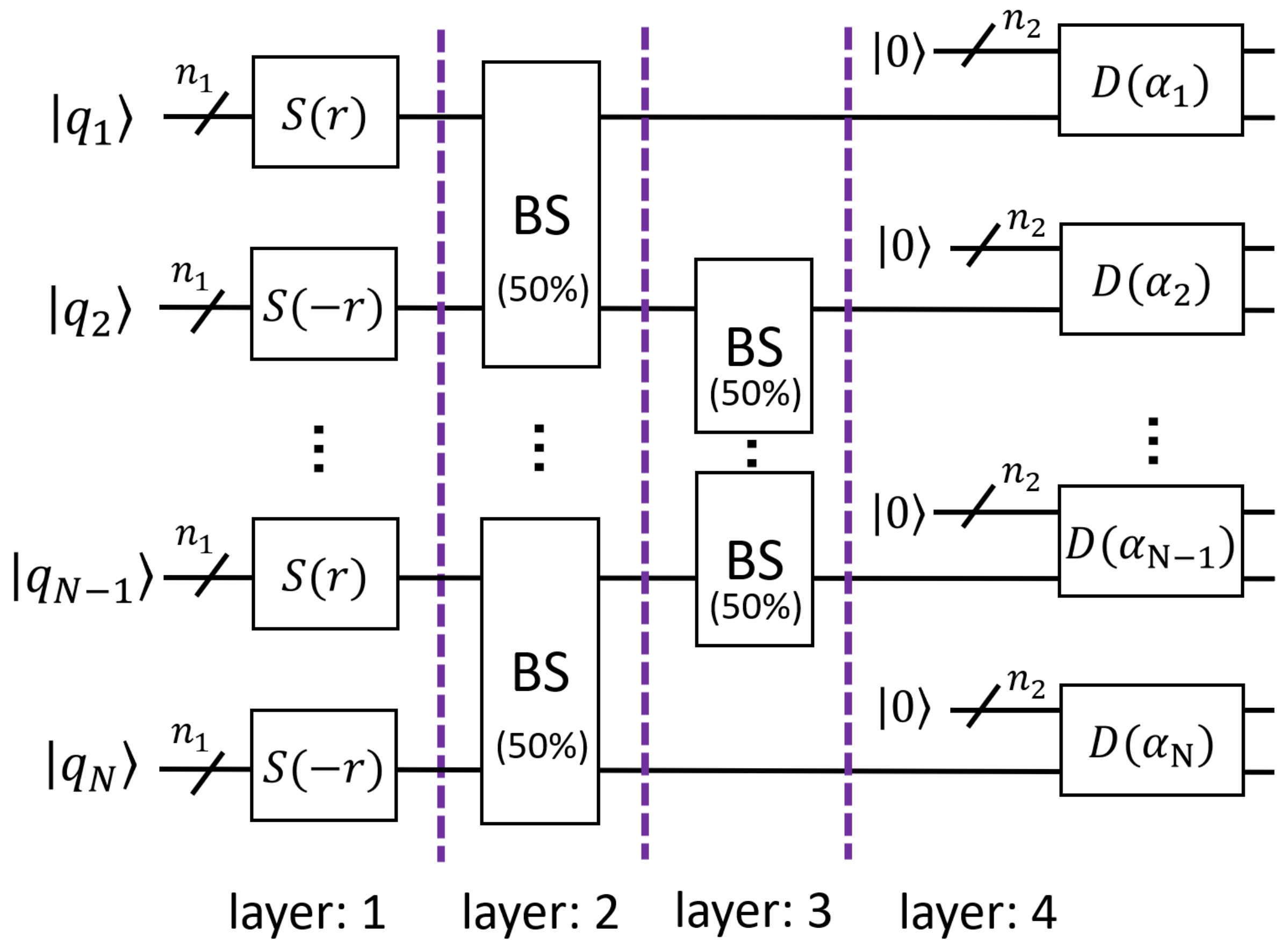}
    \caption{The circuit architecture to implement the multi-mode bosonic circuit in Fig.~\ref{fig:OriginalCircuit} with qubits in the Fock basis. 
    All unitary operations are effected via multi-qubit quantum gates. `BS' denotes a beam splitter.
    The number of qubits per mode is \(n_1\) before displacement. During applying displacement, the number of qubits per mode increases to \(n_1 + n_2\), as shown in the 4-th layer. The number of injected qubits $n_2$ is determined by the maximum anticipated displacement amplitude.}
    \label{fig:qubit-fock-arch}
\end{figure}

Starting from the vacuum state, which corresponds to $c_0 = 1$ in Eq.~\ref{fock_truncated}, we truncate the Hilbert space of each mode to a dimension $N_p+1$ and map the complex amplitudes of the Fock states into a quantum register of size $n = \lceil \log_2 (N_p+1) \rceil$. Since the $N$ modes are effectively independent at initialization, we prepare $N$ separate $n$-qubit registers and apply the target unitary operations acting on these registers.
Linear Gaussian operations, including squeezing, displacements, and beam splitting, are implemented as unitaries on the qubit register, with the simulation accuracy governed by the chosen Fock cutoff.
The quantum circuit for simulating the original multi-mode bosonic circuit in Fig.~\ref{fig:OriginalCircuit} is shown in Fig.~\ref{fig:qubit-fock-arch}.

In our protocol, the pre-displacement state $\rho$ is generated by a centered Gaussian state, satisfying the condition \(\langle \hat{\mathbf{a}}\rangle_\rho=\mathbf 0\), where \(\hat{\mathbf{a}} = (\hat{a}_1, \dots, \hat{a}_N)^T\) denotes the vector of annihilation operators. Subsequently, a final displacement layer is applied, parametrized by the complex vector \(\boldsymbol{\alpha}\in\mathbb C^N\).
Considering a single mode with an \(\hat{x}\)-quadrature displacement amplitude \(\alpha\), where \(\alpha\) is expressed in shot-noise units, we use
\begin{equation}
\mathcal{D}_{x}(\alpha)
=
\exp\!\left(-\frac{i}{2}\alpha\hat{p}\right)
\end{equation}
This convention gives
\(
\mathcal{D}_{x}^{\dagger}(\alpha)\hat{x}\mathcal{D}_{x}(\alpha)
=
\hat{x}+\alpha .
\)
The corresponding change in the mean photon number is
\begin{equation}
\langle \hat{N}\rangle_{\mathcal{D}_{x}(\alpha)\rho \mathcal{D}_{x}^{\dagger}(\alpha)}
-
\langle \hat{N}\rangle_{\rho}
=
\frac{\alpha^2}{4}
+
\alpha\,\mathrm{Re}\langle \hat{a}\rangle_{\rho}.
\end{equation}
For a centered state $\rho$ where \(\langle \hat{a}\rangle_\rho=0\), the cross-term vanishes. This implies that any nonzero displacement strictly increases the mean photon number by $|\alpha|^2/4$. Consequently, a large displacement at the end of the circuit introduces a significantly higher cutoff $N_p$ to contain the state within the truncation error bounds.
Operationally, our beam splitter is a two-mode gate, and its circuit decomposition consumes roughly quadratically more than the single-mode resources, while a displacement is single-mode and acts locally.
To reduce total simulation cost, we therefore maintain a smaller cutoff in the first three layers with each mode requiring \(n_1\) qubits. 
And then inject the necessary higher Fock levels \(n_2\) qubits to raise the cutoff only immediately before the final single-mode displacement layer, where the photon number increases significantly. To accommodate a range of potential displacement values, the number of injected qubits $n_2$ is determined by the maximum anticipated displacement amplitude.

\begin{figure}[h!]
    \centering
    \includegraphics[width=0.7\linewidth]{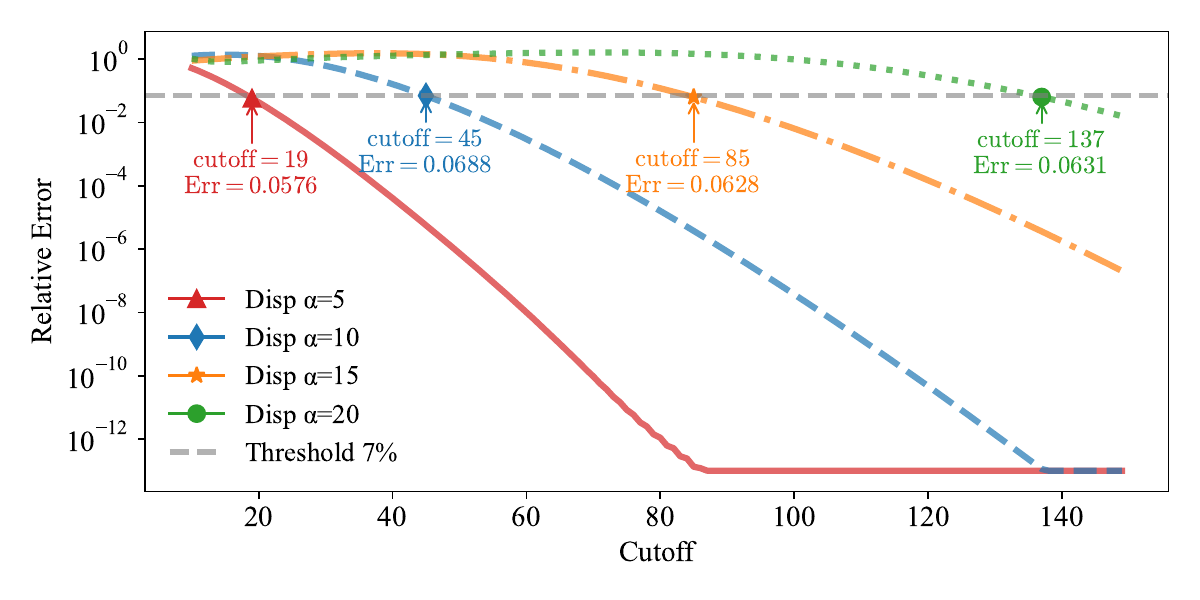}
    \caption{Relative error of the quadrature mean $\langle \hat{x} \rangle$ for the simulated final state across various Fock space cutoffs in the simplifed 2-mode case. The curves (red, blue, yellow and green) illustrate the error $\epsilon_x$ variation for displacement amplitudes $\alpha=5, 10, 15$, and $20$, respectively. Markers denote the minimum cutoff photon number $N_p$ required for each mode to satisfy a target relative error threshold of $7\%$. As displacement increases, a significantly higher cutoff is necessitated to prevent centroid shifts caused by Fock space truncation.}
    \label{fig:cutoff-no}
\end{figure}

\begin{table}[h!]
    \centering
    \caption{Resource estimation for the $N$-mode cluster state expanded by the 2-mode Fock space simulation with errorless qubits for maximum displacement value $\alpha=20$ (Cutoff $N_p$ strategy: $n_1=5, n_2=3$). Target average relative error for $\langle \hat{X_i}\rangle$ is $0.07$. The single-qubit gate count is estimated as roughly twice the CNOT count, accounting for basis changes and local operations required in standard gate decompositions.}
    \label{tab:fock_gate_requirements}
    \begin{tabular}{|c|c|c|}
        \toprule
        \textbf{Qumodes ($N$)} & \textbf{1-qubit rotations} & \textbf{CNOTs} \\
        \midrule
        128   & $2.51 \times 10^{8}$  & $1.26 \times 10^{8}$  \\
        256   & $5.04 \times 10^{8}$  & $2.52 \times 10^{8}$  \\
        512   & $1.01 \times 10^{9}$  & $5.05 \times 10^{8}$  \\
        1024  & $2.02 \times 10^{9}$  & $1.01 \times 10^{9}$  \\
        2048  & $4.05 \times 10^{9}$  & $2.02 \times 10^{9}$  \\
        4096  & $8.09 \times 10^{9}$  & $4.05 \times 10^{9}$  \\
        8192  & $1.62 \times 10^{10}$ & $8.09 \times 10^{9}$  \\
        16384 & $3.24 \times 10^{10}$ & $1.62 \times 10^{10}$ \\
        \textbf{20000} & $\boldsymbol{3.95 \times 10^{10}}$ & $\boldsymbol{1.98 \times 10^{10}}$ \\        
        32768 & $6.48 \times 10^{10}$ & $3.24 \times 10^{10}$ \\
        \bottomrule
    \end{tabular}
\end{table}

Following the universal unitary synthesis approach proposed by \cite{krol2024beyond}, 
the total number of CNOT gates required for the first three layers before displacement,
which include $N$ single-mode squeezers and $N-1$ beam splitters acting on $2n_{1}$ qubits, 
is upper bounded by
\begin{equation}
\tfrac{11}{12}16^{n_{1}}(N-1)
+ 4^{n_{1}}(3-\tfrac{61N}{24})
- \tfrac{3}{2}2^{n_{1}}N
+ 5N
- \tfrac{10}{3}.
\label{eq:cnot_bound_pre_disp}
\end{equation}
In leading order, this scales as $\mathcal{O}(16^{n_{1}}N)$, 
dominated by the beam-splitter layers.

Once displacement is applied, the number of qubits increases to $n_{1}+n_{2}$, 
and the overall CNOT count after displacement is upper bounded by
\begin{equation}
\tfrac{11}{24}4^{(n_{1}+n_{2})}N
-\tfrac{3}{2}2^{(n_{1}+n_{2})}N
+\tfrac{5}{3}N,
\label{eq:cnot_bound_post_disp}
\end{equation}
which asymptotically scales as $\mathcal{O}\!\big(4^{\,n_{1}+n_{2}}N\big)$.
Therefore, the total number of CNOT gates required for the entire circuit,
including both the pre- and post-displacement stages, is
$
\mathcal{O}\!\big((16^{n_{1}} + 4^{\,n_{1}+n_{2}})N\big)$.

We numerically simulate the qumode in the Fock basis and evaluate the simulation precision by calculating the relative error of the quadrature mean $\langle \hat{X} \rangle$ as well as $\langle \hat{X}^2 \rangle$. We define the relative error for the $i$-th qumode $\langle \hat{X_i} \rangle$ as:
\begin{equation}
\epsilon_{x_i} = \frac{|\langle \hat{X_i} \rangle_{\text{sim}} - \langle \hat{X_i} \rangle_{\text{th}}|}{|\langle \hat{X_i} \rangle_{\text{th}}|},
\end{equation}
where $\langle \hat{X_i} \rangle_{\text{th}} = \text{Re}(\alpha_i)$ is the displacement value for the $i$-th qumode as the theoretical mean. To align with our experimental precision requirements, we set a relative error threshold of $\epsilon_x \le 0.07$ which is equal to the average relative error for $\langle \hat{X_i}\rangle$ in our cluster state experiment. For cases with zero displacement where the relative error is ill-defined, we use a small displacement benchmark to determine the initial resources.

Constrained by the memory limits of classical simulation, we demonstrate a scaled-down instance of the multi-mode Bosonic circuit shown in Fig.~\ref{fig:qubit-fock-arch}. Specifically, we simulate displaced two mode squeezed states with the maximum displacement equal to 20. For a larger number of modes with larger displacement, to maintain an overall fidelity close to 0.9, each mode would require a higher number of photons. We perform simulations with the squeezing parameter $r=0.92$ and $50:50$ beam splitters.

The simulation results in Fig.~\ref{fig:cutoff-no} illustrate the resource scaling under the $\langle \hat{X} \rangle$ error metric. For a baseline displacement of $\alpha=5$, a cutoff of $N_p=19$  is required to meet the error threshold $\epsilon_x \le 0.07$, corresponding to a base register of $n_1=5$ qubits. Introducing displacement strictly increases this requirement. Specifically, to maintain the same threshold, the system requires a total of 7 qubits for both $\alpha=10$ ($N_p=45$) and $\alpha=15$ ($N_p=85$), and 8 qubits for $\alpha=20$ ($N_p=137$). 
According to the simulation result and former gate complexity approximation given by Eq.~\eqref{eq:cnot_bound_pre_disp} for the first three layers and Eq.~\eqref{eq:cnot_bound_post_disp} for the displacement layer, Table~\ref{tab:fock_gate_requirements} reports concrete counts with the implied errorless qubit gates with final error threshold $0.07$ when displacement amplitude $\alpha=20$.


\subsection{Gate-based quantum simulation for Gaussian Bosonic circuits in
continuous-variable experiments} \label{sec:covariance}

The reference \cite{PhysRevLett.134.070604} presents a qubit-based  framework for simulating Gaussian Bosonic (GB) circuits on gate-based quantum hardware platforms.
Its key advantage is exponential compression of the bosonic modes. GB circuits are fully specified by
the mean vector $\langle \hat{\mathbf z} \rangle$ and the covariance
matrix $M$. The paper assigns an $(n\!+\!1)$-qubit register to each
each object (see Eq.~\eqref{eq: encoding Gaussian bosonic}), where the number
the number of address qubits is $n=\lceil \log_2 N \rceil$
for $N$ Gaussian bosonic modes (qumodes). In this way, thus
$\langle \hat{\mathbf z} \rangle$ is encoded as a pure state
and $M$ as the density operator of a mixed state.

\begin{equation}
\begin{gathered}
\langle \hat{\mathbf z} \rangle \mapsto
\ket{\hat{\mathbf z}}_{\,n+1}
= \frac{1}{\|\langle \hat{\mathbf z} \rangle\|_2}
\sum_{m=0}^{N-1}
\Big( \langle \hat q_m \rangle \ket{0}_{\,1}
     + \langle \hat p_m \rangle \ket{1}_{\,1} \Big)
\otimes \ket{m}_{\,n}
\\[4pt]
\hat{\rho}=M/Tr{M},\qquad
(\hat{\rho})_{\alpha\beta}
= \frac{1}{\Tr{M}}
\!\left[
\tfrac{1}{2}\big\langle \hat z_\alpha \hat z_\beta + \hat z_\beta \hat z_\alpha \big\rangle
- \langle \hat z_\alpha \rangle \langle \hat z_\beta \rangle
\right],
\quad
\hat z_\alpha =
\begin{cases}
\hat q_{j}, & \alpha=2j,\\
\hat p_{j}, & \alpha=2j-1,
\end{cases}\ \ j\in\{1,\ldots,N\},
\end{gathered}
\label{eq: encoding Gaussian bosonic}
\end{equation}
where $|m\rangle_n$ denotes the computational basis state encoding mode
index $m$ on $n$ register qubits, and the
first qubit (symplectic qubit) distinguishes position ($|0\rangle_1$)
and momentum ($|1\rangle_1$) components. The normalization factors
$\|\langle \hat{z} \rangle\|_2=\sqrt{\sum_m (\langle \hat{q}_m \rangle^2+\langle \hat{p}_m \rangle^2)}$,
and normalization constant $\Tr{M}$ preserve $\Tr{\hat{\rho}}=1$ and
$\braket{\hat{\mathbf z}}=1$.

Beyond the exponential compression of modes offered by encoding
\((N=2^{n}\ \leftrightarrow\ n\!+\!1\ \text{qubits})\), this framework
provides practical benefits: (i) a constructive \emph{gate dictionary}
that compiles canonical GB operations (phase shifts, beam splitters,
single/two-mode squeezers) into few-qubit primitives acting on the
symplectic qubit and the \(n\)-qubit addressed register; (ii) structured,
shallow circuits whose depth scales with address qubits
\(n=\lceil \log_2 N\rceil\) for many global Gaussian transformations,
since multi-controlled rotations decompose into \(O(n)\) local gates; and
(iii) hardware compatibility: all required operations are standard qubit
gates, enabling mature gate-model compilation and error-mitigation stacks.

\begin{figure}[h]
    \centering
    \includegraphics[width=0.6\linewidth]{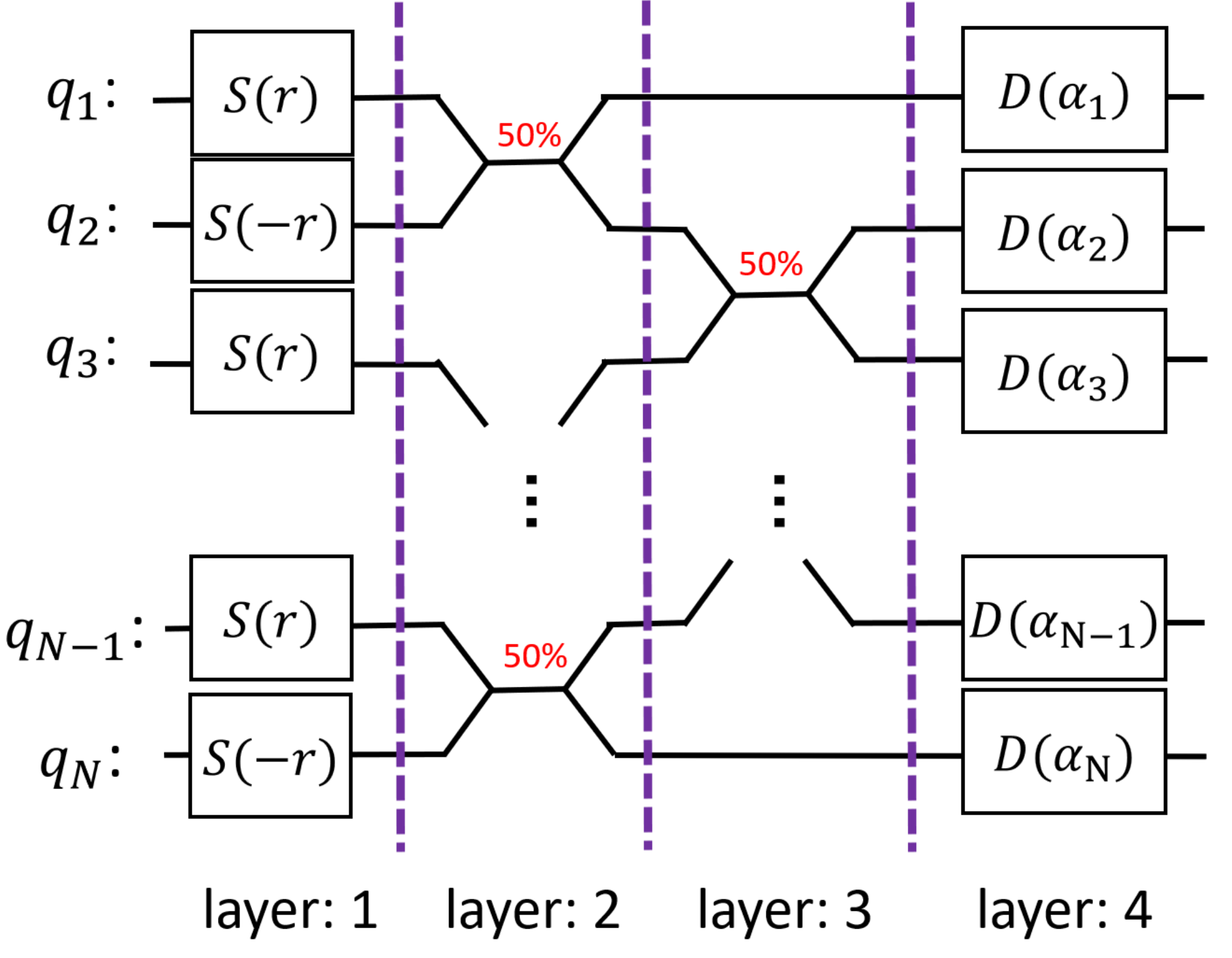}
    \caption{Circuit used in the experiment set up from the main text, decomposed into four layers and represented as spatial bosonic modes. For the qubit-based GB mapping \cite{PhysRevLett.134.070604}, Layers~1--3 are supported; Layer~4 (displacements) is not supported and thus excluded from resource counts. The corresponding initial state is the maximally mixed state  
    $\hat{\rho}_0 = \frac{\hat{I}_{2N}}{2N}$, which can be prepared relatively  
    easily on a qubit device \cite{nielsen2002quantum}.}
    \label{fig:OriginalCircuit}
\end{figure}

Let us now consider applying the qubit-based framework
\cite{PhysRevLett.134.070604} to simulate the advection system
\eqref{eq:adv eq}, assuming that the input state $|u_0\rangle$ is already available and that no resources are charged for its preparation; the comparison therefore concerns only the subsequent advection/displacement evolution and readout. The
corresponding bosonic-mode quantum circuit is depicted in
Fig.~\ref{fig:OriginalCircuit}, where for convenience we reinterpret all
qumodes as spatial quantum modes. Now, we briefly list the main problems
that prevent this from being a suitable simulator for advection equations.

\textbf{1. Inability to implement the displacement operation $\hat D(\alpha)$.} The displacement
operation (Layer~4 in Fig.~\ref{fig:OriginalCircuit}) is not supported
within this framework, as explicitly stated by the authors. The main
problem, which makes faithful simulation of the mean vector under
Hamiltonian \eqref{eq:circuit} out of reach in
\cite{PhysRevLett.134.070604}, is that the framework implements only
\emph{homogeneous linear} updates of first moments,
\[
\langle\hat z\rangle(t)=e^{t\Omega K}\,\langle\hat z\rangle(0),
\]
which, under the amplitude encoding \eqref{eq: encoding Gaussian bosonic},
correspond to linear qubit maps $|\hat z\rangle\mapsto A|\hat z\rangle$.
A displacement instead requires an affine shift
$\langle\hat z\rangle\mapsto \langle\hat z\rangle+d$, which cannot be
compiled into a linear qubit gate (and the normalization in
\eqref{eq: encoding Gaussian bosonic} makes such shifts nonlinear after
renormalization). Here $\Omega$ is the (real, antisymmetric) symplectic
form, while $K$ is real symmetric; thus $\Omega K$ is generally not
Hermitian, even though the quadratic Hamiltonian defined by $K$ is
Hermitian. In our work, applying the displacement operation corresponds
to simulating the multi-mode advection equation \eqref{eq:adv eq}
$u(t=0)\rightarrow u(t=T)$. The other operations in our main experimental
setup create the initial state $u(t=0)$. Thus, simulating the evolution
of Eq.~\eqref{eq:adv eq}, which is essential to our current work, is not
tractable within the framework in \cite{PhysRevLett.134.070604}.

\textbf{2. Inability to simulate the mean vector $\langle \hat{\mathbf z} \rangle$.} We note,
however, that despite its promising dense encoding, the method is
restricted to simulating only the covariance matrix, because simulating
the mean vector $\langle \hat{\mathbf z} \rangle$ in the qubit-based GB
framework is generally problematic. For physically relevant reference
states such as the vacuum or squeezed vacuum, the first moments vanish,
$\langle \hat q \rangle=\langle \hat p \rangle=0$. As a result, the
normalization factor in Eq.~\eqref{eq: encoding Gaussian bosonic}
collapses, and the state $\ket{\hat{\mathbf z}}$ \textit{cannot be
constructed}. A further issue with the encoding
\eqref{eq: encoding Gaussian bosonic} is the \emph{non-uniqueness induced
by normalization}: there are infinitely many distinct mean vectors that
yield the same normalized qubit description, because the overall scale
factor $\|\langle \hat{\mathbf z} \rangle\|_2$ is not encoded.
Consequently, even with full tomography of $|\hat{\mathbf z}\rangle$, one
can only reconstruct $\langle\hat{\mathbf z}\rangle$ up to an unknown
global multiplicative constant, making the correspondence between
physical mean vectors and the qubit encoding non-injective. This
limitation implies that the encoding is inherently unable to distinguish
a broad and practically important class of Gaussian states.

\textbf{3. Inability to retrieve normalization constants.} Aside from the first moments
$\langle \hat x_i\rangle,\langle \hat p_i\rangle$, which are not
simulable, in our paper we are interested in second moments of $\hat x$
and $\hat p$ like $\langle \hat x_i\hat x_j\rangle$ and
$\langle \hat p_i\hat p_j\rangle$. These objects correspond to important
quantities such as kinetic energy $\langle p_i^2\rangle\sim
\int_\mathbb{R}\abs{\frac{\partial u}{\partial x}}^2 d \vec x$, gradient
overlap $\langle p_ip_j \rangle\sim
\int_\mathbb{R}\frac{\partial u^*}{\partial x_i}\frac{\partial u}{\partial x_j}d\vec x$,
and second moments $\langle \hat x_i \hat x_j\rangle$. Let us assume we
have a density-matrix element $(\hat \rho)_{ij}$ and the amplitudes of
the mean vector $\bra{i}\ket{\hat{\textbf{z}}}$, $\bra{i}\ket{\hat{\textbf{z}}}$,
as provided by the simulation framework \cite{PhysRevLett.134.070604}.
Then, according to Eq.~\eqref{eq: encoding Gaussian bosonic}, the second
moments are expressed as
\begin{equation}
\big\langle \hat z_i \hat z_j\big\rangle=(\hat \rho)_{ij}\Tr{M}+\|\langle \hat{\mathbf z} \rangle\|_2^2\bra{i}\ket{\hat{\textbf{z}}}\bra{j}\ket{\hat{\textbf{z}}},
\end{equation}
however, the quantum states $\hat \rho$ and $\ket{\hat{\textbf{z}}}$ do
not contain information about the normalization constants $\Tr{M}$ and
$\|\langle \hat{\mathbf z} \rangle\|_2$, making estimation of the second
moments out of scope for this method.

\textbf{4. Hardness of covariance-matrix readout.} The preceding discussion fixes the
scope of the comparison: the qubit-based GB encoding stores the covariance
matrix in the normalized form given in Eq.~\eqref{eq: encoding Gaussian bosonic},
whereas our CV experiment obtains the relevant moments from homodyne
samples. To make this comparison explicit, we use the layer decomposition
shown in Fig.~\ref{fig:OriginalCircuit}, which represents the cluster-state
experiment described in Sec.~II.B of the main text.

Layer~1 prepares the squeezed input rails from vacuum. Layer~2 applies the
first $50{:}50$ beam splitter, producing two-mode squeezed correlations
within each time bin. Layer~3 implements the delay-and-second-beam-splitter
network, coupling neighboring time bins and producing the CV cluster state
used as the input to the advection step. Layer~4 applies mode-wise
displacements $D(\alpha_j)$, corresponding to the advection evolution in
Eq.~\eqref{eq:circuit}. Since displacements do not change the covariance
matrix, the covariance-readout comparison below uses the Gaussian state
generated by Layers~1--3.

The CV device does not output a covariance element directly; it produces
homodyne samples. For example, for two position quadratures $\hat x_1$ and
$\hat x_2$, the second moment is estimated from samples as
\begin{equation}
\label{eq:cv_x1x2_sample_estimator}
    \widehat{\langle \hat x_1\hat x_2\rangle}_{\rm CV}
    =
    \frac{1}{N_{\rm CVmeas}}
    \sum_{s=1}^{N_{\rm CVmeas}}
    \xi_1^{(s)}\xi_2^{(s)} ,
\end{equation}
where $\xi_i^{(s)}$ is the $s$-th homodyne sample of $\hat x_i$. After
subtracting the sample means, this gives the covariance-matrix element
$M_{12}$ used below. By Eq.~\eqref{eq: encoding Gaussian bosonic}, the
same quantity is represented in the GB encoding by the corresponding
density-matrix element,
\begin{equation}
\label{eq:M12_rho12_relation}
    M_{12}=\Tr(M)\,(\hat\rho)_{12}.
\end{equation}
Thus the readout comparison below reduces to estimating one matrix element
of $\hat\rho=M/\Tr(M)$ and translating its error back to the corresponding
covariance-matrix element.

\begin{table}[h!]
\centering
\caption{Gate counts by layer for the qubit-based GB simulation of our circuit (Layers~1--3 only). The implementation is described in \cite{PhysRevLett.134.070604}.}
\label{tab:gb_gate_counts_layer}
\begin{tabular}{|c|c|c|}
\hline
\textbf{Layer} & \textbf{CNOT count} & \textbf{Single-qubit count} \\
\hline
Layer 1 & $18$ & $32$ \\
Layer 2 & $0$  & $1$  \\
Layer 3 & $13\log_2 N - 23$ & $16\log_2 N - 29$ \\
\hline
\textbf{Total (Layers 1--3)} & $13\log_2 N - 5$ & $16\log_2 N + 4$ \\
\hline
\end{tabular}
\end{table}

For further comparison, we
consider only simulation of the density matrix in the state-preparation
stage (Layers~1--3 in Fig.~\ref{fig:OriginalCircuit}) and the readout
(retrieval of the covariance-matrix element). The corresponding qubit resources for each supported layer
(1--3) are summarized in Table~\ref{tab:gb_gate_counts_layer}.
Here $N$ denotes the number of qumodes, and
$\log_2 N$ is the number of address qubits.
In this compilation, multi-controlled rotations decompose into
$O(\log_2 N)$ two-qubit and single-qubit gates, yielding an
overall circuit depth of $O(\log N)$.

To contextualize feasibility on near-term devices,
Table~\ref{tab:gb_gate_requirements} reports concrete gate counts
and the implied per-gate fidelity target $x$ satisfying
$F = x^{G}$ with terminal fidelity $F=0.9$,
where the total operation count is taken as
the sum of all one-qubit rotations and CNOTs from
Table~\ref{tab:gb_gate_counts_layer}, similar to the analysis
made in the previous Section.

\begin{table}[h!]
    \centering
    \caption{Resource and fidelity targets for the qubit-based GB simulation (Layers~1--3 from Fig~\ref{fig:OriginalCircuit}). Terminal fidelity $F=0.9$ is translated into the single-gate fidelity requirements using the method from Appendix~\ref{sec: direct}.}
    \label{tab:gb_gate_requirements}
    \begin{tabular}{|c|c|c|c|}
        \toprule
        \textbf{Qumodes} & \textbf{1-qubit rotations} & \textbf{CNOTs} & \textbf{Single-gate fidelity required} \\
        \midrule
        128    & $116$ & $86$  & $1 - 5.21 \times 10^{-4}$ \\
        256    & $132$ & $99$  & $1 - 4.56 \times 10^{-4}$ \\
        512    & $148$ & $112$ & $1 - 4.05 \times 10^{-4}$ \\
        1024   & $164$ & $125$ & $1 - 3.65 \times 10^{-4}$ \\
        2048   & $180$ & $138$ & $1 - 3.31 \times 10^{-4}$ \\
        4096   & $196$ & $151$ & $1 - 3.04 \times 10^{-4}$ \\
        8192   & $212$ & $164$ & $1 - 2.80 \times 10^{-4}$ \\
        16384  & $228$ & $177$ & $1 - 2.60 \times 10^{-4}$ \\
        32768  & $244$ & $190$ & $1 - 2.43 \times 10^{-4}$ \\
        \bottomrule
    \end{tabular}
\end{table}

Finally, we note that even though the method
exhibits good scaling, retrieval remains the weak
spot in the qubit-based GB-circuit mapping.
Let $\hat\rho = M/\Tr(M)$ be the encoded mixed
state, and suppose we target an additive
error $\epsilon_M$ on the matrix element of $M$
according to \eqref{eq: encoding Gaussian bosonic}.
We denote by $\epsilon_\rho$ the additive error
on the matrix element $\hat\rho_{ij}$.
Because $M=\Tr(M)\,\hat\rho$, these errors satisfy
\begin{eqnarray}
\epsilon_M \;=\; \Tr(M)\,\epsilon_\rho .
\label{eq:eps_relation}
\end{eqnarray}
For Gaussian states one has $\Tr(M)=\Theta(2N)$ so
that
\begin{eqnarray}
\epsilon_\rho \sim \frac{\epsilon_M}{2N}.
\end{eqnarray}

We compare two methods by considering the task
of recovering a single density--matrix element $\rho_{ij}$.
This value can be retrieved using two simple
projective settings on the $\{\ket{i},\ket{j}\}$ subspace.
In this paper, we do not consider Heisenberg-scaling
methods achieving $\mathcal{O}(1/\epsilon)$, as those methods
require the fault-tolerant regime of computing \cite{Suzuki2020},
which in turn requires implementing error-correction codes,
making the comparison more complicated and method-dependent. Moreover, fault-tolerant regime of computing was not demonstrated in hardware.

The state-of-the-art measurement protocol \cite{tomography1element},
which does not assume coherent access, requires
$\mathcal{O}(1/\epsilon_\rho^2)$ independent copies to reach
additive error $\epsilon_\rho$, with failure probability $\delta$
controlled by a logarithmic factor $\log(1/\delta)$:
\begin{equation}
N_{\text{DVmeas}} =
\mathcal{O}\!\left(\frac{1}{\epsilon_\rho^{2}}
\log\frac{1}{\delta}\right)
=
\mathcal{O}\!\left(\frac{N^{2}}{\epsilon_M^{2}}
\log\frac{1}{\delta}\right),
\label{eq:elem_final}
\end{equation}
where $N_{\text{DVmeas}}$ denotes the number of measurements (runs)
in the Gaussian-bosonic circuit depicted in
Fig.~\ref{fig:OriginalCircuit}.
Based on the method of \cite{tomography1element} applied to the
scheme in Fig.~\ref{fig:OriginalCircuit}, and using the covariance
normalization of the Layers~1--3 cluster-state instance,
$\Tr(M)\approx 2\sqrt{2}\,N$, one can estimate the required
number of measurements as
\begin{equation}
N_{\text{DVmeas}} \approx \frac{5(\Tr\{M\})^2}{\epsilon_M^2}
\approx\frac{40N^2}{\epsilon_M^2},
\label{eq:DVsamples exact}
\end{equation}
where we neglected the $\log(1/\delta)$ term.

As a comparison, we estimate how many runs
$N_{\text{CVmeas}}$ of the continuous-variable experimental scheme
proposed in our paper, which uses photonic bosonic
modes, are needed to achieve the same accuracy.
Using the standard result from statistics for correlated
Gaussian random variables, we have
\begin{equation}
\mathrm{std}\big(\widetilde{M}_{ij}\big)\approx\sqrt{\frac{M_{ii}M_{jj}+M_{ij}^{2}}{N_\text{CVmeas}}},
\label{eq:cv_std}
\end{equation}
where $\widetilde{M}_{ij}=\sum_{s=0}^{N_\text{CVmeas-1}}\frac{(\xi_i-\langle\xi_i\rangle)(\xi_j-\langle\xi_j\rangle)}{N_\text{CVmeas}-1}$
is a sample-based estimator of $M_{ij}$.
Applying this to the Monte Carlo sampling error induced
by homodyne measurements in our scheme gives
\begin{equation}
N_{\text{CVmeas}}\approx\frac{M_{ii}M_{jj}+M_{ij}^{2}}{\epsilon_{M}^{2}}
=\mathcal{O}\!\left(\frac{1}{\epsilon_M^{2}}\right),
\label{eq:error_measurement_of_cov_matrix_element_CV}
\end{equation}
and by estimating the covariance-matrix elements
$M_{ii},M_{ij}$ for the circuit in Fig.~\ref{fig:OriginalCircuit}, we calculate
\begin{equation}
N_{\text{CVmeas}}\approx \frac{70}{\epsilon_M^2}.
\label{eq:CVsamples exact}
\end{equation}

Thus, even though the Gaussian-bosonic qubit approach admits
polylogarithmic depth, the sampling burden
\eqref{eq:DVsamples exact} governs the end-to-end cost.
To reach the same additive tolerance $\epsilon_M$, the DV route
incurs an intrinsic factor of $N^2$ in the number of
copies compared to estimating a single entry by
CV homodyne sampling \eqref{eq:CVsamples exact}.
This $N^2$ penalty on the DV side originates from the
normalization by $\Tr(M)=\Theta(N)$ when translating accuracy
from $M$ to $\rho$, and it offsets the exponential advantage
gained in the evolution stage: the measurement stage becomes
the bottleneck.

\begin{table}[h!]
\centering
\caption{The approximate number of experimental runs required to achieve error $\epsilon_M=0.01$ depending on the number of qumodes $N$ for estimating one element of the covariance matrix for the Gaussian bosonic circuit in Fig.~\ref{fig:OriginalCircuit} using the continuous-variable measurement approach (CVmeas) and the qubit-based discrete measurement approach introduced in this Section (DVmeas). For DV we also report the parallel--adjusted cost $\mathcal{M}_{\mathrm{DV}\parallel}$, which divides $N_{\mathrm{DVmeas}}$ by the number of simultaneously executable runs $N/\log_2 N$.}
\begin{tabular}{||r||r||r||r||}
\hline\hline
\textbf{$N$} &
\textbf{$N_{\mathrm{CVmeas}}$} &
\textbf{$N_{\mathrm{DVmeas}}$} &
\textbf{$\mathcal{M}_{\mathrm{DV}\parallel}$} \\
\hline\hline
128      & $7.00\times 10^{5}$  & $6.55\times 10^{9}$   & $3.58\times 10^{8}$   \\
256      & $7.00\times 10^{5}$  & $2.62\times 10^{10}$  & $8.19\times 10^{8}$   \\
512      & $7.00\times 10^{5}$  & $1.05\times 10^{11}$  & $1.84\times 10^{9}$   \\
1{,}024  & $7.00\times 10^{5}$  & $4.19\times 10^{11}$  & $4.10\times 10^{9}$   \\
2{,}048  & $7.00\times 10^{5}$  & $1.68\times 10^{12}$  & $9.01\times 10^{9}$   \\
4{,}096  & $7.00\times 10^{5}$  & $6.71\times 10^{12}$  & $1.97\times 10^{10}$  \\
8{,}192  & $7.00\times 10^{5}$  & $2.68\times 10^{13}$  & $4.26\times 10^{10}$  \\
16{,}384 & $7.00\times 10^{5}$  & $1.07\times 10^{14}$  & $9.18\times 10^{10}$  \\
32{,}768 & $7.00\times 10^{5}$  & $4.29\times 10^{14}$  & $1.97\times 10^{11}$  \\
\hline\hline
\end{tabular}
\label{table:Gausian-bosonic approximate}
\end{table}

A complementary comparison should account for the number
of physical units (detectors) available in each layout,
since the DV and CV implementations do not deploy the
same number of measurement channels (i.e., physical objects).
In the DV layout, the number of qubits and detectors is
exponentially smaller than the number of modes in the
CV layout, scaling as $\log N$ for DV versus $N$ for CV.
If one were to endow the DV platform with as many detectors
as there are CV modes, one could, in principle,
parallelize the computation; to make the baseline comparison
fair while keeping the CV scheme unchanged, we therefore
divide the DV sample count by the number of parallel runs
available on the DV device.
A single DV run uses $\log N$ qubits, while the total
hardware budget is $N$ qubits, so up to $N/\log N$ runs
can be executed concurrently.
Defining the parallel--adjusted DV cost for single--entry
estimation by
\begin{equation}
\begin{aligned}
&\mathcal{M}_{\text{DV}\parallel}
:=
\frac{N_{\text{DVmeas}}}{N/\log N}
=
\mathcal{O}\!\left(\frac{N\log N}{\epsilon_M^{2}}\log\frac{1}{\delta}\right)
\approx
\frac{40\,N\log_2 N}{\epsilon_M^2},
\end{aligned}
\end{equation}
we see that even after allowing maximal DV parallelism,
the CV scheme retains the more favorable scaling in \(N\)
for achieving the same accuracy; thus the overall supremacy
of our CV method is \(\mathcal{O}(N\log N)\).
We conclude our comparison by providing an approximate
number of experimental runs needed to simulate
Eq.~\eqref{eq:adv eq} by both methods in
Table~\ref{table:Gausian-bosonic approximate}.

\subsection{Computational runtime comparison}

We now combine the measurement counts in
Table~\ref{table:Gausian-bosonic approximate} with the gate counts
in Table~\ref{tab:gb_gate_counts_layer} to estimate computational runtime of estimating covariance matrix element $M_{ij}$. We highlight that the embedding \eqref{eq: encoding Gaussian bosonic} doesn't contains information about the $\Tr{M}$, so \textbf{the readout analysis below possible only if the $\Tr{M}$ is known}.

For a single DV run on \(N\) bosonic modes
(with \(n=\log_2 N\)), the circuit uses
\((16n+4)\) single-qubit gates, \((13n-5)\) two-qubit gates,
and one measurement and reset.
Therefore, the per-run latency is
\begin{equation}
T_{\mathrm{run}}(N)
=
(16\log_2 N+4)\,t_{1\mathrm{Q}}
+
(13\log_2 N-5)\,t_{2\mathrm{Q}}
+
t_{\mathrm{meas}}+t_{\mathrm{reset}}.
\label{eq:dv_per_run_time}
\end{equation}
The total serial runtime is
\(T_{\mathrm{serial}}=N_{\mathrm{DVmeas}}\,T_{\mathrm{run}}(N)\).
If we allow full detector-level parallelism, i.e.,
\(N/\log_2 N\) runs executed concurrently, the parallelized
runtime becomes
\(T_{\parallel}=\mathcal{M}_{\mathrm{DV}\parallel}\,T_{\mathrm{run}}(N)\),
where
\(\mathcal{M}_{\mathrm{DV}\parallel}=N_{\mathrm{DVmeas}}/(N/\log_2 N)\).

Using these expressions, we provide total runtime estimates
for different qubit-based quantum-computing architectures in
Tables~\ref{table: measurements superconductor}--
\ref{table:dv_runtime_photonics_0.9375}, including state-of-the-art
superconducting devices, cold atoms, trapped ions, and photonic
qubit-based quantum computers.

\begin{table}[h!]
\centering
\renewcommand{\arraystretch}{1.25}
\caption{Estimated DV runtimes on Google Willow~\cite{Acharya2025}. Per-run latency \(T_{\mathrm{run}}(N)\) uses Eq.~\eqref{eq:dv_per_run_time} with \(t_{1\mathrm{Q}}=25~\mathrm{ns}\), \(t_{2\mathrm{Q}}=32~\mathrm{ns}\), \(t_{\mathrm{meas}}=500~\mathrm{ns}\), \(t_{\mathrm{reset}}=400~\mathrm{ns}\). Serial and parallel totals are obtained by multiplying \(T_{\mathrm{run}}(N)\) by \(N_{\mathrm{DVmeas}}\) and by \(\mathcal{M}_{\mathrm{DV}\parallel}\), respectively, using the measurement counts from Table~\ref{table:Gausian-bosonic approximate}.}
\begin{tabular}{||r||r|r||r|r||}
\hline\hline
\textbf{$N$} & \textbf{$n=\log_2 N$} & \textbf{$T_{\mathrm{run}}(N)$ (Willow)} & \textbf{$T_{\mathrm{serial}}$} & \textbf{$T_{\parallel}$} \\
\hline\hline
$128$      & $7$  & $6.552\,\mu\mathrm{s}$  & $11.9\,\mathrm{h}$        & $39.1\,\mathrm{min}$ \\
$256$      & $8$  & $7.368\,\mu\mathrm{s}$  & $2.2\,\mathrm{d}$         & $1.7\,\mathrm{h}$    \\
$512$      & $9$  & $8.184\,\mu\mathrm{s}$  & $9.9\,\mathrm{d}$         & $4.2\,\mathrm{h}$    \\
$1{,}024$  & $10$ & $9.000\,\mu\mathrm{s}$  & $43.6\,\mathrm{d}$        & $10.2\,\mathrm{h}$   \\
$2{,}048$  & $11$ & $9.816\,\mu\mathrm{s}$  & $191\,\mathrm{d}$         & $1.0\,\mathrm{d}$    \\
$4{,}096$  & $12$ & $10.632\,\mu\mathrm{s}$ & $2.3\,\mathrm{yr}$        & $2.4\,\mathrm{d}$    \\
$8{,}192$  & $13$ & $11.448\,\mu\mathrm{s}$ & $9.7\,\mathrm{yr}$        & $5.6\,\mathrm{d}$    \\
$16{,}384$ & $14$ & $12.264\,\mu\mathrm{s}$ & $41.6\,\mathrm{yr}$       & $13.0\,\mathrm{d}$   \\
$32{,}768$ & $15$ & $13.080\,\mu\mathrm{s}$ & $177.9\,\mathrm{yr}$      & $29.8\,\mathrm{d}$   \\
\hline\hline
\end{tabular}
\label{table: measurements superconductor}
\end{table}

\begin{table}[h!]
\centering
\renewcommand{\arraystretch}{1.25}
\caption{Estimated DV runtimes on a cold-atom neutral-atom platform. Per-run latency $T_{\mathrm{run}}(N)$ uses Eq.~\eqref{eq:dv_per_run_time} with single- and two-qubit gate times $t_{1\mathrm{Q}}=250~\mathrm{ns}$ and $t_{2\mathrm{Q}}=416~\mathrm{ns}$ from Ref.~\cite{radnaev2025universalneutralatomquantumcomputer}, and dead time $t_{\mathrm{meas}}+t_{\mathrm{reset}}=2.9~\mathrm{ms}$ from Ref.~\cite{Finkelstein2024}. Serial and parallel totals are obtained by multiplying $T_{\mathrm{run}}(N)$ by $N_{\mathrm{DVmeas}}$ and by $\mathcal{M}_{\mathrm{DV}\parallel}$, respectively, using the measurement counts from Table~\ref{table:Gausian-bosonic approximate}.}
\begin{tabular}{||r||r|r||r|r||}
\hline\hline
\textbf{$N$} & \textbf{$n=\log_2 N$} & \textbf{$T_{\mathrm{run}}(N)$ (cold atoms)} & \textbf{$T_{\mathrm{serial}}$} & \textbf{$T_{\parallel}$} \\
\hline\hline
$128$     & $7$  & $2.96\,\mathrm{ms}$ & $225\,\mathrm{d}$               & $12.3\,\mathrm{d}$     \\
$256$     & $8$  & $2.97\,\mathrm{ms}$ & $2.5\,\mathrm{yr}$              & $28\,\mathrm{d}$       \\
$512$     & $9$  & $2.98\,\mathrm{ms}$ & $9.9\,\mathrm{yr}$              & $64\,\mathrm{d}$       \\
$1{,}024$ & $10$ & $2.99\,\mathrm{ms}$ & $40\,\mathrm{yr}$               & $142\,\mathrm{d}$      \\
$2{,}048$ & $11$ & $3.00\,\mathrm{ms}$ & $1.6\times 10^{2}\,\mathrm{yr}$ & $313\,\mathrm{d}$      \\
$4{,}096$ & $12$ & $3.01\,\mathrm{ms}$ & $6.4\times 10^{2}\,\mathrm{yr}$ & $2.0\,\mathrm{yr}$     \\
$8{,}192$ & $13$ & $3.02\,\mathrm{ms}$ & $2.6\times 10^{3}\,\mathrm{yr}$ & $4.1\,\mathrm{yr}$     \\
$16{,}384$& $14$ & $3.03\,\mathrm{ms}$ & $1.0\times 10^{4}\,\mathrm{yr}$ & $8.8\,\mathrm{yr}$     \\
$32{,}768$& $15$ & $3.04\,\mathrm{ms}$ & $4.1\times 10^{4}\,\mathrm{yr}$ & $19\,\mathrm{yr}$      \\
\hline\hline
\end{tabular}
\label{table: measurements coldatom}
\end{table}

\begin{table}[h!]
\centering
\label{table: measurements ion}
\caption{Estimated DV runtimes on a trapped-ion platform. Per-run latency $T_{\mathrm{run}}(N)$ uses Eq.~\eqref{eq:dv_per_run_time}, with single- and two-qubit gate times $t_{1\mathrm{Q}}=110~\mu\mathrm{s}$ and $t_{2\mathrm{Q}}\approx 900~\mu\mathrm{s}$ and a cooling overhead $t_{\mathrm{cool}}\approx 3~\mathrm{ms}$ \cite{chen2024benchmarking}, together with a conservative trapped-ion measurement time $t_{\mathrm{meas}}\approx 125~\mu\mathrm{s}$ \cite{reens2022high}. Serial and parallel totals $T_{\mathrm{serial}}$ and $T_{\parallel}$ are obtained by multiplying $T_{\mathrm{run}}(N)$ by $N_{\mathrm{DVmeas}}$ and by $\mathcal{M}_{\mathrm{DV}\parallel}$, respectively, using the measurement counts from Table~\ref{table:Gausian-bosonic approximate}.}
\renewcommand{\arraystretch}{1.25}
\begin{tabular}{||r||r|r||r|r||}
\hline\hline
\textbf{$N$} & \textbf{$n=\log_2 N$} & \textbf{$T_{\mathrm{run}}(N)$ (trapped ions)} & \textbf{$T_{\mathrm{serial}}$} & \textbf{$T_{\parallel}$} \\
\hline\hline
$128$     & $7$  & $143.27\,\mathrm{ms}$ & $29.8\,\mathrm{yr}$              & $1.6\,\mathrm{yr}$              \\
$256$     & $8$  & $163.62\,\mathrm{ms}$ & $1.4\times 10^{2}\,\mathrm{yr}$  & $4.2\,\mathrm{yr}$              \\
$512$     & $9$  & $183.97\,\mathrm{ms}$ & $6.1\times 10^{2}\,\mathrm{yr}$  & $10.8\,\mathrm{yr}$             \\
$1{,}024$ & $10$ & $204.33\,\mathrm{ms}$ & $2.7\times 10^{3}\,\mathrm{yr}$  & $26.6\,\mathrm{yr}$             \\
$2{,}048$ & $11$ & $224.68\,\mathrm{ms}$ & $1.2\times 10^{4}\,\mathrm{yr}$  & $64.2\,\mathrm{yr}$             \\
$4{,}096$ & $12$ & $245.03\,\mathrm{ms}$ & $5.2\times 10^{4}\,\mathrm{yr}$  & $1.5\times 10^{2}\,\mathrm{yr}$ \\
$8{,}192$ & $13$ & $265.39\,\mathrm{ms}$ & $2.3\times 10^{5}\,\mathrm{yr}$  & $3.6\times 10^{2}\,\mathrm{yr}$ \\
$16{,}384$& $14$ & $285.74\,\mathrm{ms}$ & $9.7\times 10^{5}\,\mathrm{yr}$  & $8.3\times 10^{2}\,\mathrm{yr}$ \\
$32{,}768$& $15$ & $306.10\,\mathrm{ms}$ & $4.2\times 10^{6}\,\mathrm{yr}$  & $1.9\times 10^{3}\,\mathrm{yr}$ \\
\hline\hline
\end{tabular}
\label{tab:dv_runtime_ions}
\end{table}

\begin{table}[h!]
\centering
\renewcommand{\arraystretch}{1.25}
\caption{Estimated DV photonic runtimes on $p_{\mathrm{2Q}}= 0.6048$. Serial and parallel totals are obtained by multiplying \(T_{\mathrm{run}}(N)\) by \(N_{\mathrm{DVmeas}}\) and by \(\mathcal{M}_{\mathrm{DV}\parallel}\), respectively, using the measurement counts from Table~\ref{table:Gausian-bosonic approximate}.}
\begin{tabular}{||r||r|r||r|r||}
\hline\hline
\textbf{$N$} & \textbf{$n=\log_2 N$} & \textbf{$T_{\mathrm{run}}(N)$ ($p_{\mathrm{2Q}}= 0.6048$)} & \textbf{$T_{\mathrm{serial}}$} & \textbf{$T_{\parallel}$} \\
\hline\hline
$64$      & $6$  & $2.2\,\mathrm{yr}$ & $3.6\times10^{9}\,\mathrm{yr}$       & $3.4\times10^{8}\,\mathrm{yr}$ \\
$256$     & $8$  & $1.1\times10^{6}\,\mathrm{yr}$ & $2.8\times10^{16}\,\mathrm{yr}$       & $8.7\times10^{14}\,\mathrm{yr}$    \\
$1{,}024$ & $10$ & $5.0\times10^{11}\,\mathrm{yr}$ & $2.1\times10^{23}\,\mathrm{yr}$      & $2.1\times10^{21}\,\mathrm{yr}$    \\
$2{,}048$ & $11$ & $3.5\times10^{14}\,\mathrm{yr}$ & $5.8\times10^{26}\,\mathrm{yr}$     & $3.1\times10^{24}\,\mathrm{yr}$   \\
$4{,}096$ & $12$ & $2.4\times10^{17}\,\mathrm{yr}$ & $1.6\times10^{30}\,\mathrm{yr}$     & $4.7\times10^{27}\,\mathrm{yr}$    \\
$16{,}384$& $14$ & $1.1\times10^{23}\,\mathrm{yr}$ & $1.2\times10^{37}\,\mathrm{yr}$   & $1.1\times10^{34}\,\mathrm{yr}$    \\
\hline\hline
\end{tabular}
\label{table:dv_runtime_photonics_0.6048}
\end{table}

\begin{table}[h!]
\centering
\renewcommand{\arraystretch}{1.25}
\caption{Estimated DV photonic runtimes on $p_{\mathrm{2Q}}= 0.9375$. Serial and parallel totals are obtained by multiplying \(T_{\mathrm{run}}(N)\) by \(N_{\mathrm{DVmeas}}\) and by \(\mathcal{M}_{\mathrm{DV}\parallel}\), respectively, using the measurement counts from Table~\ref{table:Gausian-bosonic approximate}.}
\begin{tabular}{||r||r|r||r|r||}
\hline\hline
\textbf{$N$} & \textbf{$n=\log_2 N$} & \textbf{$T_{\mathrm{run}}(N)$ ($p_{\mathrm{2Q}}= 0.9375$)} & \textbf{$T_{\mathrm{serial}}$} & \textbf{$T_{\parallel}$} \\
\hline\hline
$64$      & $6$  & $0.898\,\mu\mathrm{s}$ & $24.5\,\mathrm{min}$       & $2.3\,\mathrm{min}$ \\
$256$     & $8$  & $4.772\,\mu\mathrm{s}$ & $1.4\,\mathrm{d}$       & $1.1\,\mathrm{h}$    \\
$1{,}024$ & $10$ & $25.516\,\mu\mathrm{s}$ & $123.9\,\mathrm{d}$      & $1.2\,\mathrm{d}$    \\
$2{,}048$ & $11$ & $59.035\,\mu\mathrm{s}$ & $3.1\,\mathrm{yr}$     & $6.2\,\mathrm{d}$   \\
$4{,}096$ & $12$ & $136.601\,\mu\mathrm{s}$ & $29.1\,\mathrm{yr}$     & $31.1\,\mathrm{d}$    \\
$16{,}384$& $14$ & $731.440\,\mu\mathrm{s}$ & $2490.4\,\mathrm{yr}$   & $2.1\,\mathrm{yr}$    \\
\hline\hline
\end{tabular}
\label{table:dv_runtime_photonics_0.9375}
\end{table}

\begin{table}[h!]
\centering
\renewcommand{\arraystretch}{1.25}
\caption{Runtime estimates for the continuous-variable photonic implementation of the cluster-state advection experiment. The benchmark is the time-domain two-rail CV cluster-state experiment with mode-resolved displacement and BHD A/B homodyne readout described in Sec.~II.B of the main text. Here $T_{\mathrm{run}}(N)$ denotes the acquisition time for one $N$-qumode experimental run, $T_{\mathrm{serial}}=N_{\mathrm{CVmeas}}T_{\mathrm{run}}(N)$ uses $N_{\mathrm{CVmeas}}=7.0\times10^{5}$ shots for $\epsilon_M=0.01$ from Eq.~\eqref{eq:CVsamples exact}, and $T_{\parallel}=T_{\mathrm{serial}}/2$ accounts for the two homodyne rails measured in parallel.}

\begin{tabular}{||r||r|r||r|r||}
\hline\hline
\textbf{$N$} & \textbf{$n=\log_2 N$} & \textbf{$T_{\mathrm{run}}(N)$ (Our work)} & \textbf{$T_{\mathrm{serial}}$} & \textbf{$T_{\parallel}$} \\
\hline\hline
$64$      & $6$  & $0.286\,\mu\mathrm{s}$ & $0.2\,\mathrm{s}$       & $0.1\,\mathrm{s}$ \\
$256$     & $8$  & $1.147\,\mu\mathrm{s}$ & $0.8\,\mathrm{s}$       & $0.4\,\mathrm{s}$    \\
$1{,}024$ & $10$ & $4.586\,\mu\mathrm{s}$ & $3.2\,\mathrm{s}$      & $1.6\,\mathrm{s}$    \\
$2{,}048$ & $11$ & $9.173\,\mu\mathrm{s}$ & $6.4\,\mathrm{s}$     & $3.2\,\mathrm{s}$   \\
$4{,}096$ & $12$ & $18.346\,\mu\mathrm{s}$ & $12.8\,\mathrm{s}$     & $6.4\,\mathrm{s}$    \\
$16{,}384$& $14$ & $73.382\,\mu\mathrm{s}$ & $51.4\,\mathrm{s}$   & $25.7\,\mathrm{s}$    \\
\hline\hline
\end{tabular}
\label{table:cv_runtime_photonics}
\end{table}

By contrast, in a DV photonic implementation with $n=log_2 N$ spatial modes and an array of single-photon detectors (e.g. superconducting nanowire single-photon detectors, SNSPDs) measuring all modes in parallel, one run is essentially set by the repetition period of the optical pump, gate implementation, and the detector dead time
\begin{equation}
\label{eq:photonics_dv_run_time}
    T^{\mathrm{SPD}}_{\mathrm{run}}(N)\approx t_{\mathrm{prop}}+t_{\mathrm{gate}}+t_{\mathrm{dead}}
\end{equation}
On the integrated photonic platform of Ref.~\cite{DV_photonics_chip_exp}, the detector dead time and the pump repetition rate are tightly linked, since the inverse dead time sets the maximum sustainable count rate and the optical pump is engineered not to exceed this limit. 
In their Hong–Ou–Mandel experiments the device operates at a pump repetition rate $f_{\mathrm{rep}}\approx 125~\mathrm{MHz}$, corresponding to a repetition period $t_{\mathrm{prop}} + t_{\mathrm{dead}} \approx 1/f_{\mathrm{rep}} \approx 8~\mathrm{ns}$. 
Time-resolved traces of the on-chip photon-number-resolving detectors indicate an effective response window of order $1\text{–}2~\mathrm{ns}$, which is comfortably shorter than $T_{\mathrm{rep}}$ and therefore compatible with shot-to-shot operation at this rate. Neglecting non-idealities such as finite detection efficiency and dark counts, one can treat the per-attempt latency of a DV photonic experiment as essentially bounded by $1/f_{\mathrm{rep}}$, and this latency does not scale with the number of spatial modes $n$, provided that a sufficient number of detection channels is available.


A crucial difference with respect to deterministic superconducting qubits arises from the probabilistic nature of two-qubit gates in DV photonic schemes. 
In standard linear optical quantum computing, measurement-induced Bell-type entangling operations, commonly referred to as fusion gates, underpin the construction of two-qubit interactions. 
In the canonical Browne–Rudolph scheme, a type-II fusion gate succeeds with probability $p_{\mathrm{2Q}}= 0.5$ in the idealized lossless limit, whereas single-qubit linear-optical transformations are deterministic. 
Boosted schemes employing ancillary photons can increase this probability beyond 0.5 \cite{DV_photonics_bell_theo}. 
In theory, boosted qubit-level fusion gates can achieve a success probability of $p_{\mathrm{2Q}}=1-2^{-(k+1)}$ when supplied with $2(2^k-1)$ ancilla photons \cite{DV_photonics_bell_theo_newest}. 
Experimentally, the current state of the art reports a success probability of 60.48\% \cite{2025_DV_Bell_measure_success_exp}.

\begin{figure}[h!]
    \centering

    \begin{subfigure}{0.274\linewidth}
        \centering
        \includegraphics[width=\linewidth]{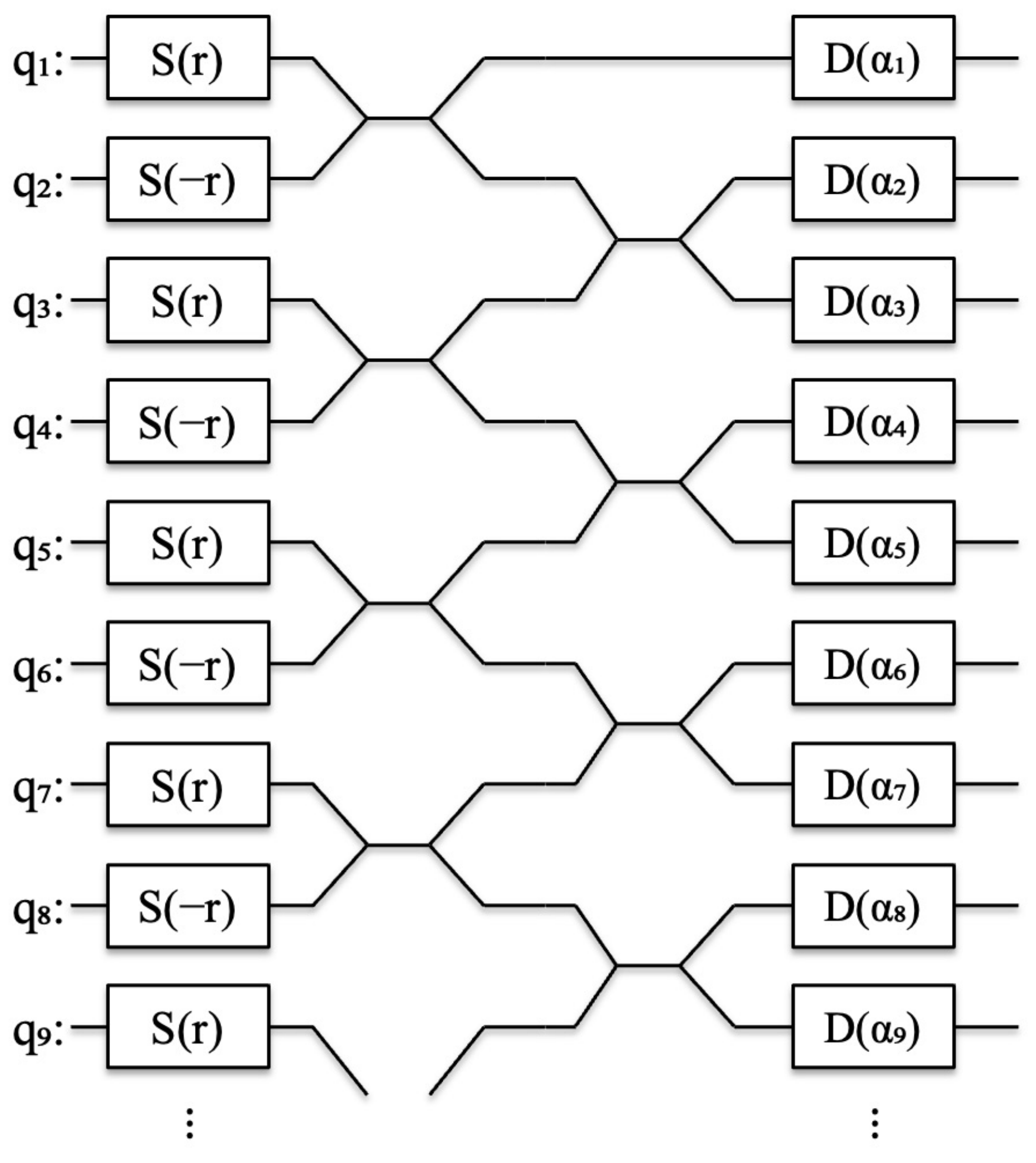}
        \caption{Full $N$-mode circuit.}
        \label{fig:lightcone_full_circuit}
    \end{subfigure}
    \hfill
    \begin{subfigure}{0.315\linewidth}
        \centering
        \includegraphics[width=\linewidth]{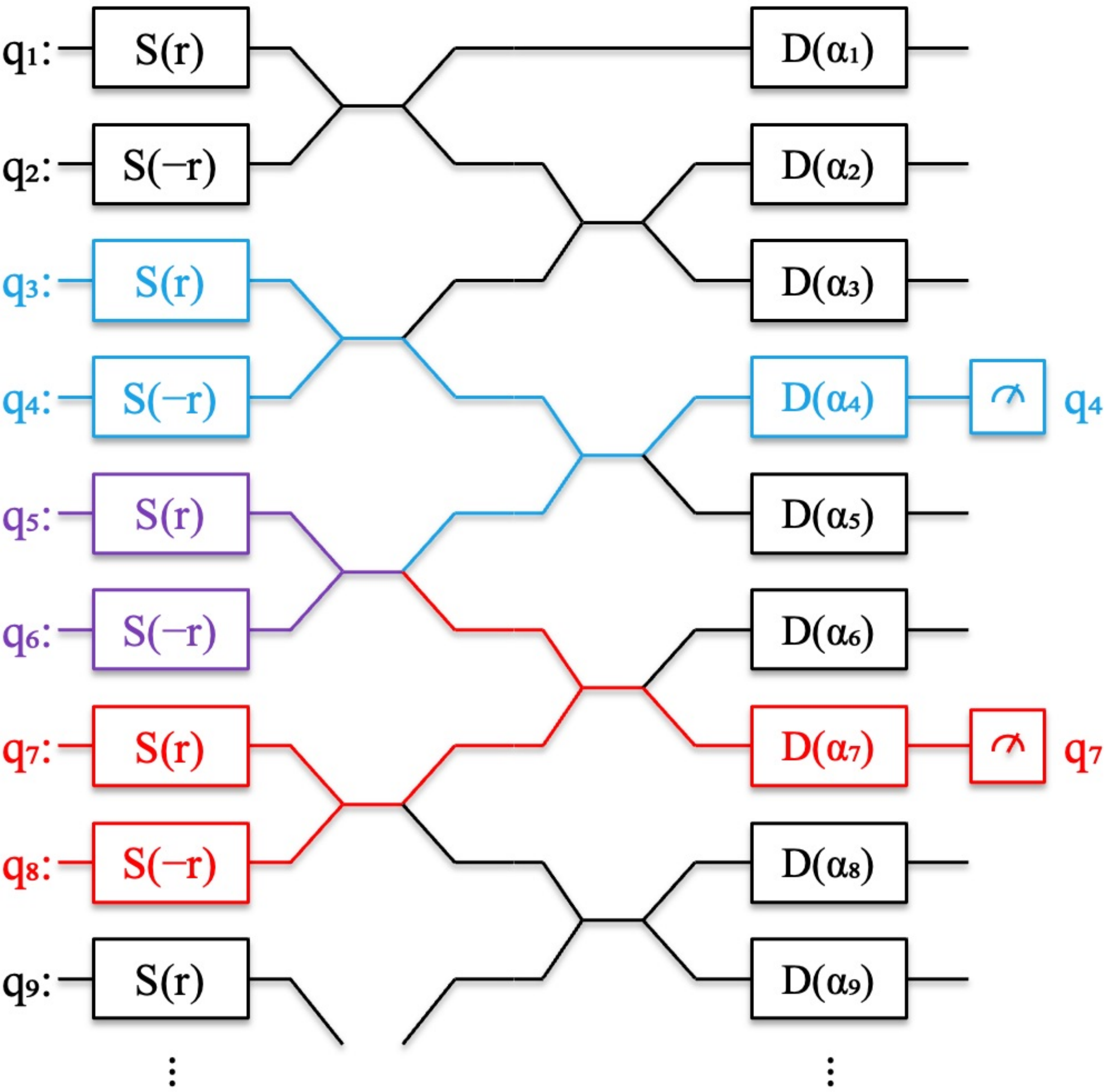}
        \caption{Backward light cone for $\langle \hat X_4 \hat X_7\rangle$.}
        \label{fig:lightcone_highlighted}
    \end{subfigure}
    \hfill
    \begin{subfigure}{0.32\linewidth}
        \centering
        \includegraphics[width=\linewidth]{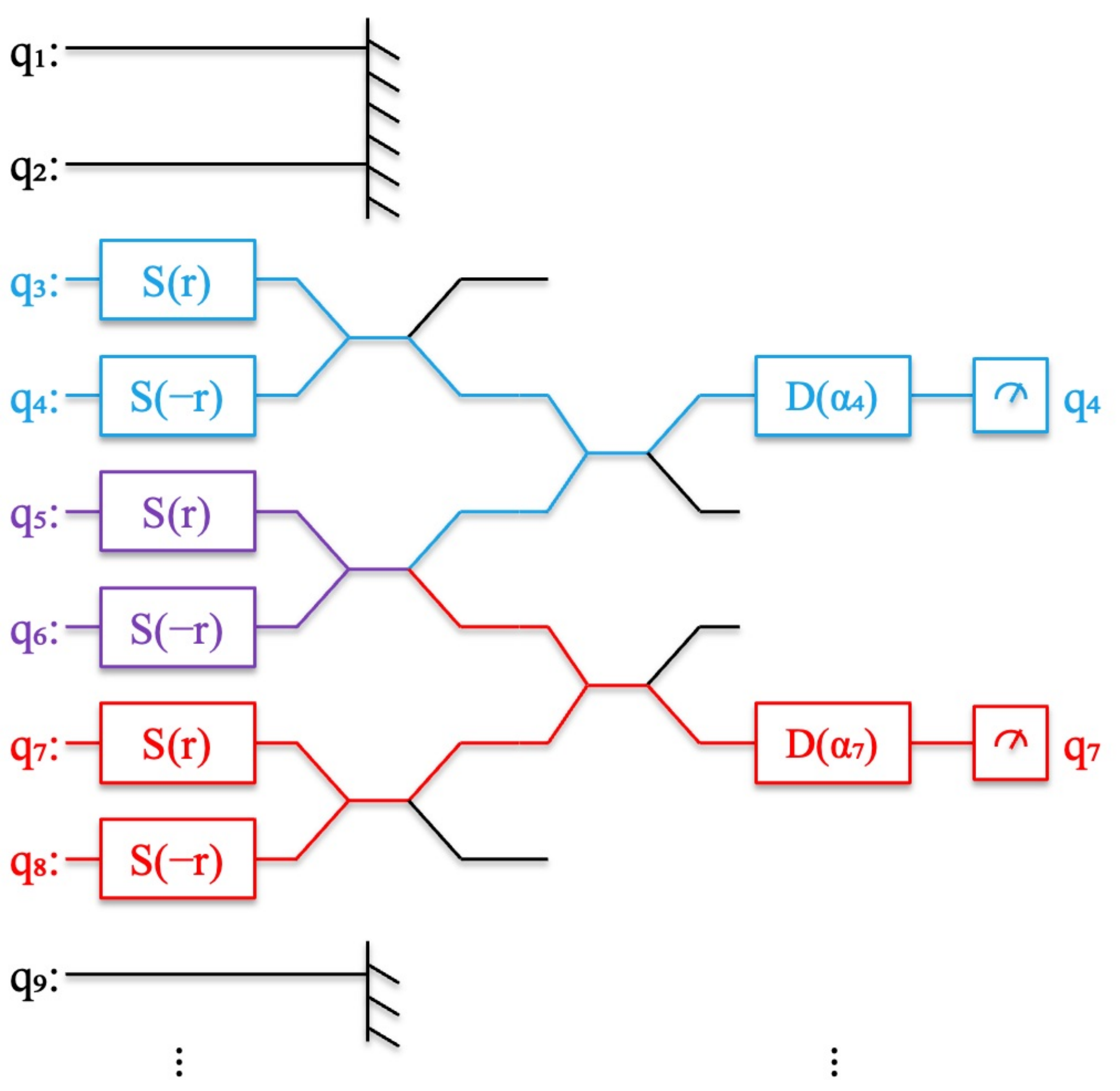}
        \caption{Clipped circuit on $q_3,\ldots,q_{8}$.}
        \label{fig:lightcone_clipped}
    \end{subfigure}

    \caption{Problem-specific reduction of the quantum scheme in Fig.~\ref{fig:OriginalCircuit}.
    Panel (a) shows the full circuit.
    Panel (b) highlights the backward light cones relevant for evaluating
    the covariance element $\langle \hat X_4 \hat X_7\rangle$:
    blue and red denote the parts associated with $q_4$ and $q_7$,
    respectively, while purple denotes their overlap.
    Panel (c) shows the corresponding clipped circuit, which contains
    only the modes $q_3,\ldots,q_{8}$ with the same color coding.}
    \label{fig:problem_specific_lightcone}
\end{figure}

In our circuit, one DV run contains $(13n-5)$ two-qubit gates that must succeed within a single run of the circuit. 
Since all heralding conditions must be simultaneously satisfied, the probability that a full run succeeds is $p_{\mathrm{2Q}}^{13n-5}$, and the expected number of experimental attempts required to obtain one completely successful realization is $1/p_{\mathrm{2Q}}^{13n-5}$. 
Under these assumptions, and neglecting the comparatively negligible execution time of single-qubit gates, the expected wall-clock time needed to realize one successful DV photonic run is thus

\begin{equation}
\label{eq:photonics_dv_gate_run_time}
    t_{\mathrm{gate}} \approx \frac{1}{p_{\mathrm{2Q}}^{13n-5}} \cdot \frac{1}{f_{\mathrm{rep}}}
\end{equation}

Using the highest experimentally demonstrated success probability, $p_{\mathrm{2Q}}= 0.6048$, and neglecting the resource consumption associated with ancillary photons, detectors, and temporal overhead, the estimated runtimes are summarized in Table~\ref{table:dv_runtime_photonics_0.6048}.

Regarding the boosted fusion scheme described in Ref.~\cite{DV_photonics_bell_theo_newest}, setting $k=3$, corresponding to ancilla states of the form $\bigotimes_{q=1}^{k=3} \bigl| \mathrm{GHZ}^{(2q)}_\nu \bigr\rangle$, yields a theoretical success probability of 93.75\%. Under the same assumptions of disregarding ancillary-state generation cost and detection overhead, the corresponding estimated runtimes are reported in Table~\ref{table:dv_runtime_photonics_0.9375}.

As shown, the runtime decreases substantially as the fusion-gate success probability increases. However, these estimates do not account for the substantial overhead required to generate the ancillary GHZ states. Current state-of-the-art experiments can produce 18-qubit GHZ states, but only at a rate of 0.2 Hz \cite{2018_GHZ_DV+photonics_exp}, corresponding to an effective success probability of $2.8\times10^{-9}$. When incorporated into the boosted-fusion protocol, the success probability of the ancillary state is effectively absorbed into the overall fusion-gate success probability, thereby dramatically increasing the total runtime.

\section{Problem-specific reduction by locality and symmetry}
\label{sec:problem_specific_lightcone}

The resource estimates in the previous sections were derived for the task
of implementing the full multi-mode circuit as a general quantum transformation.
Accordingly, those estimates were agnostic both to the particular input
state and to the specific output quantities to be measured. In other
words, the gate counts were obtained at the level of synthesizing the
entire evolution operator. However, in a problem-specific setting, once
the observables of interest are fixed, it may be possible to replace the
full model by a smaller effective model that reproduces the same outputs.

In the circuit of Fig.~\ref{fig:OriginalCircuit}, the relevant
simplification follows from back-propagating the observable in the
Heisenberg picture. For a local covariance element, such as
$\langle \hat X_i \hat X_j\rangle$, one can propagate the corresponding
operator backward through the short depth circuit; under this
procedure, its support expands only inside a bounded backward light cone,
rather than across the full system, see \cite{RevModPhys.84.621}. As a
result, the target covariance element is determined only by a clipped
subcircuit involving the modes inside this light cone, instead of the entire $N$-mode circuit.

For example, consider the correlation element
$\langle \hat X_4 \hat X_7\rangle$. To evaluate this quantity, it is not
necessary to simulate the full circuit in Fig.~\ref{fig:OriginalCircuit}.
Instead, one traces the measured modes $q_4$ and $q_8$ backward through
the circuit and retains only the gates and modes lying in their backward
light cone. The resulting reduced circuit contains exactly the part of
the dynamics that can affect the target observable. In the example shown
in Fig.~\ref{fig:problem_specific_lightcone}, this light cone contains
the eight modes $q_3,\ldots,q_{8}$. Therefore,
$\langle \hat X_4 \hat X_7\rangle$ can be obtained from the clipped
eight-mode circuit rather than from the full $N$-mode circuit, see Fig.~\ref{fig:problem_specific_lightcone} (c).

The light-cone argument above reduces the circuit size needed for any
fixed covariance-matrix element. A second simplification comes from the
symmetry of the circuit itself. In Fig.~\ref{fig:OriginalCircuit}, the
beam splitters are applied with identical parameters, while the final
displacement layer does not affect the covariance matrix. Therefore, for
the vacuum input relevant here, many entries of the final covariance
matrix $M$ are symmetry-equivalent.

It can be shown that only $K=5$ distinct covariance values are needed to
specify $M$ completely. Moreover, once $N\ge 3$, both the number $K$ and
the values of these five entries become independent of the total number
of qumodes $N$. Hence the full covariance matrix $M(N)$ can be
reconstructed from a small representative instance of the same
experiment, for example $N=3$, by evaluating these five values once and
placing them into the corresponding symmetry-related entries.

From this perspective, the present problem provides a useful benchmark for
future quantum simulation methods. The constant-coefficient advection
equation is simple enough to allow a transparent analysis, but it already
exposes several issues that are central to larger-scale quantum simulation:
state representation, hardware-native evolution, observable readout,
sampling cost, locality, and symmetry. The problem-specific reductions
identified here therefore do not diminish the significance of the full
resource comparison. This viewpoint is constructive for the development of quantum computation
as a whole.  In this sense, the present analysis has value beyond the
particular advection example: it identifies a practical route for designing
quantum simulations that are observable-driven, structure-aware, and
hardware-compatible.

\end{document}